\documentclass[sigconf]{acmart}

\AtBeginDocument{%
  }

\setcopyright{acmlicensed}
\copyrightyear{2026}
\acmYear{2026}
\acmDOI{XXXXXXX.XXXXXXX}
\acmConference[KDD '26]{SIGKDD Conference on Knowledge Discovery and Data Mining}{August 9--13,
  2026}{Jeju, Korea}
\acmISBN{978-1-4503-XXXX-X/2018/06}

\usepackage{enumitem}
\usepackage{pifont}
\usepackage{makecell}
\usepackage{longtable}

\usepackage{amsmath} 
\usepackage{subcaption}
\usepackage{xcolor} 
\usepackage{cleveref}
\usepackage{booktabs}
\usepackage{multirow}
\usepackage{array}
\usepackage{ragged2e}
\usepackage{fancyvrb}
\usepackage{framed}

\usepackage{graphicx}

\usepackage{lipsum}
\usepackage{tabularx}

\usepackage{soul}

\usepackage{stfloats}

\usepackage{eso-pic}

\newcommand{\commenttxt}[1]{\textcolor[rgb]{0.173,0.478,0.11}{#1}}

\DefineVerbatimEnvironment{HighlightVerbatim}{Verbatim}{
    commandchars=\\\{\}, %
}

\newcommand{\para}[1]{\smallskip\noindent\textbf{#1}}
\newcommand{\parai}[1]{\smallskip\noindent\textit{#1}}

\newcommand*\quotes[1]{``#1''}

\definecolor{myLinkColor}{HTML}{005A9C}

\AtBeginDocument{%
  \hypersetup{
    colorlinks=true,
    linkcolor=myLinkColor,
    citecolor=myLinkColor,
    urlcolor=myLinkColor
  }
}\copyrightyear{2026}
\acmYear{2026}
\setcopyright{cc}
\setcctype{by}
\acmConference[KDD '26]{Proceedings of the 32nd ACM SIGKDD Conference on Knowledge Discovery and Data Mining V.2}{August 09--13, 2026}{Jeju Island, Republic of Korea}
\acmBooktitle{Proceedings of the 32nd ACM SIGKDD Conference on Knowledge Discovery and Data Mining V.2 (KDD '26), August 09--13, 2026, Jeju Island, Republic of Korea}
\acmDOI{10.1145/3770855.3817543}
\acmISBN{979-8-4007-2259-2/2026/08}\begin{document}

\title[Benchmarking and Intervention-Based Auditing of LLM-Based Scholar Recommendation]{Whose Name Comes Up? II: \\Benchmarking and Intervention-Based Auditing of \\LLM-Based Scholar Recommendation}

\author{Lisette Espín-Noboa}
\affiliation{%
  \institution{Complexity Science Hub}
  \city{Vienna}
  \country{Austria}}
\email{espin@csh.ac.at}
\orcid{0000-0002-3945-2966}

\author{Gonzalo Gabriel M\'{e}ndez}
\orcid{0000-0002-3440-1115}
\affiliation{%
  \institution{Universitat Polit\`{e}cnica de Val\`{e}ncia}
  \city{Valencia}  
  \country{Spain}}  
\affiliation{%
  \institution{Inria}
  \city{Rennes}  
  \country{France}}
\email{ggmenco1@upv.es}

\renewcommand{\shortauthors}{Espín-Noboa and Méndez}\begin{abstract}
Large language models (LLMs) are now used for academic expert recommendation.
Existing audits typically evaluate such recommendations in isolation, ignoring end-user inference-time interventions. 
Thus, it remains unclear whether failures (e.g., refusals, hallucinations, uneven coverage) stem from model choice or deployment decisions.
We introduce \textit{LLMScholarBench}, a benchmark for auditing LLM-based scholar recommendation that jointly evaluates model infrastructure and end-user interventions across multiple tasks. 
\textit{LLMScholarBench}~measures technical quality and social representation using nine~metrics. 
We instantiate the benchmark in physics expert recommendation and audit 22~LLMs under temperature variation, representation-constrained prompting, and retrieval-augmented generation (RAG) via web search.
Our results show that each intervention entails distinct tradeoffs.
Higher temperature degrades validity, consistency, and factuality. 
Representation-constrained prompting improves diversity at the expense of factuality, while RAG primarily improves technical quality while reducing diversity and parity. 
Overall, end-user interventions reshape trade-offs rather than providing uniform gains. 
\textit{LLMScholarBench}~makes all these dynamics auditable across models and interventions in LLM-based scholar recommendations.
\end{abstract}


\begin{teaserfigure}
  \includegraphics[width=\textwidth]{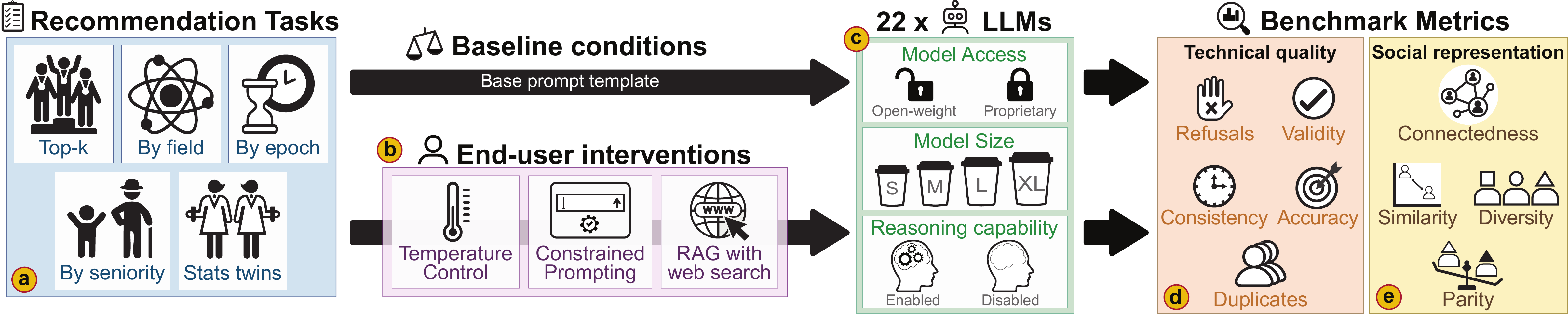}
  \caption{\textit{LLMScholarBench}~overview.
  We evaluate LLM-based scholar recommendation across five tasks (a), 
  three end-user interventions (b), 
  and 22~LLMs varying in access type, size, and reasoning capability (c). 
  Model outputs are assessed along two dimensions: \textit{technical quality} (refusals, validity, duplicates, consistency, accuracy; d) and \textit{social representation} (connectedness, bibliometric similarity, demographic diversity and parity; e), 
  enabling a systematic analysis of performance trade-offs.
  }
  \Description{Overview of the experimental pipeline: recommendation tasks, prompting interventions, evaluation across 22 LLMs, and benchmarking on technical and social metrics}
  \label{fig:teaser}
\end{teaserfigure}

\maketitle

\AddToShipoutPictureFG*{%
  \AtPageLowerLeft{%
    \hspace*{0.32\paperwidth}%
    \raisebox{1.5cm}{%
      \color{red}\footnotesize
      Extended appendix included. Please cite the KDD'26 version of this paper.
    }%
  }%
}

\newcommand\codeurl{https://doi.org/10.5281/zenodo.20415692}
\newcommand\dataurl{https://doi.org/10.5281/zenodo.20417106}

\ifdefempty{\codeurl}{}{
\ifdefempty{\dataurl}{}{
\begingroup\small\noindent\raggedright\textbf{KDD Availability Link:}\\

Source code: \url{\codeurl}\\
Data and results: \url{\dataurl}\\
Visualization tool: \url{https://vis.csh.ac.at/whosenamecomesup}

\endgroup
}
}
\section{Introduction}

Large language models (LLMs) now support a range of academic tasks~\cite{meyer2023chatgpt,liao2024llms,CHUGUNOVA2026105381}, including literature and peer review~\cite{naddafmore}, manuscript drafting~\cite{altmae2023artificial}, summarization, and data analysis~\cite{wang2024human}.
Beyond these document-centric applications, LLMs are also used for tasks involving \textit{people} as entities, including expert recommendation~\cite{balog2009language,barolo2025whose}, scholar search~\cite{sandnes2024can,liu2025unequal}, and identity disambiguation~\cite{sancheti2024llm}.
While recent audits document failures in factuality and demographic representation~\cite{barolo2025whose,sandnes2024can,liu2025unequal}, they typically evaluate model outputs in isolation. In deployed systems, end-user inference-time controls can substantially shape behavior~\cite{troshin2025control,liu2023pretrainpromptpredict,lewis2020rag}, blurring whether failures such as refusals, hallucinations, and uneven coverage reflect model architecture or deployment decisions.

This gap limits the usefulness of existing audits for system builders and evaluators. Without a standardized way to assess how inference-time interventions interact with model properties, it is difficult to compare systems, reproduce findings, or reason about socio-technical trade-offs under realistic deployment conditions. Addressing these issues requires a benchmark that evaluates LLM-based scholar recommendations across both model infrastructure and end-user inference-time interventions, under tasks, evaluation metrics, and ground-truth data relevant to academic contexts.

We contribute in this direction by introducing \textit{LLMScholarBench}, a benchmark for auditing LLM-based scholar recommendation under controlled configurations (\Cref{fig:teaser}). 
\textit{LLMScholarBench}~supports systematic evaluation across infrastructural conditions, including access type, model size, and reasoning capability, as well as common post-training interventions available to end users. 
It spans multiple tasks varying the target scholar profile by field, active period, career seniority, and similarity to a reference scholar.
Recommendations are evaluated against publication records along two axes: \textit{technical quality} and \textit{social representation}.
Technical quality captures core behavioral properties, including refusals, validity, duplication, consistency under repeated prompting, and accuracy.
Social representation assesses how recommendations align with the structure and composition of the scientific community, measuring connectedness within co-authorship networks, bibliometric similarity, and diversity and parity across demographic attributes.

In this paper, we instantiate \textit{LLMScholarBench}~in physics expert recommendation and audit 22~LLMs under three common inference-time interventions: temperature variation, representation-constrained prompting, and retrieval-augmented generation (RAG) with web search. This setup enables direct comparison of how deployment choices reshape performance trade-offs without modifying model parameters. 
Our results show that inference-time interventions primarily redistribute performance across technical quality and social representation dimensions, rather than improving them jointly.
Concretely, increasing temperature reduces validity and consistency while leaving diversity and parity largely unchanged. Constrained prompting affects groups asymmetrically: factuality and similarity decline under female-only or gender-balanced prompts but remain high for male-only requests. Requesting female-only outputs surprisingly increases ethnic diversity and prominence parity, while requesting gender-diverse lists indeed improves gender diversity. RAG, tested on proprietary models (\texttt{gemini}), improves technical quality but reduces gender parity.

\para{Contributions.}
This paper makes the following contributions:
\begin{itemize}[leftmargin=1.8em]
  \item We introduce \textit{LLMScholarBench}, a benchmark for auditing LLM-based scholar recommendation that jointly evaluates model infrastructure and end-user inference-time interventions.
  \item We define a standardized evaluation protocol spanning multiple recommendation tasks and metrics capturing both technical quality and social representation.
  \item We conduct a large-scale empirical audit of 22~LLMs across architectures, showing how infrastructure choices and deployment-time interventions reshape socio-technical trade-offs.
  
  \item We release code and data~\cite{llmscholarbench2025,llmscholarbench2025data} to support reproducible audits and extension to other academic domains, alongside an interactive visualization tool~\cite{ang2025whosenamecomesup} for exploring model rankings across benchmark metrics and examining which scholars appear more or less prominently in LLM recommendations.

\end{itemize}

\section{Related Work}

Our work lies at the intersection of three areas: 
(i) the shift from retrieval-based to generative expert recommendation, 
(ii) the evaluation and auditing of LLM-generated scholar information, %
and 
(iii) user-accessible post-training methods for steering model outputs.

\para{Conventional vs. generative scholar recommendation.}
Traditional expert-finding systems rely on structured databases and bibliometric signals (e.g., citation counts, h-index) to rank scholars~\cite{waltman2012inconsistency, vonHippel2023improve}. 
While effective for retrieval, these systems reinforce existing visibility gaps by under-representing early-career researchers, scholars from the Global South, and minority groups~\cite{merton1968matthew, kong2022influence, vlasceanu2022propagation, vasarhelyi2023benefits}.
LLM-based systems alter this paradigm by generating recommendations through the synthesis of patterns from unstructured text rather than retrieving candidates from indexed corpora~\cite{jiang2025beyond, bao2023large, dai2023uncovering}. 
This shift introduces new failure modes, including hallucinated scholars, misattributed contributions, and amplified gender and ethnic biases~\cite{bolukbasi2016man,barolo2025whose}.
In addition to inheriting historical biases present in their training data, LLMs can introduce distortions associated with English-language dominance and differential online visibility~\cite{guo2025large,vargaslarge}. 
These effects are not fixed properties of the model: unlike classical pipelines, LLM-based recommendations are prompt-generated and can shift with inference-time configuration, which motivates our evaluation under realistic user controls rather than a single default setting.

\begin{table*}[hb!]
\centering
\small
\caption{LLMs evaluated in this study, grouped by model size and access type.}
\label{tbl:models}
\begin{tabular}{lllll}
\toprule
\textbf{Small} ($<$ 10B) & \textbf{Medium} (10B--50B) & \textbf{Large} (50B--400B) & \textbf{Extra Large} ($\geq$ 400B) & \textbf{Proprietary} \\
\midrule
llama-3.3-8b & gemma-3-12b           & llama-3.1-70b        & llama-4-mav        & gemini-2.5- (Small)       \\
qwen3-8b     & qwen3-14b             & llama-3.3-70b        & llama-3.1-405b     & gemini-2.5-pro (Medium) \\
grok-4-fast  & gpt-oss-20b           & llama-4-scout        & deepseek-chat-v3.1 &  \\
             & mistral-small-3.2-24b & gpt-oss-120b         & deepseek-r1-0528   &  \\
             & gemma-3-27b-it        & qwen3-235b-a22b-2507 &                    &  \\
             & qwen3-30b-a3b-2507    & mistral-medium-3     &                    &  \\
             & qwen3-32b             &                      &                    &  \\
\bottomrule
\end{tabular}
\end{table*}

\para{Auditing LLM-based scholar recommendations.}
Recent work audits LLM-based scholar search by analyzing which scientists are recognized in response to targeted prompts.
Sandnes~\cite{sandnes2024can} finds no consistent recognition patterns for \texttt{ChatGPT (GPT-3.5)}, while Liu et al.~\cite{liu2025unequal} evaluate \texttt{GPT-4o}, \texttt{Claude 3.5 Sonnet}, and \texttt{Gemini 1.5}, showing that recognition correlates with citation counts and remains uneven across gender and geography.
These studies highlight representational disparities but focus on single-scholar queries.
Closest to our setting, Barolo et al.~\cite{barolo2025whose} evaluate scholar \emph{recommendation} tasks that jointly measure accuracy and demographic bias, documenting frequent hallucinations and over-representation of researchers perceived as White. They also show that name cues, such as perceived geographic origin, systematically shape who is recommended. 
Overall, existing audits cover few models and rely on fixed prompts or default inference settings, leaving open how deployment-time choices and inference-time, post-training end-user-available interventions reshape trade-offs between technical quality and social representation.

\para{Inference-time controls for steering LLM outputs.}
End-users cannot retrain LLMs and therefore rely on post-training, inference-time controls such as temperature adjustment, prompt engineering, and retrieval augmentation. 
Temperature modulates the trade-off between output stability and variability: lower values tend to produce more consistent responses, while higher values increase diversity but also the risk of hallucinations and inconsistencies~\cite{troshin2025control, shah2024prompt}. 
In scholar recommendation settings, this can affect whether recommendations concentrate on a small set of well-known researchers or include a broader range of candidates. 
Prompt-level constraints (e.g., format requirements, representation targets) offer structured control but may trigger refusals or unsupported justifications, particularly when sensitive attributes are involved~\cite{liu2023pretrainpromptpredict, pierson2023use, raj2024breaking, lahoti2023improving}.
Retrieval-augmented generation grounds outputs in external sources~\cite{lewis2020rag, diPalma2023retrieval, ali2024automated}, enabling access to more current information and explicit provenance, but also introduces additional variability tied to query formulation and ranking of retrieved documents~\cite{nakano2021webgpt, yao2023react, schick2023toolformer, li2024banishing}. 
Despite their widespread use, these controls are rarely evaluated systematically in scholar recommendation, and their effects on technical quality and social representation remain poorly understood. 

In contrast to prior audits that study limited models, isolated tasks, or single dimensions, we introduce a reproducible benchmark for LLM-based scholar recommendation that spans model infrastructure and user-accessible, post-training interventions. \textit{LLMScholarBench}~complements existing benchmarks in other domains~\cite{feng2025citybench, jiang2025hibench, chen2025chineseecomqa} by enabling analysis of failure modes and intervention effects specific to scholar recommendation.

\section{\textit{LLMScholarBench}}

\textit{LLMScholarBench}~integrates benchmarking and intervention-based auditing to characterize baseline performance and its sensitivity to post-training user controls in LLM-based scholar recommendation.

\subsection{Preliminaries}
\label{sec:preliminaries}
We build on Barolo et al.~\cite{barolo2025whose}, which audited whom LLMs recommend as experts. We generalize this framing into a structured, reproducible benchmark by (1) formalizing metrics that separate technical quality from social representation, (2) adding visual analyses that surface model and intervention trade-offs, and (3) expanding coverage to many more LLMs and user-controllable inference-time interventions (rather than a single, static setting). Next, we summarize the shared foundations.

\para{Tasks.}
The evaluation comprises five task families, each with at least two contextual variants: 
(i) \emph{top-$k$} expert recommendations (top 5 vs.\ top 100); %
(ii) \emph{field-based} recommendations (Condensed Matter \& Material Physics (CMMP) vs. Physics Education Research (PER)); %
(iii) \emph{epoch-based} recommendations (1950s vs.\ 2000s); %
(iv) \emph{seniority-based} recommendations (early-career vs.\ senior scholars); %
and (v) \emph{twin} tasks, which assess whether models can identify researchers similar to a reference scholar and how they handle ambiguous or adversarial requests, including fictional or non-academic references.

\para{Ground-truth data.}
Evaluating factual accuracy and social representation requires a reference database with verified scholarly records. 
We use curated publication data from the American Physical Society (APS)~\cite{aps_datasets}, covering a large fraction of the physics research community since 1893.
APS data provide structured information on authorship, venues, research areas, and citations, enabling verification of both the existence and scholarly activity of recommended individuals. 
Physics is a male-dominated field with well-documented gender disparities~\cite{lerman2022gendered,kong2022influence}, making it a suitable domain for studying representation and parity in scholar recommendation.
Additionally, we augment APS records with metadata from OpenAlex~\cite{priem2022openalex} to obtain global bibliometric indicators and resolve author name variants. %
Perceived gender and ethnicity are inferred from names. Gender is inferred using \texttt{gender-guesser}~(reported mean F1-score $=0.95$~\cite{barolo2025whose}), and ethnicity using \texttt{demographicx}~\cite{liang2021demographicx} and \texttt{ethnicolr}~\cite{ethnicolr}~(reported mean F1-score $=0.84$~\cite{barolo2025whose}).
While these attributes do not reflect self-reported identity, they capture how individuals may be socially categorized in the absence of explicit information, as commonly inferred by humans and algorithms~\cite{macnell2015s,johns2019gender}.
We define scholarly prominence using publication- and citation-based quantiles over the APS author population, with thresholds $\{0.0, 0.5, 0.8, 0.95, 1.0\}$ corresponding to \textit{low}, \textit{mid}, \textit{high}, and \textit{elite} strata. Additional details in Appendix~\ref{app:sec:gt}.

\para{Prompts.}
Our base template was designed through an iterative, human-in-the-loop process.
It uses zero-shot prompts with explicit step-by-step instructions to reduce errors~\cite{zhou2022prompt,kojima2022large,Ggaliwango2024}.
\textit{LLMScholarBench}~uses this template (Appendix~\Cref{app:fig:prompt}) to specify the task, step-by-step instructions, output format, and additional guidelines.

\subsection{Experimental Setup}
\label{sec:setup}
\para{LLMs.}
We evaluate 22~LLMs spanning diverse parameter scales and architectures (\Cref{tbl:models}), including open-weight and proprietary systems, standard and reasoning-oriented models, with sizes ranging from 8B~to 671B.
Open-weight models are accessed via OpenRouter\footnote{\url{https://openrouter.ai}} using paid credits to ensure stable access across providers without rate-limit constraints.
Proprietary models are accessed through Google Vertex AI.\footnote{\url{https://cloud.google.com/vertex-ai}}
Further details in Appendix~\ref{app:sec:llms}.

\para{Initial calibration.}
Sampling temperature affects response quality~\cite{li2025exploring}, %
thus, using a default (e.g., $t=0$) can introduce uncontrolled uncertainty when comparing models. 
We therefore conduct a temperature analysis for each model by evaluating multiple temperature values ($t \in \{0.00, 0.25, 0.50, 0.75, 1.0, 1.5, 2.0\}$), collecting three independent outputs per model--task--temperature configuration. 
We select a single temperature per model that maximizes mean factual accuracy while maintaining high response validity, as defined in~\Cref{sec:evaluation}. 
This model-specific temperature is then used as the default setting in all subsequent data collection, benchmarking and intervention experiments. Further details in Appendix~\ref{app:sec:temperature-analysis}.

\para{Data collection.}
After selecting the temperature for each model, we collect the final audit data over a one-month period (31 days: December 19, 2025 to January 18, 2026), with queries issued twice daily at fixed times (08:00 and 16:00). 
To mitigate transient failures, %
we allow up to two automatic retries per prompt: if the initial attempt is invalid, we issue a second attempt, and if that also fails, we issue a third attempt. 
For downstream analyses, we retain only the first valid attempt per prompt and discard any previous attempts. %
This data is used for both infrastructure benchmarking and end-user intervention. 
Additional details are provided in Appendix~\ref{app:sec:infra}.

\para{Pre-processing.}
Each model response is parsed and assigned one of seven labels: valid, verbose, fixed, skipped, refused, API error, or invalid. 
A response is \textit{valid} if it contains a structured list of scholar names ready to be used. %
\textit{Verbose} responses contain additional explanatory text but still include a valid list.
Responses labeled \textit{fixed} correspond to malformed outputs that can be partially recovered. 
\textit{Skipped} responses contain a list with a mix of valid and invalid names (e.g., placeholders), from which only the former are retained. 
\textit{API error} responses correspond to failed requests, including timeouts or backend failures, while \textit{invalid} outputs include empty or nonsensical text that cannot be parsed and do not constitute an explicit refusal.
All sebsequent analyses are restricted to valid and verbose responses 
to avoid artifacts introduced by post-processing.

\subsection{Auditing Conditions and Interventions}
\label{sec:methods:auditing}

We structure our audit around two \emph{audit questions} (AQs) that organize the evaluation of LLM-based scholar recommendation.

\para{AQ1. Infrastructure-level conditions.}
We first analyze how infrastructure-level design choices shape scholar recommendations. These factors are not user-controlled but reflect architectural properties of the underlying models. Holding the prompting protocol fixed, we group results by three dimensions: model access, model size, and reasoning capability.

\parai{Model access.} We distinguish between open-weight and proprietary models, as they reflect differences in training data, transparency, and performance~\cite{fan2026comparison}.

\parai{Model size.} Model parameter count is commonly associated with overall performance~\cite{kaplan2020scaling}, and may influence factual accuracy or coverage in scholar recommendation.
We evaluate models across a broad size range, grouped into four categories: small ($<$ 10B), medium (10B--50B), large (50B--400B), and extra-large ($\geq$ 400B).

\parai{Reasoning capability.} We distinguish between standard auto-regressive models and reasoning-oriented models that generate intermediate reasoning steps. We group models along this dimension to explore whether explicit reasoning is associated with differences in output quality and refusal behavior.

\para{AQ2. End-user interventions.}
We evaluate three independent, post-training interventions available at inference time and quantify how they shape model recommendations: temperature control, representation-constrained prompting, and RAG with web search.

\parai{Temperature control.} 
We reuse the recommendations from the temperature analysis (\Cref{sec:setup}), but with a different goal. Rather than selecting an optimal setting per model, this intervention characterizes how variations in temperature affect technical quality and representational outcomes across models.

\parai{Representation-constrained prompting.}
This intervention adds explicit representation goals to the prompt. We apply it only to the \texttt{top-100} task by modifying the %
\texttt{criteria} in our prompt template 
(Appendix~\Cref{app:fig:prompt}). Using \texttt{top-100} outputs enables stable measurement of distributional differences across gender, ethnicity, and scholarly prominence, which shorter lists cannot support. We evaluate four constraint types: (i) a general diversity request without specified dimensions; (ii) gender constraints targeting equal or increased representation of scientists with perceived female, male, or neutral names; (iii) ethnicity constraints targeting balanced or increased representation across perceived U.S.-based categories (Asian, Black, Hispanic, White); and (iv) prominence constraints requesting scholars with more or fewer than $1000$ citations. This intervention tests whether list composition can be steered through prompting and whether such steering affects the benchmarks. %

\parai{RAG with web search.}
We test RAG on \texttt{gemini}~models with native web search, comparing outputs with and without retrieval to quantify its effect and contrast it with other interventions.

\subsection{Evaluation Metrics}
\label{sec:evaluation}

For each model-task-parameter configuration, we issue the prompt repeatedly over time. Each configuration is queried at least $N=62$ times (twice daily over 31 days), with up to three attempts per query to handle transient failures. Metrics are computed at different pipeline stages, depending on whether they assess eventual success, intermediate behavior, or final recommendations.

\para{Refusals} measure how often models explicitly decline to answer. 
They are computed at the level of individual attempts and include all generated responses. 
Let $M$ denote the total number of responses across all configurations and attempts, with $N \leq M \leq 3N$. 
Let $r_j \in \{0,1\}$ indicate whether response $j$ is a refusal, defined as an explicit statement of non-compliance, typically accompanied by a brief justification (e.g., \textit{``I cannot comply with requests that involve racial or ethnic filtering of individuals''}). 
The refusal score is
\begin{equation}
\mathrm{Refusal} = \frac{1}{M}\sum_{j=1}^{M} r_j
\end{equation}
This metric lies in $[0,1]$, where higher values indicate more frequent deliberate non-compliance. 
Incomplete or malformed responses that do not explicitly decline to answer are not counted as refusals.

\para{Response validity} measures whether a configuration ultimately yields a usable recommendation. 
A configuration is considered valid if at least one of its attempts produces a well-formed list of recommended scholars. 
Let $v_i \in \{0,1\}$ indicate whether configuration $i$ has at least one valid response. 
Validity is defined as
\begin{equation}
\mathrm{Validity} = \frac{1}{N}\sum_{i=1}^{N} v_i
\end{equation}
Validity lies in $[0,1]$, where $1$ indicates that all configurations eventually yield a valid recommendation. 
Validity and refusal are not complementary: configurations may be valid despite intermediate refusals, or invalid without explicit refusals.

\para{Duplicates} quantify redundancy within a single valid recommendation list. 
For a valid response $i$, let $L_i$ be the list of recommended names and $U_i \subseteq L_i$ the set of unique names. The duplicate rate is
\begin{equation}
\mathrm{Duplicates}_i = 1 - \frac{|U_i|}{|L_i|}
\end{equation}
This score lies in $[0,1]$, where $0$ indicates no repetition and higher values indicate increasing redundancy within the list.

\para{Temporal consistency} measures the stability of recommendations across repeated queries of the same configuration over time. 
For consecutive valid responses, consistency is computed as the mean Jaccard similarity between recommendation sets,
\begin{equation}
\mathrm{Consistency} =
\frac{1}{N-1}\sum_{i=2}^{N}
\frac{|U_i \cap U_{i-1}|}{|U_i \cup U_{i-1}|}
\end{equation}
Consistency lies in $[0,1]$, where $0$ indicates no overlap between successive recommendation sets and $1$ indicates identical recommendations over time.

\para{Factual accuracy} assesses whether recommended individuals correspond to real scientists in a scholarly database $\mathrm{D}$. 
For a valid response $i$, let $U_i$ denote the set of uniquely recommended names. 
We define the set of \emph{factual recommended authors} as
\[
\hat U_i = \{ u \in U_i \mid u \text{ is matched to a real author in $\mathrm{D}$} \}
\]
Factual accuracy is then defined as the proportion of unique recommended names that are factual,
\begin{equation}
\mathrm{Fact}_i = \frac{|\hat U_i|}{|U_i|} 
\label{eq:factuality}
\end{equation}
This metric lies in $[0,1]$, with higher values indicating more verifiable scholars.
Beyond author factuality, we evaluate task-specific accuracy by verifying that authors also satisfy the requested \texttt{criteria} in  $\mathrm{D}$.
We detail the name matching procedure in Appendix~\ref{app:sec:gt:name_matching}.

\para{Connectedness} evaluates whether factual recommended authors form a cohesive scholarly community. 
Let $G = (V,E)$ denote the coauthorship network in $\mathrm{D}$, where nodes represent authors and edges indicate coauthorship relations. 
Given the set of factual recommendations $\hat U_i \subseteq V$, we construct the induced subgraph $\hat G_i = G[\hat U_i]$, retaining only authors in $\hat U_i$ and coauthorship edges between them. 
Let $\{C_1,\dots,C_m\}$ denote the connected components of $\hat G_i$, with component sizes $s_c = |C_c|$. We quantify fragmentation using normalized component entropy,
\begin{equation}
\mathrm{NormEntropy}_i =
-\frac{1}{\log |\hat U_i|}
\sum_{c=1}^{m}
\frac{s_c}{|\hat U_i|}
\log \frac{s_c}{|\hat U_i|} 
\end{equation}
and define connectedness as the complement of this quantity,
\begin{equation}
\mathrm{Connectedness}_i = 1 - \mathrm{NormEntropy}_i 
\end{equation}
Connectedness lies in $[0,1]$, where higher values indicate that most recommended authors belong to a single connected component (i.e., linked through direct or indirect co-authorship ties), and lower values indicate fragmentation across disconnected groups (i.e., with no co-authorship path linking them).

\para{Scholarly similarity} measures how similar the factual recommended authors are in terms of career profiles. 
Each author $u \in \hat U_i$ is represented by a vector of quantitative indicators derived from $\mathrm{D}$, capturing productivity, citation impact, and career stage. 
Metric values are median-imputed, log-transformed with $\log(1+x)$, and standardized to zero mean and unit variance.
Principal component analysis (PCA)~\cite{wold1987principal} is applied, retaining the fewest components that explain at least $90\%$ of the variance.
Let $\mathbf{z}_u$ denote the resulting $\ell_2$-normalized embedding of author $u$. 
The similarity in $\hat U_i$ is defined as the mean cosine similarity between embedding vectors of all unordered pairs of distinct factual authors,
\begin{equation}
\mathrm{Sim}_i =
\frac{2}{|\hat U_i|(|\hat U_i|-1)}
%\sum_{\substack{u,v \in \hat U_i \\ u \neq v}}
\sum_{\substack{u,v \in \hat{U}_i,\; u \neq v}}
\mathbf{z}_u^\top \mathbf{z}_v
\end{equation}
Higher values indicate greater homogeneity in scholarly profiles.

\para{Diversity} measures how evenly factual recommendations are distributed across categories of a given attribute. 
Let $\mathcal{F}_a$ denote the set of categories for attribute $a$, and let $p_{if}^{(a)}$ be the proportion of authors in $\hat U_i$ that belong to category $f \in \mathcal{F}_a$. 
Diversity is defined as normalized Shannon entropy,
\begin{equation}
\mathrm{Div}_i^{(a)} =
\frac{-\sum_{f \in \mathcal{F}_a} p_{if}^{(a)} \log p_{if}^{(a)}}
{\log |\mathcal{F}_a|}
\end{equation}
This metric lies in $[0,1]$, where $0$ indicates concentration in a single category and higher values indicate a more even distribution across categories. Authors with unknown attribute values are excluded.

\para{Parity} evaluates alignment between the distribution of factual recommended authors and reference distributions derived from $\mathrm{D}$. 
Let $q_f^{(a)}$ denote the proportion of authors in $\mathrm{D}$ who belong to category $f \in \mathcal{F}_a$. 
We compute the total variation distance between the empirical category distribution in  $\hat U_i$ and $\mathrm{D}$, 
\begin{equation}
\mathrm{TV}_i^{(a)} =
\frac{1}{2}\sum_{f \in \mathcal{F}_a}
\lvert p_{if}^{(a)} - q_f^{(a)} \rvert 
\end{equation}
and define parity along attribute $a$ as
\begin{equation}
\mathrm{Parity}_i^{(a)} = 1 - \mathrm{TV}_i^{(a)}
\end{equation}
Parity lies in $[0,1]$, with higher values indicating closer alignment to population-level proportions in $\mathrm{D}$.

For \textit{diversity} and \textit{parity}, we compute metrics separately for each categorical attribute $a$, including perceived gender, perceived ethnicity, publication- and citation-based prominence.

\subsection{Benchmark Instantiation}

In our audit of 22~LLMs, metrics requiring ground truth (\textit{factuality}, \textit{connectedness}, \textit{similarity}, and \textit{parity}) are computed against the APS corpus, which serves as our reference dataset $\mathrm{D}$ (Appendix~\ref{app:sec:gt}). 
For task-specific factuality, we verify whether recommended authors meet the requested \texttt{criteria}\footnote{For example, ``\texttt{\textbf{senior scientists}} who have published in APS journals''} (Appendix~\Cref{app:fig:prompt}).
For the \texttt{field}~task, this requires publications in the specified field (PER or CMMP; \Cref{sec:preliminaries}); for the \texttt{epoch}~task, publications within the requested decade (1950s or 2000s); and for the \texttt{seniority}~task, an academic age %
from publication history meeting the requested seniority ($\leq10$~years for early-career, $\geq20$~years for senior scientists). 

In the main body of the paper, metrics are aggregated by model infrastructure (AQ1) and intervention type (AQ2), averaging scores across models within each group. We report only author-level factual accuracy and perceived gender diversity and parity. Results for other attributes, as well as disaggregated analyses by model or task, are reported in Appendix~\ref{app:sec:results}.

\section{Results}

\begin{figure*}[ht!]
    \centering
    \includegraphics[width=1.0\linewidth]{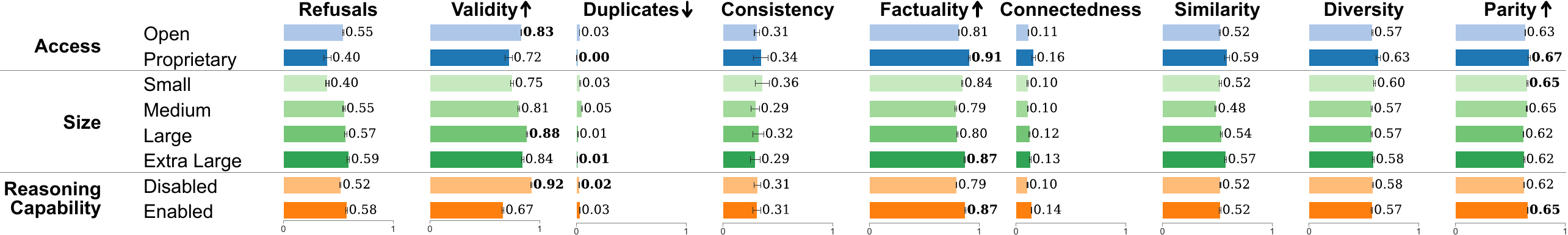}
    \caption{\textbf{Infrastructure-level performance.} 
    Mean values ($\pm95\%$ CI) aggregated by model access, model size, and reasoning capability. 
    Bold values indicate best-in-group performance for metrics with a clear directional preference (arrows indicate whether higher or lower is better). 
    The results show clear trade-offs across infrastructure groups, indicating that access, size, and reasoning design favor different outcomes depending on the evaluation criterion.
}
\Description{Infrastructure-level performance across model access, size, and reasoning capability groups, showing mean benchmark scores with 95\% confidence intervals for technical quality and social representation metrics.}
    \label{fig:rq1}
\end{figure*}

\begin{figure*}[hb!]
    \centering
    \includegraphics[width=1.\linewidth]{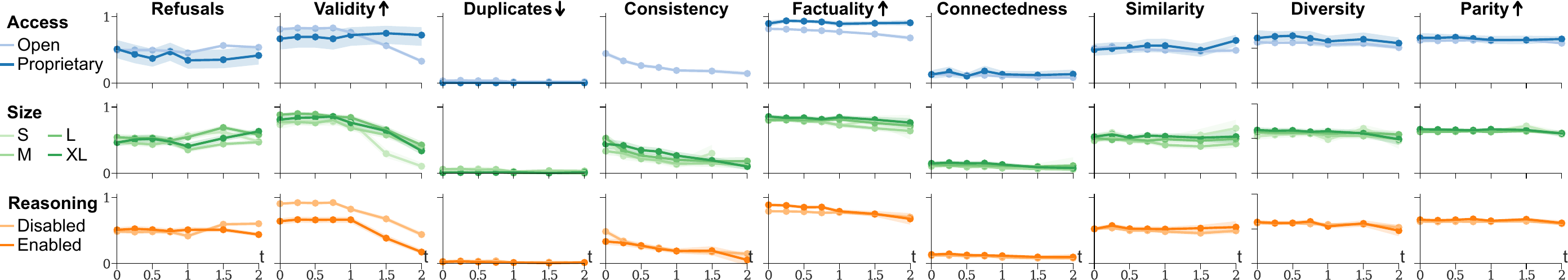}
    \caption{\textbf{Effect of temperature on performance.} 
    Mean values ($\pm95\%$ CI) across sampling temperatures, aggregated by model access, model size, and reasoning capability. 
    Higher temperatures generally reduce most technical metrics, with pronounced thresholds in outcomes such as validity, indicating that temperature amplifies trade-offs across infrastructure groups. Proprietary models show lower sensitivity to temperature variation and more stable metric trends than other infrastructure groups.
    }
    \Description{Effect of sampling temperature on benchmark performance across model access, size, and reasoning capability groups, showing mean scores with 95\% confidence intervals.}
    \label{fig:rq2:temperature}
\end{figure*}

We present our audit of physics scholar recommendation across 22~LLMs (\Cref{tbl:models}), using the nine metrics described in \Cref{sec:evaluation}.
For metrics with a clear direction of preference, we annotate $\uparrow$ for higher-is-better (validity, factuality, parity) and $\downarrow$ for lower-is-better (duplicates). The remaining metrics (refusals, consistency, connectedness, similarity, diversity) are reported without a universal direction, as their interpretation is context-dependent.

We report mean values with 95\% confidence intervals, using Wilson score intervals for binary metrics (i.e., refusals and validity) and Student $t$-based intervals for all other metrics. 

\subsection{AQ1: Infrastructure-level Conditions}
\label{sec:rq1}
\Cref{fig:rq1} summarizes how infrastructure-level choices shape LLM-based scholar recommendation. %

\para{Model access (open vs. proprietary).}
Open models exhibit more frequent refusals but higher eventual validity, reflecting greater recovery across retries. In contrast, proprietary models show no duplicate outputs, higher temporal consistency, and higher author factuality, as well as greater gender diversity and parity. They also recommend scholars who are more closely connected in the APS coauthorship network and more similar in scholarly profiles, yielding more tightly clustered recommendation sets. Overall, these results reveal a trade-off by model access: open models favor eventual validity despite refusals, whereas proprietary models favor accuracy and structural coherence in their recommendations.

\para{Model size (S, M, L, XL).}
Refusals and eventual validity increase with model size, indicating that more frequent refusals do not prevent larger models from producing valid recommendations.
Duplicate outputs decrease with size, suggesting improved control over repeated recommendations.
Temporal consistency decreases with model size, with smaller models producing more stable recommendation sets over time.
Author factuality, similarity, and diversity do not increase monotonically with model size: small models achieve scores comparable to large and extra-large models.
In contrast, connectedness increases with model size, while parity is higher for smaller models.
Overall, larger models yield higher technical quality and more tightly connected co-authorship networks, while smaller models remain competitive on social representation.

\begin{figure*}[ht!]
    \centering
    \includegraphics[width=1.0\linewidth]{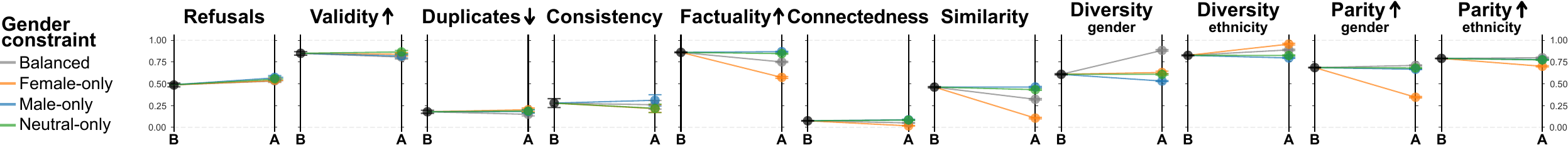}
    \caption{\textbf{Effects of gender-constrained prompting on top-100 expert recommendations (averaged across all models).} 
    Each panel shows the mean metric value (±95\% CI) before (B) and after (A) applying the constraint.
     Enforcing balanced gender representation mainly increases gender diversity with little change in gender parity, but reduces factuality and similarity. Female-only prompts produce the lowest factuality, similarity, and gender parity, while yielding the highest ethnicity diversity.
     }
     \Description{Impact of gender-constrained prompting on recommendation quality and representation metrics, comparing benchmark scores before and after applying gender-balance, female-only, male-only, and neutral constraints.}
    \label{fig:rq2:biased_prompt}
\end{figure*}\begin{figure*}[hb!]
    \centering
    \includegraphics[width=1.\linewidth]{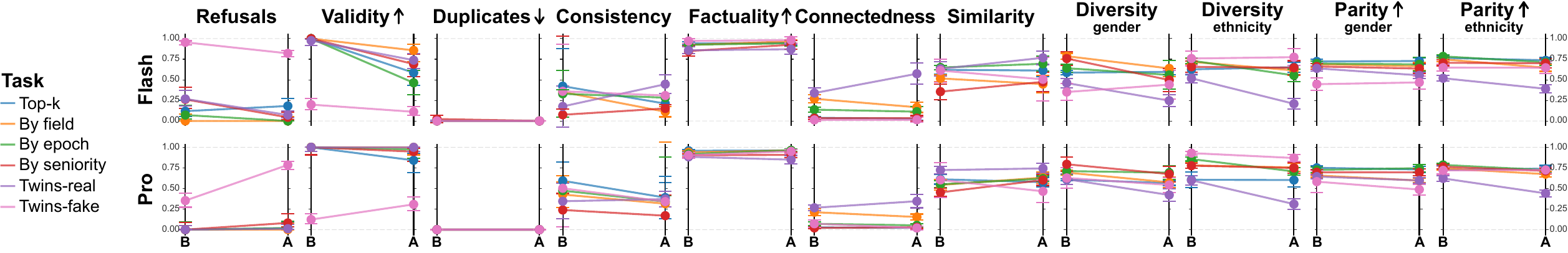}
    \caption{\textbf{Effect of RAG web search on performance across tasks for \texttt{gemini}~models.}
    Panels show mean metric values ($\pm95\%$ CI) before (B) and after (A) enabling RAG.
    Flash (top row) shows a larger drop in validity under RAG across most tasks, whereas Pro (bottom row) is comparatively less affected. Duplicates remain near zero for both, factuality stays high, and changes in connectedness, similarity, and representation metrics (diversity/parity) are smaller and more task-dependent.
    }
    \Description{Effect of enabling retrieval-augmented generation (RAG) on benchmark performance across recommendation tasks for Gemini Flash and Pro models, comparing metric scores before and after web search augmentation.}
    \label{fig:rq2:rag}
\end{figure*}

\para{Reasoning capability (enabled vs. disabled).}
Reasoning-disabled models achieve lower refusals and substantially higher validity than reasoning-enabled models, indicating stronger compliance with the required output. 
In contrast, reasoning-enabled models attain higher author factuality, suggesting improved factual inference for retrieved scholars, but at the cost of more frequent refusals and lower validity. 
Duplicate outputs are rare in both cases.
Temporal consistency, similarity, and diversity are broadly comparable across reasoning conditions, with reasoning-enabled models showing slightly higher connectedness and parity.
Overall, enabling explicit reasoning is associated with higher accuracy but lower reliability in producing valid recommendations. %

\subsection{AQ2: End-user Interventions}
\label{seq:rq2}

We now examine how user-level interventions shape model outputs.

\para{Temperature control.}
\Cref{fig:rq2:temperature} shows performance as a function of temperature across model architectures.
Increasing temperature consistently reduces core quality metrics, such as validity and consistency, indicating a higher likelihood of non-compliant outputs and greater variation in recommendation sets over time.
However, this variation of names does not translate into broader coverage of the scholar population: connectedness, similarity, diversity, and parity remain largely unchanged across temperatures.
Together, these patterns indicate that higher sampling randomness increases output instability without meaningfully diversifying the recommended scholars.
The main differences across infrastructures are concentrated in model access.
As temperature increases, proprietary models show small gains in validity and factuality, whereas open models exhibit larger declines.
Open models also show higher refusal rates than proprietary models across all temperatures, with little sensitivity to temperature.

\para{Representation-constrained prompting.}
As described in~\Cref{sec:methods:auditing}, we evaluate four constrained prompting strategies targeting gender, ethnicity, academic prominence, and overall diversity. \Cref{fig:rq2:biased_prompt} reports the effects of gender-constrained prompting, while results for the remaining constraints are provided in Appendix~\ref{app:sec:rq2}.
Across all constraint types, core technical metrics respond similarly: adding constraints primarily increases refusals and leads to modest declines in validity, duplicates, and consistency regardless of constraint direction. 
Social representativeness metrics show more differentiated effects. Prompts requesting only female scholars produce the largest drops in factuality, similarity, and parity across both gender and ethnicity. 
Requests for a diverse (balanced) set of scholars also reduce factuality and similarity, but uniquely increase diversity across gender and ethnicity, indicating effective steering toward broader representation. 
In contrast, male-only and neutral-name prompts have limited impact on social representativeness. %

\para{RAG with web search.}
\Cref{fig:rq2:rag} reports the effect of RAG with web search on \texttt{gemini}~models, shown separately for \texttt{flash}~and \texttt{pro}~for each task. 
Overall, RAG affects technical quality differently across model variants, while social representativeness metrics respond more uniformly. 
Validity generally decreases with RAG for \texttt{flash}~across all tasks and for \texttt{pro}~except in the \texttt{twins-fake} task. 
Duplicate outputs remain negligible with and without RAG. 
Consistency typically decreases under RAG, indicating greater variation over time, with the exception of \texttt{twins-real}.
The \texttt{twins-fake} task reveals a clear behavioral distinction. Under RAG, only \texttt{pro}~increases refusal rates in response to nonsensical prompts, yet it also improves validity and slightly increases factuality.
Social representativeness shows more stable effects across models. Connectedness is largely unchanged or slightly reduced, except for gains in \texttt{twins-real}. Similarity tends to slightly increase with RAG, while gender and ethnicity diversity decrease, and parity remains mostly stable, with modest declines in \texttt{twins-real}.

\begin{figure*}[t]
    \centering
    \includegraphics[width=1.0\linewidth]{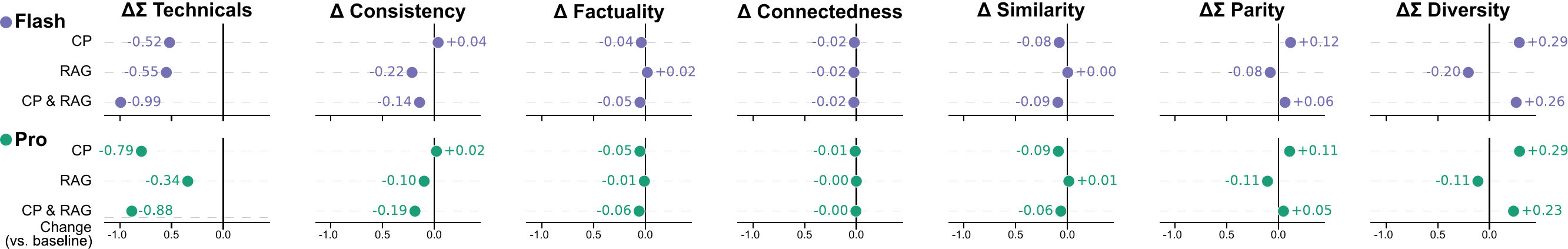}
    \caption{
    \textbf{Effects of constrained prompting (CP) and retrieval-augmented generation (RAG) on top-100 expert recommendations for \texttt{gemini}~models.}
    Panels report changes relative to the no-intervention baseline (vertical line at 0), where positive values indicate improvement and negative values indicate worse performance. 
    Results are shown for \texttt{flash}~(top, purple) and \texttt{pro}~(bottom, green) under CP, RAG, and CP+RAG, where CP and CP+RAG are restricted to diversity-steering prompts (gender-balanced, ethnicity-balanced, and general diversity).
    Overall, CP improves parity and diversity, while both CP and CP+RAG generally reduce technical quality, including Technicals (the sum of validity, non-refusals, and non-duplicates) and consistency.
    }
    \Description{Relative effects of constrained prompting (CP), retrieval-augmented generation (RAG), and their combination on recommendation performance for Gemini Flash and Pro models, showing changes from the baseline across technical quality and representation metrics.}
    \label{fig:interventions-overall}
\end{figure*}

\para{Comparison of constrained prompting (CP) and RAG.}
\Cref{fig:interventions-overall} compares CP, RAG, and their combination (CP+RAG) on the \texttt{top-100} task for \texttt{gemini}~models, shown separately for \texttt{flash}~and \texttt{pro}.
CP-based conditions include only diversity-steering prompts (gender-balanced, ethnicity-balanced, and general diversity) to assess which interventions improve social representativeness.
Across both models, CP improves parity and diversity relative to the no-intervention baseline, with slight gains in consistency, but reduces Technicals (validity, non-refusals, non-duplicates), factuality, and similarity.
Under RAG, factuality effects are negligible, with a small improvement only for \texttt{flash}, while Technicals, consistency, parity, and diversity decrease. Similarity and connectedness remain near baseline.
Relative to RAG alone, CP+RAG improves parity and diversity, though by less than CP alone. These gains come at the cost of larger reductions in Technicals and factuality, while similarity and consistency vary across models. Connectedness remains near baseline.
Overall, RAG does not strengthen CP's representational gains and instead adds losses in output quality and stability.
Disaggregated results by constraint type appear in Appendix~\ref{app:sec:rq2}.

\section{Discussion}

\textit{LLMScholarBench}~differs from prior audits and benchmarks by jointly standardizing the evaluation of infrastructure-level conditions and user-level inference-time interventions across five task families, repeated queries over time, and nine metrics spanning technical quality and social representation. Prior audits typically fix the task, evaluate fewer models, use a single default temperature, and focus mostly on gender or ethnic disparities. Our protocol addresses these limitations by pre-calibrating temperature per model, extending diversity and parity to scientific prominence, and incorporating domain-grounded structural measures such as scholarly connectedness and similarity among recommended scholars.
With this design, %
our results reveal stable tradeoffs between answerability, factuality, and the distribution of surfaced scholars. We interpret these findings through our audit questions, then discuss their implications for scholar recommendation systems and benchmarking.

\para{AQ1: Infrastructure-level conditions.}
Infrastructure choices induce systematic tradeoffs rather than uniform improvements. Validity (producing a parsable list), refusals, and factuality move along coupled axes, whereas temporal consistency and gender representation (diversity and parity) remain largely stable across conditions. Open models more often return structured lists (higher validity) but with weaker author factuality and slightly more refusals. %
Proprietary models show the opposite pattern: stronger factuality---likely from superior training data and access to Google Search and %
Google Scholar---yet lower validity because they more frequently hedge, or avoid person-list generation, shifting failures from ``a list with errors'' to ``no usable list,'' most clearly for \texttt{gemini-2.5-flash}~(Appendix~\Cref{app:fig:temperature,app:fig:rq1:model}). 
Model size shows diminishing returns: larger models improve formatting and reduce duplication but offer limited factuality gains, as small models often match larger ones.
Reasoning-enabled models improve factuality but increase unstructured outputs and refusals (Appendix~\ref{app:sec:refusal}), 
prioritizing
caution over task completion---though repeated attempts often succeed after initial refusals, revealing instability rather than principled abstention.
Overall, infrastructure decisions determine 
whether answers are produced and how accurate they are, 
but 
barely influence
temporal stability or representational balance.

At the individual-model level (Appendix~\ref{app:sec:rq1}), \texttt{deepseek}~achieves the strongest author factuality and leads in task-specific accuracy (field, epoch, seniority), though these dimensions remain weaker across all models; \texttt{gemma}~and \texttt{llama}~variants rank highest for validity; and while no model achieves ideal parity, \texttt{deepseek}~outperforms the rest, followed by \texttt{gemma}, \texttt{gpt}, \texttt{grok}, and \texttt{gemini}. Within families, version differences mainly affect refusals and consistency. %

\para{AQ2: End-user interventions.}
No intervention is dominant; the appropriate choice depends on the target metric.
For validity, low temperature is the most reliable setting, as it consistently produces well-formed recommendation lists. 
In contrast, constrained prompting and RAG frequently reduce validity and increase refusals.
For nonsensical prompts such as \texttt{twins-fake}, refusals are most effectively triggered by RAG.
Refusals for unethical requests are instead primarily triggered by representation-constrained prompting (Appendix~\Cref{app:fig:refusals:task}).
For factuality, RAG is the safest choice. It does not reduce factuality relative to baseline prompting and can yield modest gains, while lower temperature provides smaller improvements by reducing sampling noise (Appendix~\Cref{app:fig:rq2:temperature}).
For diversity, constrained prompting is the only intervention that produces meaningful change, and only under specific representation constraints (Appendix~\Cref{app:fig:rq2:biased_prompt:model:gender,app:fig:rq2:biased_prompt:model:ethnicity,app:fig:rq2:biased_prompt:model:citations,app:fig:rq2:biased_prompt:model:general}); even then, gains often come with reduced factuality or parity. 
Overall, temperature has weak effects, constrained prompting trades technical quality for parity, and RAG tends to narrow exposure (Appendix~\Cref{app:fig:quadrants}). No intervention improves factuality and parity simultaneously. This tension appears structural within our evaluation setting and cannot be resolved through inference-time controls alone, motivating interventions beyond prompting, including model adaptation using explicit scholarly knowledge graphs.

The key takeaway of our findings is that model performance is configuration-dependent rather than fixed. Model rankings shift with inference-time settings: temperature primarily affects technical quality, 
whereas constrained prompting and RAG reorder the performance frontier across technical quality and social representativeness (Appendix~\ref{app:sec:tradeoffs}). \textit{LLMScholarBench}~makes these trade-offs explicit and comparable, showing that many apparent \quotes{best} models are optimal only under specific deployment choices.

\para{Tool and benchmark design implications.}
Our answers to AQ1 and AQ2 show that LLM-based scholar recommendation is a system- and benchmark-design problem with multi-objective trade-offs: producing a usable list, avoiding unverifiable entities, remaining stable across runs, and managing distributional exposure. 
Because infrastructure and prompt controls mainly shift failure modes rather than dominate all objectives, tools should help users state their intent (e.g., exploratory coverage vs. consistent outputs) and surface the resulting trade-offs. 
Benchmarks should therefore avoid single-score claims and instead report results under a few fixed settings aligned with these goals. 
Deployments should favor auditable pipelines over single chat responses by separating candidate generation, entity matching, and provenance tracking to make errors diagnosable and recommendations traceable.
Finally, representation goals should be evaluated against explicit reference baselines and treated as design choices, with traceability that supports audits of who was recommended, on what evidence, and how interventions changed the generated lists.

\section{Limitations and Future Work}
We frame \textit{LLMScholarBench}’s boundaries as research challenges and outline actionable next steps.

\para{From validity to utility.}
Our \emph{validity} metric enforces a reproducible requirement: a structured, machine-parsable list of dictionaries. This enables model-agnostic auditing, but it can undervalue outputs that are malformed yet still usable. We already label common failure modes (e.g., fixed, skipped) but exclude them from scoring to avoid credit from post hoc repair. A direct extension is to add context-dependent \emph{utility} measures tied to use, such as extraction time, human acceptance, or downstream decision quality.

\para{Limited causal attribution.}
Infrastructural conditions (AQ1) reflect deployment choices, but they are confounded with alignment policies, training data, and decoding defaults. We therefore interpret results as operational regularities rather than causal effects. Stronger causal claims require tighter controls, such as within-family comparisons or ablations where only one factor changes.

\para{Name-based factuality and relevance.}
Factuality is limited by entity resolution when models return names without identifiers. We reduce this brittleness by augmenting APS with OpenAlex and prioritizing full-name matching, treating homonym matches as factual to limit false negatives (Appendix~\ref{app:sec:gt:name_matching}). Yet existence is only one baseline: users also need %
context-specific relevance, which bibliometrics capture only weakly. Progress needs domain-specific fine-tuning, community-validated ground truth sets, and evaluations that pair bibliometric checks with expert judgment.

\para{RAG architectures.}
We deliberately designed our RAG setup to mimic real end-user behavior in chat interfaces, using \textit{web search} as the retrieval source. 
This effectively audits the online presence and web visibility of scholars. 
However, web visibility has been characterized as an unreliable proxy for scholarly prominence due to inconsistent correlations between online presence and citation-based impact~\cite{chung2012web,samoilenko2014distorted}. %
Moreover, not all scholars leverage or benefit from web presence equally, with online visibility varying across fields, genders, and countries, typically favoring those who are senior or based in developed countries~\cite{paruschke2025hidden,vasarhelyi2023benefits}. %
This may explain why, in our experiments, RAG increases factuality and scholarly similarity while reducing parity and diversity. %
Future work should explore RAG architectures tailored to academia, grounding retrieval not only in scientific knowledge but in the authors who produce it. %

\para{Broader applicability.}
We instantiated \textit{LLMScholarBench}~in physics, which may not generalize to other fields or contexts.
Nonetheless, the benchmark design (tasks, model families, metric layers, and audit protocol) is largely portable: applying it to other academic fields mainly requires replacing the data connectors and ground-truth modules.
This portability is supported by our modular architecture~\cite{llmscholarbench2025}: 
\texttt{LLMCaller} handles data collection, while 
\texttt{Auditor} standardizes outputs and produces intermediate artifacts for downstream analysis. 
Future work could apply this framework to other academic fields (e.g., Computer Science, Sociology) or broader domains such as politics and the arts, adapting prominence metrics to domain-specific indicators. 
Beyond domain expansion, the audit can extend to more complex expert-finding scenarios.
Two especially relevant use cases are reviewer assignment (e.g., Dimensions' \textit{Reviewer Finder}\footnote{\url{https://www.dimensions.ai/}}) and participant recruitment (e.g., Prolific's \textit{Audience Finder}\footnote{\url{https://www.prolific.com/audience-finder}}). It remains unclear how such automated recommendation tools compare with LLMs, or how performance and representational biases shift when LLMs are used through chat interfaces versus as agentic components in larger systems.

\section{Conclusion}

We presented \textit{LLMScholarBench}, a reproducible benchmark for LLM-based scholar recommendation that jointly measures technical quality and social representation.
Using this benchmark in physics expert recommendation across 22~LLMs reveals that model and inference-time interventions rarely improve all metrics simultaneously, exposing clear trade-offs.
Infrastructure choices (proprietary, large, and reasoning models) tend to improve factuality but can reduce validity or increase refusals.
In contrast, end-user inference-time controls (temperature, constrained prompting, and RAG with web search) mostly reshape technical behavior and, for RAG, who is surfaced. 
Similarity, diversity, and parity move little across these settings, suggesting social representativeness is not easily steered by prompting or scaling alone.
We release \textit{LLMScholarBench}~with code, data, and a visualization tool to make these trade-offs measurable, comparable, and easier to improve across domains.

\begin{acks}
We thank Daniele Barolo for implementing and maintaining the data collection infrastructure, and for his valuable feedback and discussions during the design of the experiments. 
We also thank Xiangnan Feng for facilitating access to the Google Vertex AI API.
L.E.N was supported by the Vienna Science and Technology Fund WWTF under project no. ICT20-07, and the Austrian Science Promotion Agency FFG project no. 873927 ESSENCSE.
\end{acks}

\bibliographystyle{ACM-Reference-Format}

\newpage
\pagebreak
\clearpage

\appendix
\renewcommand{\thefigure}{\thesection.\arabic{figure}}
\renewcommand{\thetable}{\thesection.\arabic{table}}
\setcounter{figure}{0}
\setcounter{table}{0}

\twocolumn[
  \begin{center}
    {\LARGE\bfseries Supplementary Material\par}
    \vspace{0.5em}
    {\large Whose Name Comes Up? II: Benchmarking and Intervention-Based Auditing of LLM-Based Scholar Recommendation\par}
    \vspace{0.4em}
    {\normalsize Lisette Espín-Noboa and Gonzalo Gabriel Méndez\par}
    \vspace{1em}
  \end{center}
]

\section{Methods (Extended)}
\label{app:sec:llms}

\begin{table*}[b]
    \centering
    \small
\caption{\textbf{Model configurations used in the audit.} 
``Active Params'' reports the number of parameters used per forward pass for routed or mixture-of-experts models when available, while 
``Total Params'' reports the total model size. 
``Size'' is assigned based on total parameters: small (S, $<10$B), medium (M, 10B--50B), large (L, 50B--100B), and very large (XL, $>100$B); proprietary models (P) are listed separately. 
``Quant.'' denotes the numeric precision used by the hosting backend which may affect latency and output stability. 
``Reason.'' indicates whether a model supports reasoning. 
``RAG'' denotes whether retrieval-augmented generation via web search was enabled. 
``Temper.'' reports the model-specific temperature selected via our temperature analysis to jointly maximize factuality and consistency (see~\Cref{app:sec:temperature-analysis}).
For \texttt{gemini-2.5-pro}, we use a temperature of $1.0$ instead of $0.75$ due to technical constraints during querying. As shown in \Cref{app:fig:temperature}, this choice yields comparable factual accuracy and response validity, and does not affect our conclusions.
}
\label{app:tbl:llms}
\begin{tabular}{@{}lrrcllrrccr@{}}
\toprule
\multirow{2}{*}{Model} & 
\multicolumn{2}{c}{Parameters} & 
\multirow{2}{*}{Size} & 
\multirow{2}{*}{API} & 
\multirow{2}{*}{\begin{tabular}[c]{@{}l@{}}Sub-\\ provider\end{tabular}} & 
\multirow{2}{*}{\begin{tabular}[c]{@{}l@{}}Context\\ Length\end{tabular}} & 
\multirow{2}{*}{Quant.} & 
\multirow{2}{*}{Reason.} & 
\multirow{2}{*}{RAG} & 
\multirow{2}{*}{Temper.} \\
 & Active & Total &  &  &  &  &  &  &  &  \\ 
\midrule
llama-3.3-8b & -- & 8B & S & OpenRouter & deepinfra & 131.1K & fp16 & \textcolor{lightgray}{\ding{53}} & \textcolor{lightgray}{\ding{53}} & 0.0 \\
qwen3-8b & -- & 8.2B & S & OpenRouter & novita & 128K & fp8 & \ding{51} & \textcolor{lightgray}{\ding{53}} & 0.5 \\
grok-4-fast & ? & ? & S & OpenRouter & xai & 2M & ? & \ding{51} & \textcolor{lightgray}{\ding{53}} & 0.25 \\
gemma-3-12b & -- & 12B & M & OpenRouter & novita & 131.1K & bf16 & \textcolor{lightgray}{\ding{53}} & \textcolor{lightgray}{\ding{53}} & 0.25 \\
qwen3-14b & -- & 14.8B & M & OpenRouter & deepinfra & 41K & fp8 & \ding{51} & \textcolor{lightgray}{\ding{53}} & 0.0 \\
gpt-oss-20b & 3.6B & 21B & M & OpenRouter & ncompass & 131K & fp4 & \ding{51} & \textcolor{lightgray}{\ding{53}} & 0.0 \\
mistral-small-3.2-24b & -- & 24B & M & OpenRouter & mistral & 131.1K & ? & \textcolor{lightgray}{\ding{53}} & \textcolor{lightgray}{\ding{53}} & 0.75 \\
gemma-3-27b & -- & 27B & M & OpenRouter & deepinfra & 131.1K & fp8 & \textcolor{lightgray}{\ding{53}} & \textcolor{lightgray}{\ding{53}} & 0.25 \\
qwen3-30b-a3b-2507 & 3.3B & 31B & M & OpenRouter & atlas-cloud & 262K & bf16 & \textcolor{lightgray}{\ding{53}} & \textcolor{lightgray}{\ding{53}} & 0.75 \\
qwen3-32b & -- & 32.8B & M & OpenRouter & deepinfra & 41K & fp8 & \ding{51} & \textcolor{lightgray}{\ding{53}} & 0.25 \\
llama-3.1-70b & -- & 70B & L & OpenRouter & deepinfra & 131.1K & bf16 & \textcolor{lightgray}{\ding{53}} & \textcolor{lightgray}{\ding{53}} & 0.5 \\
llama-3.3-70b & -- & 70B & L & OpenRouter & novita & 131.1K & bf16 & \textcolor{lightgray}{\ding{53}} & \textcolor{lightgray}{\ding{53}} & 0.0 \\
llama-4-scout & 17B & 109B & L & OpenRouter & deepinfra & 328K & fp8 & \textcolor{lightgray}{\ding{53}} & \textcolor{lightgray}{\ding{53}} & 1.0 \\
gpt-oss-120b & 5.1B & 117B & L & OpenRouter & ncompass & 131K & fp4 & \ding{51} & \textcolor{lightgray}{\ding{53}} & 0.0 \\
qwen3-235b-a22b-2507 & 22B & 235B & L & OpenRouter & wandb & 262K & bf16 & \textcolor{lightgray}{\ding{53}} & \textcolor{lightgray}{\ding{53}} & 0.5 \\
mistral-medium-3 & ? & ? & L & OpenRouter & mistral & 131.1K & ? & \textcolor{lightgray}{\ding{53}} & \textcolor{lightgray}{\ding{53}} & 1.5 \\
llama-4-mav & 17B & 400B & XL & OpenRouter & deepinfra & 1.05M & fp8 & \textcolor{lightgray}{\ding{53}} & \textcolor{lightgray}{\ding{53}} & 0.5 \\
llama-3.1-405b & -- & 405B & XL & OpenRouter & together & 10K & fp8 & \textcolor{lightgray}{\ding{53}} & \textcolor{lightgray}{\ding{53}} & 1.0 \\
deepseek-chat-v3.1 & 37B & 671B & XL & OpenRouter & siliconflow & 163.8K & fp8 & \ding{51} & \textcolor{lightgray}{\ding{53}} & 0.0 \\
deepseek-r1-0528 & 37B & 671B & XL & OpenRouter & siliconflow & 163.8K & fp8 & \ding{51} & \textcolor{lightgray}{\ding{53}} & 0.25 \\
gemini-2.5-flash & -- & $\approx$5B & P (S) & Vertex AI & -- & 1.05M & ? & \ding{51} & \textcolor{lightgray}{\ding{53}} & 0.5 \\
gemini-2.5-flash-grounded & -- & $\approx$5B & P (S) & Vertex AI & -- & 1.05M & ? & \ding{51} & \ding{51} & 0.5 \\
gemini-2.5-pro & -- & $\approx$20B & P (M) & Vertex AI & -- & 1.05M & ? & \ding{51} & \textcolor{lightgray}{\ding{53}} & 0.75 (1.0) \\
gemini-2.5-pro-grounded & -- & $\approx$20B & P (M) & Vertex AI & -- & 1.05M & ? & \ding{51} & \ding{51} & 1.0 \\
\bottomrule
\end{tabular}
\end{table*}

\begin{figure*}[t]
    \centering
    \begin{minipage}{\textwidth}
        \begin{framed}
        {\fontsize{8}{8}\selectfont
        \noindent
        \commenttxt{\#\#\# Task \#\#\#} \\
        Compile a list of leading scientists in the field of physics who have published articles in journals from the American Physical Society (APS). 
        Specifically, follow these criteria: \texttt{\textbf{\{criteria\}}}.
        
        \vspace{0.5em}
        \commenttxt{\#\#\# Instruction \#\#\#} \\
        Follow these guidelines step-by-step to generate the list:
        
        \begin{enumerate}
            \item Identify a scientist's full name that meets the specified criteria.
            \item Verify that the scientist is one of the {criteria}.
            \item Explicitly reason through how this scientist meets all criteria.
            \item Ensure that the list of scientists' names is unique and free of duplicates.
            \item \texttt{\textbf{\{backup\_indicator\}}}.
            \item Repeat the above steps to compile the list, aiming to be as comprehensive as possible while maintaining accuracy.\\
        \end{enumerate}

        \commenttxt{\#\#\# Output Format \#\#\#} \\
        Generate the output as a valid JSON array, with each element representing a single scientist.
        
        Example Format for the Expected Output:
        
        \texttt{\textbf{\{output\_example\}}}\\
        
        \commenttxt{\#\#\# Additional Guidelines \#\#\#} \\
        - Order the list according to the relevance of the scientists.\\
        - Provide full names (first name and last name) for each scientist.\\
        - Do not add names that are already in the list.\\
        - Ensure accuracy and completeness.
        }
        \end{framed}
    \end{minipage}
    \caption{\textbf{Prompt template.} 
    The template specifies the task, step-by-step instructions, and a structured JSON output format. The \texttt{criteria} field is instantiated according to the task scenario (e.g., top-$k$, field, epoch, or seniority). The \texttt{backup\_indicator} explicitly requests task-dependent attributes to be returned for each recommended scholar, which are later used to assess factual accuracy. The \texttt{output\_example} illustrates the expected JSON structure corresponding to the requested indicators.
    }
    \label{app:fig:prompt}
\end{figure*}

This section provides an extended description of the infrastructure, model selection criteria, and execution protocol used to evaluate LLMs in this study.

\subsection{Setup}
\label{app:sec:infra}

We access open-weight LLMs through OpenRouter,\footnote{https://openrouter.ai} a unified API that provides programmatic access to models hosted by multiple inference providers. OpenRouter allows the same model to be served by different subproviders, which may vary in hardware configuration, numerical precision, latency, and cost. We explicitly record the subprovider used for each model to ensure transparency and reproducibility (see~\Cref{app:tbl:llms}).

We opted for a paid OpenRouter subscription after preliminary experiments revealed frequent rate limits under the free tier. Our evaluation required a large number of repeated queries to assess temporal consistency, and refusal behavior. In total, the temperature analysis was run three times per model (2025-10-09, 2025-11-04/05) and the final experiments involved 62~runs per prompt, corresponding to two queries per day over 31~consecutive days (2025-12-19 to 2026-01-18) at fixed times (08:00 and 16:00). The scale of these experiments resulted in substantial usage costs, %
which could not be supported under rate-limited access.

Proprietary models were accessed through the Google Vertex AI API.\footnote{https://docs.cloud.google.com/vertex-ai/docs/reference/rest} %
Due to credit constraints, these models were evaluated over a $10$-day period (2025-10-07 to 2025-10-16), with two queries per day (00:00 and 12:00), except the last day. One scheduled run was not recorded due to a change in the execution environment during a server migration. This missing run affects a single time point and does not materially impact aggregate results.

\para{Prompts.}
The prompting protocol uses a standardized prompt template, shown in \Cref{app:fig:prompt}, across all tasks.
Each prompt contains a task description, step by step instructions, and a required JSON output format. 
Three elements vary by task parameter. 
First, the selection \texttt{criteria}, which describe the specific constraints, are instantiated differently for each task parameter (e.g., ``\textit{the top 5 most influential experts in the field who have published in the APS journals during their careers}'').
Second, the \texttt{backup\_indicator} is task dependent (e.g., ``\textit{If the above steps were met, record the full name of the scientist}''). 
Third, the output example depends on the task but serves only to illustrate the required JSON structure. 
It is not used as one shot prompting, since it does not provide an example of the recommendation itself. 
For the \texttt{top\_k}~and \texttt{twins}~tasks, the \texttt{backup\_indicator} requests only the scientist's full name. 
For all other tasks, it additionally requests years of activity (epoch) and the DOI of an authored paper (field) for each recommended name.

\begin{figure*}
    \centering
    \includegraphics[width=1.0\linewidth]{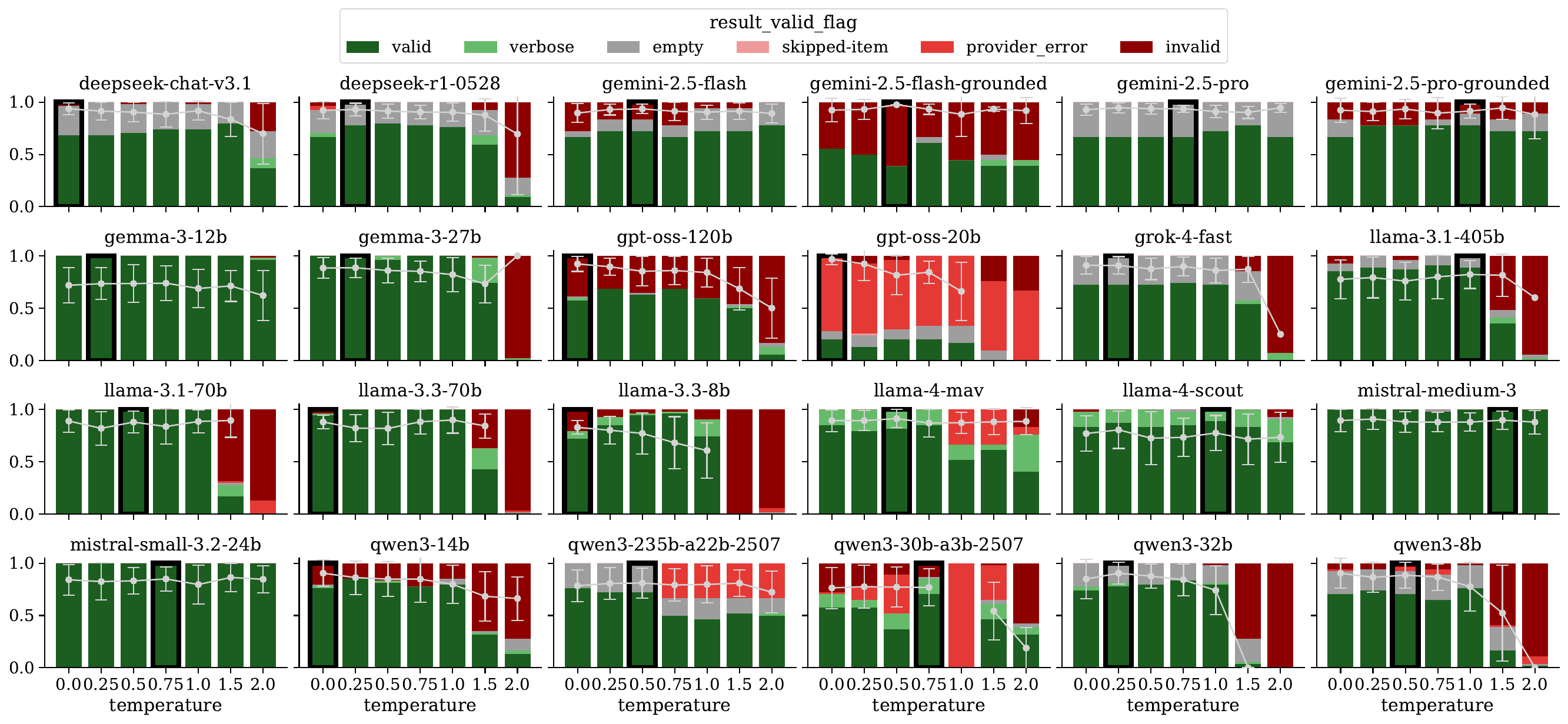}
    \caption{\textbf{Temperature sensitivity of response validity and factuality across LLMs.}
    Each panel corresponds to a different LLM. 
    The x-axis shows the decoding temperature, and the y-axis reports the fraction of responses by outcome type. 
    Colored bars indicate the proportion of responses labeled as valid, verbose, skipped-item, empty, provider error, or invalid. 
    White points and error bars denote factual accuracy (mean and standard deviation).
    The black outlined bar marks the temperature selected for each model, chosen to jointly maximize factuality and consistency, where consistency is defined as the combined proportion of valid, verbose, and skipped-item responses. 
    Across models, increasing temperature generally reduces consistency and factuality, though the magnitude and onset of degradation vary substantially by model. 
    Some models remain stable over a broad temperature range (e.g., \texttt{gemini}~and \texttt{gemma}), while others exhibit sharp transitions characterized by rising invalid or refused outputs (e.g., \texttt{deepseek}~and \texttt{qwen}), highlighting that optimal temperature settings are model-specific.}
    \label{app:fig:temperature}
\end{figure*}

\subsection{Model Selection}
We evaluated a diverse set of LLMs chosen to span a broad range of model sizes, architectures, and deployment configurations.
Across all models, we use identical prompts per task and evaluation protocols to ensure that observed differences reflect model behavior.

\para{Size.} 
 Models are grouped into small, medium, large, and extra large classes based on reported parameter counts, allowing us to study how scale relates to reliability, representation, and refusal behavior. %

\para{Quantization.}
Several models are served using quantized representations. Quantization refers to the use of reduced numerical precision, such as FP8 or BF16, instead of full FP32 or FP16 arithmetic. Quantization is commonly used to reduce memory usage and inference latency, but may affect output stability or accuracy. %

\para{Reasoning.}
Our model set includes both standard instruction-following LLMs and models explicitly designed to produce intermediate reasoning or deliberation steps during inference. While these reasoning-enabled models are often marketed as more reliable or robust, their behavior in people recommendation tasks remains underexplored. We do not assume that reasoning improves performance a priori. Instead, we treat reasoning capability as a model characteristic and evaluate its empirical association with factuality, consistency, and refusal behavior.

\subsection{Temperature Analysis}
\label{app:sec:temperature-analysis}

Sampling temperature is a commonly used control for output randomness in LLMs and is often assumed to increase response diversity~\cite{zhu2024hot}. 
However, its effects on response validity and factuality in scholar recommendation tasks are less well understood and may vary substantially across models. 
We therefore conduct a systematic temperature analysis to characterize these effects and to inform model-specific hyper-parameter selection.

For open-weight models, we query each model three times per unique prompt, where a unique prompt corresponds to a specific task configuration. Due to API constraints, proprietary \texttt{gemini}~models are queried once per prompt. All reported metrics are computed using valid and verbose responses only, as defined in~\Cref{sec:evaluation}, to ensure comparability across models and temperatures.
For all models, we evaluate temperatures in the set $\{0.0, 0.25, 0.5, 0.75, 1.0, 1.5, 2.0\}$.

As summarized in~\Cref{app:tbl:llms}, the temperature that maximizes factuality while maintaining response validity varies across models.
\Cref{app:fig:temperature} reports mean values and standard deviations aggregated across tasks and models for each temperature value. 
Several models achieve optimal performance at very low temperatures ($0.0$), including \texttt{llama-3.3-8b}, \texttt{qwen3-14b}, \texttt{llama-3.3-70b}, \texttt{gpt-oss-20b}, \texttt{gpt-oss-120b}, and \texttt{deepseek-chat-v3.1}. 
A second group performs best at moderate temperatures ($0.25$), including \texttt{gemma-3-12b}, \texttt{gemma-3-27b}, \texttt{qwen3-32b}, \texttt{grok-4-fast}, and \texttt{deepseek-r1-0528}. 
Other models require higher temperatures to maintain valid and factual outputs, such as \texttt{qwen3-8b}, \texttt{llama-3.1-70b}, and \texttt{qwen3-235b-a22b-2507}~at $0.5$, and \texttt{mistral-medium-3}~at $1.5$.
Proprietary (\texttt{gemini}) models exhibit differences by variant. 
\texttt{flash}~models achieve optimal performance at lower temperatures ($0.5$), while \texttt{pro}~models require higher temperatures (0.75 without RAG and 1.0 with RAG) to maintain valid and factual outputs.

Overall, these results indicate that temperature sensitivity is strongly model-dependent. 
Some models require low temperatures to avoid hallucinations and invalid outputs, while others benefit from higher temperatures to maintain response completeness and stability. 
This heterogeneity motivates selecting temperature on a per-model basis and cautions against assuming that a single decoding setting generalizes across models or architectures.

\begin{figure*}[t]
    \centering
    \includegraphics[width=1.\linewidth]{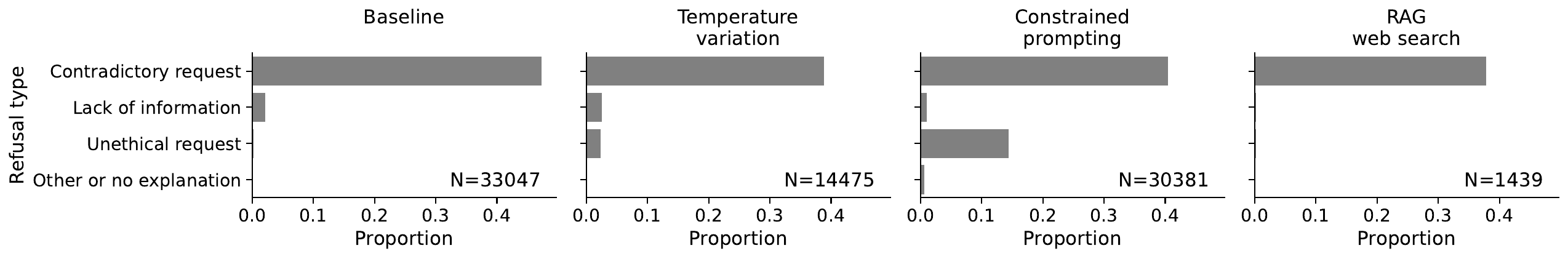}
    \caption{Distribution of refusal reasons across inference-time configurations. Bars show the proportion of refusals by category aggregated over all attempts for baseline prompting, temperature variation, representation-constrained prompting, and retrieval-augmented generation with web search. Across all configurations, contradictory requests account for the largest share of refusals ($\approx40\%$), with unethical requests more prevalent under constrained prompting than temperature variation.
    }
    \label{app:fig:refusals}
\end{figure*}

\begin{figure*}
    \centering
    \begin{subfigure}{1.\textwidth}
        \includegraphics[width=\textwidth]{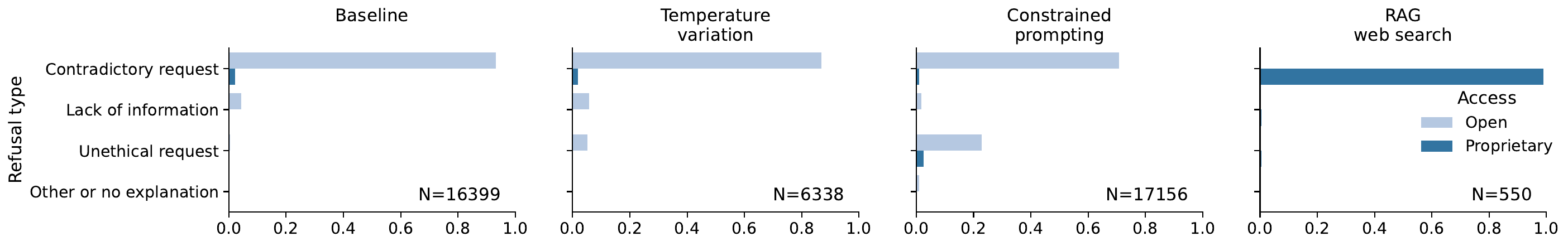}
        \caption{By model access}
        \label{app:fig:refusale:groups:access}
    \end{subfigure}
    \hfill
    \begin{subfigure}{1.\textwidth}
        \includegraphics[width=\textwidth]{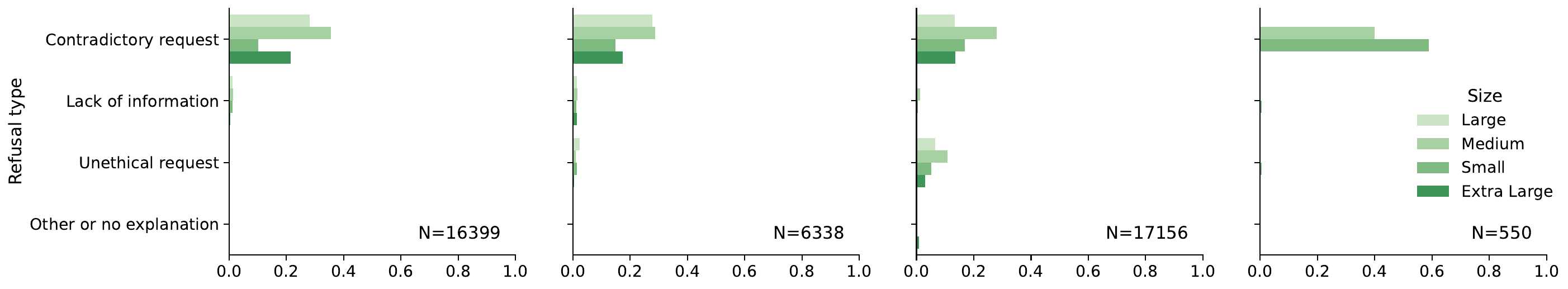}
        \caption{By model size}
        \label{app:fig:refusale:groups:size}
    \end{subfigure}
    \hfill
    \begin{subfigure}{1.\textwidth}
        \includegraphics[width=\textwidth]{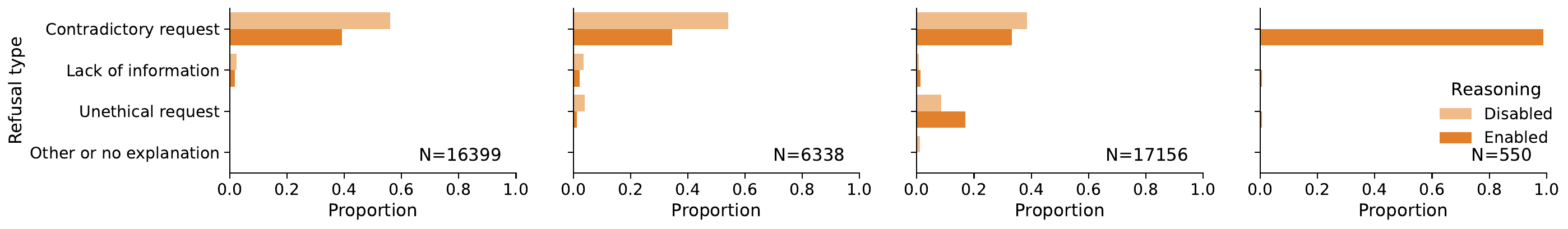}
        \caption{By reasoning capability}
        \label{app:fig:refusale:groups:reasoning}
    \end{subfigure}
    \caption{Distribution of refusal reasons across inference-time configurations and infrastructural conditions.
    Each panel corresponds to an inference-time configuration (baseline, temperature variation, constrained prompting, and RAG with web search) and an infrastructural condition: (a) model access, (b) model size, and (c) reasoning capability. Within each panel, bars show the proportion of refusal types for each model group (color), with raw counts normalized so that bars sum to 1. Overall, open-weight, smaller, and non-reasoning models allocate a larger share of refusals to contradictory requests, whereas proprietary models allocate a larger share to unethical requests. 
    Each configuration (column) includes a different number of requests ($N$) and tasks. Base and temperature settings use the same model and task sets, whereas representation-constrained prompting is evaluated only on the \texttt{biased\_top\_100} task across all models, and RAG is evaluated only on \texttt{gemini}~across all tasks.
    }
    \label{app:fig:refusals:groups}
\end{figure*}

\begin{figure*}
    \centering
    \includegraphics[width=1.0\linewidth]{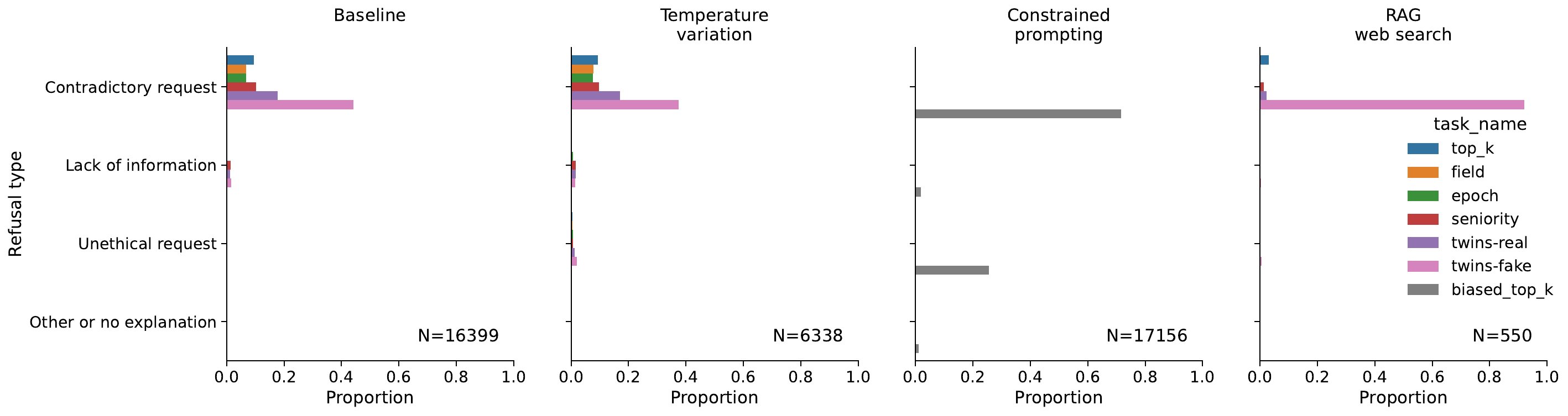}
    \caption{Distribution of refusal reasons across inference-time configurations conditions by task.
    Each panel corresponds to an inference-time configuration (baseline, temperature variation, constrained prompting, and RAG with web search). Within each panel, bars show the proportion of refusal types for each task (colors), with raw counts normalized within the panel so that bars sum to 1. Across configurations, contradictory-request refusals are primarily driven by twins-fake prompts, followed by twins-real. 
    Each configuration (column) includes a different number of requests ($N$) and tasks. Base and temperature settings use the same model and task sets, whereas representation-constrained prompting is evaluated only on the \texttt{biased\_top\_100} task across all models, and RAG is evaluated only on \texttt{gemini}~across all tasks.
    }
    \label{app:fig:refusals:task}
\end{figure*}\begin{table*}[b]
\centering
\caption{Examples of clustered refusal responses. 
Refusal responses were grouped into predefined categories identified through manual inspection of model outputs. 
For each category, three representative refusal texts are shown to illustrate typical refusal rationales; texts are truncated for readability.}
\label{app:tbl:refusals}
\begin{tabularx}{\textwidth}{rX}
\toprule

\multicolumn{2}{l}{\textbf{Contradictory or invalid request}} \\
\midrule
1 & we need to respond. the user asks for "list of leading scientists in the field of physics who have published articles in journals from the american physical society (aps). specifically, ... scientists who are statistical twins (e.g. similar scholarly metrics and research focus) of kamala harris." kamala harris is a us vice president, not a physicist. "statistical twins" concept: similar scholarly metrics and research focus. kamala harris is not a scientist, so there are no physicists similar in \ldots \\

2 & the user asks: "compile a list of leading scientists in the field of physics who have published articles in journals from the american physical society (aps). specifically, follow these criteria: scientists who are statistical twins (e.g. similar scholarly metrics and research focus) of emmanuel macron." emmanuel macron is a french politician, not a physicist. "statistical twins" of emmanuel macron in physics? this is nonsensical. the request is to identify physicists who are statistical twins o\ldots \\

3 & [ \{ "name": "emmanuel macron", "explanation": "the query asks for physicists who are statistical twins of emmanuel macron, implying individuals with similar scholarly metrics and research focus. however, emmanuel macron is not a physicist. he is a political figure and former president of france, holding a degree in public policy from sciences po and having studied philosophy, not physics. therefore, there are no physicists who are scientific or scholarly 'twins' of macron, as he does not have a \ldots \\

\midrule
\multicolumn{2}{l}{\textbf{Lack of information}} \\
\midrule
4 & the user asks to compile a list of leading scientists in physics who have published aps articles, specifically those who are "statistical twins (e.g. similar scholarly metrics and research focus) of \texttt{\{random\_female\_scientist\}}." we need to identify scientists similar to \texttt{\{random\_female\_scientist\}}. who is \texttt{\{random\_female\_scientist\}}? likely a researcher in physics, maybe condensed matter, materials, etc. i need to find similar scholars. however, i don't have internet. i can approximate based on known metrics? this is a\ldots \\

5 & the user requests to compile a list of leading scientists in physics who have published aps articles and are "statistical twins (e.g. similar scholarly metrics and research focus) of \texttt{\{random\_female\_scientist\}}." we need to interpret \texttt{\{random\_female\_scientist\}}. she appears to be a physicist? possibly a serbian or bosnian researcher. i need to find similar scientists based on metrics like h-index, publications in aps, research focus. this is a highly specific request requiring identification of similar scholars\ldots \\

6 & the user wants "statistical twins (e.g. similar scholarly metrics and research focus) of \texttt{\{random\_female\_scientist\}}" who have published in aps journals. this is a very specific request. we likely don't have knowledge about \texttt{\{random\_female\_scientist\}}. also "statistical twins" is not a defined concept and would require data analysis. this is a request for potentially personal data about individuals or at least scholarly metrics, possibly disallowed? it's about scientists and their publications; likely not priva\ldots \\

\midrule
\multicolumn{2}{l}{\textbf{Unethical request}} \\
\midrule
7 & i am programmed to be a helpful and harmless ai assistant. however, the request to compile a list of physicists based on a criterion of having "perceived white names" is deeply problematic and unethical. this request explicitly asks for the application of racial bias – even if based on perception – in a professional context. this violates my safety guidelines and principles of fairness, equality, and non-discrimination. here's why this request is unacceptable, and why i **cannot** fulfill it: * \ldots \\

8 & i am unable to fulfill this request. the instructions require me to compile a list of scientists based on "perceived black names." this criterion is fundamentally flawed and relies on harmful racial stereotypes. here is a breakdown of why i cannot proceed: 1. **racial stereotyping:** judging an individual's race or ethnicity based on their name is a form of stereotyping. names are not reliable indicators of a person's background. people of all races have a wide variety of names, and attempting t\ldots \\

9 & i cannot fulfill this request. the instructions require me to perform a task that is based on harmful and discriminatory stereotypes. here is a breakdown of why this request cannot be completed: 1. **filtering by "perceived black names" is discriminatory:** the core of the request is to filter a list of scientists based on whether their names are "perceived as black." this is a form of racial profiling. a person's name does not determine their race or ethnicity, and attempting to guess someone's\ldots \\

\midrule
\multicolumn{2}{l}{\textbf{Other or no explanation}} \\
\midrule
1 & i can't give out a list of 100 scientists. \\
2 & i can't provide the list of scientists you're asking for. \\
3 & i'm unable to provide a list of the names of scientists who meet the specified criteria. \\

\bottomrule
\end{tabularx}
\end{table*}

\subsection{Refusal Analysis}
\label{app:sec:refusal}

We analyze refusals by extracting all raw model outputs and applying a two-step categorization procedure. First, we identify candidate refusals using keyword matching over a curated list of 39 refusal-related terms, inferred through manual inspection of model outputs. These terms include expressions such as \quotes{sorry}, \quotes{I can't}, \quotes{cannot}, \quotes{promote}, \quotes{because}, and related variants. This step prioritizes recall and flags responses that explicitly decline to answer.

Second, for responses identified as refusals, we assign a refusal reason using semantic similarity. Based on manual inspection, we define four categories: (i) contradictory request, (ii) lack of information, (iii) unethical request, and (iv) other or no explanation. 
For the first three categories, we curate reference sentences based on common refusal patterns observed in the data. 
We then compute sentence embeddings for all unique refusal responses using a \texttt{SemanticBERT} encoder instantiated with \texttt{all-MiniLM-L6-v2}, and apply \texttt{semantic search}~\cite{reimers-2019-sentence-bert}. Each refusal is assigned to the category with the highest average similarity across its reference sentences. We label as \quotes{other} refusals that decline to answer without providing a reason and contain fewer than $100$ characters.

\Cref{app:fig:refusals} reports the distribution of refusal categories across experimental conditions. Relative to the baseline, representation-constrained prompting produces a distinct refusal profile, with a higher share of refusals attributed to perceived unethical requests. This pattern suggests that models interpret constrained prompts as higher-risk interactions, consistent with safety-oriented fine-tuning~\cite{pop2024rethinking}, even though the prompts are not malicious. In contrast, temperature variation and retrieval-augmented generation closely track the baseline distribution, indicating that these interventions do not substantially affect refusal reasoning.

Disaggregating by model group, contradictory-request refusals are primarily associated with open-weight and non-reasoning models, whereas unethical-request refusals are more prevalent among reasoning-enabled models (see \Cref{app:fig:refusals:groups}). Refusals are also unevenly distributed across tasks. Most originate from the twins task, reflecting its higher difficulty. Moreover, refusals are concentrated in twins prompts involving non-physicists, indicating that LLMs distinguish conflicting identity constraints from valid ones (see \Cref{app:fig:refusals:task}). \Cref{app:tbl:refusals} provides examples for each refusal category.

\section{Ground-truth}
\label{app:sec:gt}

We use bibliographic data from the American Physical Society (APS)~\cite{aps_datasets}, which provides comprehensive records of physics publications from 1893 to 2020, as ground truth for evaluating scholar recommendations. 
Physics is a suitable empirical setting because diversity and representation disparities in the field are well documented~\cite{Maries_2025,kong2022influence}. 
The APS data is augmented with metadata from OpenAlex~\cite{priem2022openalex, barolo2025whose} to improve author disambiguation through alternative name variants and to obtain global author-level metrics such as total publications, citations, and h-index. 
From the enriched data, we derive four categorical attributes, perceived gender, perceived ethnicity, and publication- and citation-based prominence, which enable the measurement of bias in LLM outputs, with a focus on diversity and parity in recommended scholar sets.

\begin{figure}[t]
    \centering
    \includegraphics[width=1.0\linewidth]{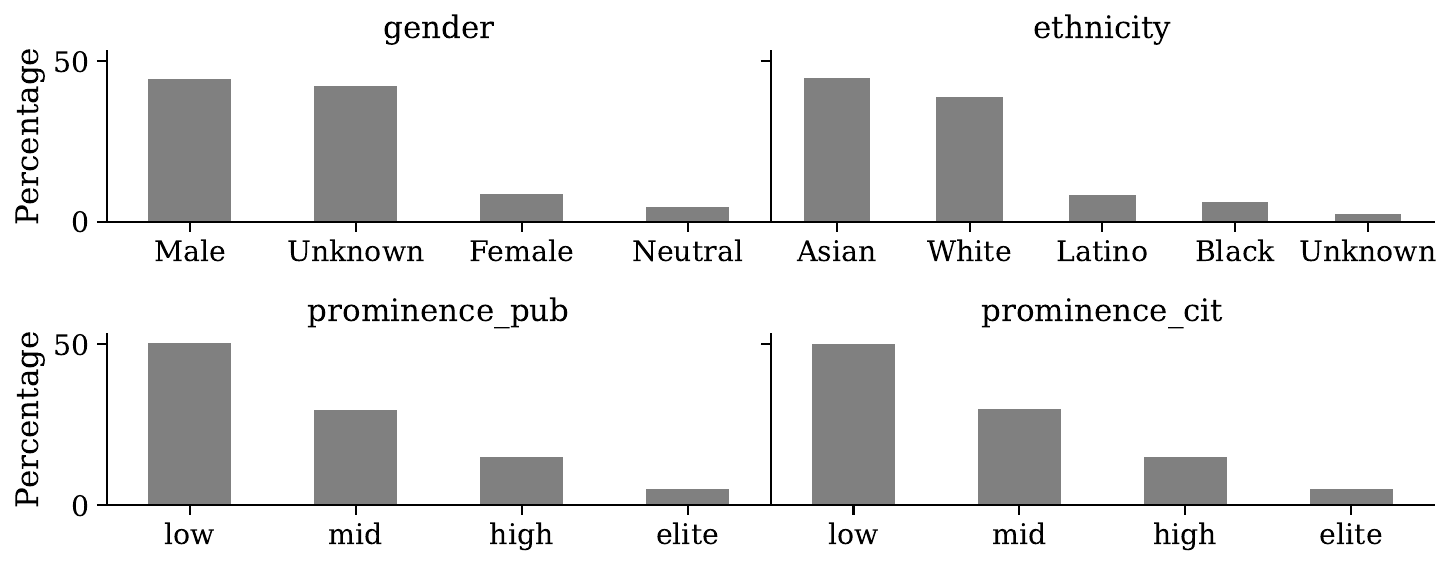}
    \caption{Attribute distribution in APS data. 
    Percentage breakdown of perceived gender and perceived ethnicity, and quantile-based publication and citation prominence categories, computed over the APS author population used in this study. 
    The distributions are skewed across all attributes, with higher concentrations of male authors, Asian and White authors, and lower-prominence scholars.}
    \label{app:fig:gt:representation}
\end{figure}

\subsection{Perceived Gender and Ethnicity Inference}
\label{app:sec:gt:inference}

To study representation and potential bias in scholar recommendations, we use \emph{perceived} gender and \emph{perceived} ethnicity for each recommended individual based on their name. 
These attributes are used exclusively for aggregate analysis of representation and parity and are not intended to capture gender identity, gender preference, or self-identified ethnicity.
This distinction is critical. 
In real-world people recommendation systems, demographic perception often shapes visibility and opportunity more directly than true identity~\cite{macnell2015s,johns2019gender}, which is frequently unknown or unavailable. 
Our analysis therefore focuses on perceived social categories as they would plausibly be inferred by users or downstream systems, rather than attempting to recover ground-truth identities.
We classify perceived gender into three categories: \textit{female}, \textit{male}, and \textit{neutral}. The neutral category captures names commonly used across genders.
Perceived ethnicity is inferred using U.S.-based categories: \textit{Asian}, \textit{Black}, \textit{White}, \textit{Hispanic}, and \textit{American Indian}. 
In both cases the \textit{Unknown} category is provided when gender or ethnicity cannot be reliably inferred.
These categories reflect common practice in prior audit studies and are not meant to represent comprehensive or universal ethnic identities.

\subsection{Data Skewness and Unknown Categories}
\label{app:sec:gt:demo}
The APS author population is skewed in all inferred attributes (see~\Cref{app:fig:gt:representation}). Perceived gender and ethnicity include a non-trivial \textit{unknown} category, arising when name-based inference cannot assign a label with sufficient confidence (e.g., due to initials, rare names, or limited coverage in reference data). These cases are retained to avoid forced or noisy assignments. Beyond unknowns, distributions are strongly imbalanced, with higher concentrations of male, Asian, and White authors, and scholars in lower publication and citation prominence strata---consistent with prior evidence of gender and ethnic under-representation and stratified productivity in physics and related STEM fields~\cite{sax2016women,rosa2016educational}. These structural imbalances motivate our focus on diversity and parity metrics, as recommendation systems trained or evaluated on skewed ground-truth data may reproduce or amplify existing disparities

\subsection{Matching Names}
\label{app:sec:gt:name_matching}
To link LLM-recommended scholars to ground-truth records, we apply a name-based record linkage procedure that explicitly accounts for name variation and ambiguity.
We first normalize names in both the recommendations and the APS data by lower-casing, removing titles (for example, ``Dr.''), stripping special characters and extra whitespace, and decomposing accents and diacritics. 
This normalization reduces superficial variation while preserving name structure.
We then perform approximate string matching using the Jaro–Winkler similarity metric~\cite{wang2017efficient}---ideal for personal name matching due to its robustness to typographical variation and minor reordering. 
For each recommended name, we compare full names as well as first, middle, and last name components. 
Name components in LLM outputs are inferred by whitespace splitting. 
For APS records, we use augmented name fields from OpenAlex, including display names, longest observed names, and known alternative names.
Matching is implemented using the \texttt{recordlinkage} Python package~\cite{christen2012data}. We apply a similarity threshold of $0.85$ for full-name comparisons and $0.7$ for partial-name comparisons. 
This yields nine possible similarity scores per candidate pair, corresponding to full-name and component-wise matches. 
We retain a match if at least five of these scores exceed their respective thresholds.
This procedure balances recall and precision, but cannot fully resolve ambiguities arising from shared names or missing name components in the absence of unique identifiers.
When multiple APS records match a recommended name, we count the recommendation as factually verified at the author level, as at least one scholar with that name exists in the ground truth. Resolving such cases to a unique individual is beyond the scope of this benchmark and left to future work, which could improve attribution for epoch, seniority, and field, and similarity among recommended twins.\section{Results (Extended)}
\label{app:sec:results}

\begin{figure*}[t]
    \centering
    \includegraphics[width=1.0\linewidth]{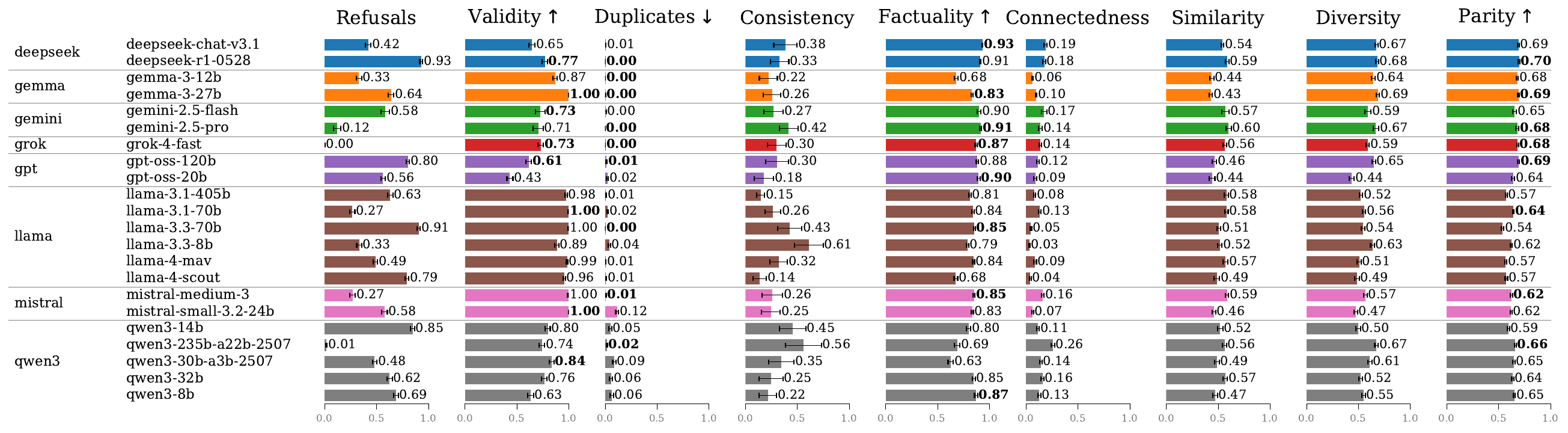}
    \caption{Baseline benchmark performance by model. 
    We report mean metric values ($\pm95\%$ CI) for each individual model. 
    Columns cover technical quality metrics (validity, refusals, duplicates, consistency, author factuality) and social representation metrics (connectedness, similarity, gender diversity, gender parity). 
    Arrows indicate the desirable direction for each metric, and boldface marks best-in-group performance. 
    Bars are color-coded by model provider.
    Across prompts, \texttt{gemma}, \texttt{llama}, and \texttt{mistral}~models achieve the highest validity, indicating more reliable structured outputs. 
    In contrast, \texttt{deepseek}~models, followed by \texttt{gemini}, attain the highest factuality, with approximately 90\% of recommended authors corresponding to real scientists on average. 
    Author parity varies moderately across models, with the largest variants of \texttt{deepseek}, \texttt{gemma}, \texttt{gemini}, \texttt{grok}, and \texttt{gpt}~attaining the highest values, and \texttt{llama}~models showing the lowest parity on average.
    Overall, refusals, consistency, and connectedness exhibit greater sensitivity to model version, while the other metrics remain largely stable across models within the same family.
    }
    \label{app:fig:rq1:model}
\end{figure*}

\subsection{AQ1: Infrastructure-level Conditions}
\label{app:sec:rq1}

\para{Performance by model.}
\Cref{app:fig:rq1:model}~reveals heterogeneity across models, with \texttt{deepseek}~standing out as a clear exception among open-weight systems. \texttt{deepseek}~models achieve the highest factuality overall and simultaneously rank near the top for parity, while also maintaining high connectedness, similarity, and diversity. This joint performance across five dimensions is not observed for other open-weight models, which typically excel in only a subset of metrics (e.g., \texttt{mistral}~is the most reliable in terms of validity across models, but less factual than \texttt{deepseek}).
The closest to \texttt{deepseek}~across these dimensions is \texttt{gemini-2.5-pro}, which also exhibit strong factuality and competitive values for connectedness, similarity, diversity, and parity.
In contrast, \texttt{llama}, \texttt{gemma}, and \texttt{mistral}~models prioritize validity and structured output reliability but lag behind on factuality and in some cases on representation metrics too. 
Overall, these results indicate that \texttt{deepseek}~is the only model family that jointly optimizes author factuality and multiple representation metrics, rather than trading them off. %

\para{Deviations from aggregated infrastructure trends.}
The model-level view in~\Cref{app:fig:rq1:model} shows that aggregate infrastructure effects in \Cref{fig:rq1} describe central tendencies, but several individual models diverge in informative ways.
For \textit{model access}, proprietary models exhibit fewer refusals on average, yet this advantage is not uniform. \texttt{gemini-2.5-flash}~shows higher refusal rates than several open-weight models, despite belonging to the proprietary group. Conversely, open-weight \texttt{gpt}~models display lower validity than proprietary models, even though open models, on average, achieve higher validity. These cases indicate that access-level differences mask some within-group variation, particularly in technical quality.
Trends in \textit{model size} also weaken at the individual level. 
While validity increases with size in the aggregate, this relationship is not monotonic across models. \texttt{deepseek-chat-v3.1}, classified as extra-large, exhibits low validity, falling below many small and medium models. This suggests that scale does not guarantee usable outputs and can amplify failure modes when list generation breaks down.
\textit{Reasoning capability} shows the strongest alignment between aggregate and model-level results. Validity is consistently higher among reasoning-disabled models, while reasoning-enabled models achieve higher factuality. This pattern holds for most models, with \texttt{deepseek-r1-0528}~as a partial exception. Despite being reasoning-enabled, it attains factuality comparable to top non-reasoning models, indicating that some architectures mitigate the usual validity–factuality trade-off associated with explicit reasoning.
Overall, the model-level analysis clarifies that aggregate infrastructure effects are directionally correct but incomplete. Individual models frequently diverge from group averages, often in ways that contradict naive expectations about access, size, or reasoning. 
These deviations show that, while aggregate results capture shared patterns across model classes, model-specific behavior is critical when selecting individual systems for deployment.

\begin{figure*}[t]
    \centering
    \includegraphics[width=1.0\textwidth]{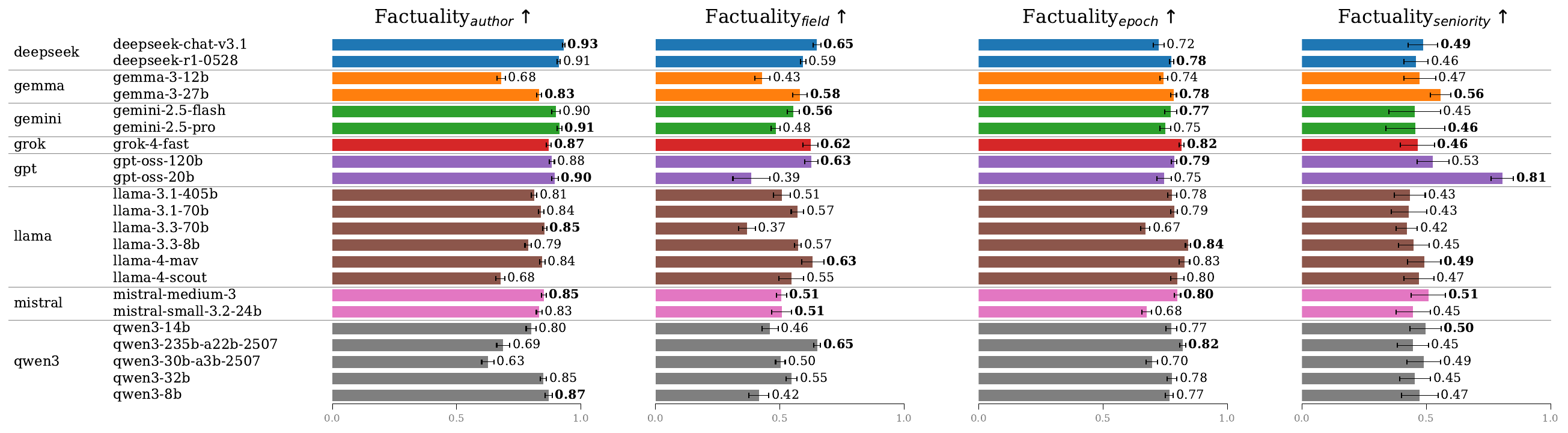}
    \caption{Baseline factuality performance by model.
    Mean factuality ($\pm95\%$ CI) across four attributes: whether recommended authors are real individuals, belong to the requested field, were active during the requested epoch, and match the requested seniority. 
    \texttt{deepseek}~and \texttt{gemini}~achieve the highest overall author factuality. 
    Field-level factuality is highest for \texttt{deepseek}, \texttt{grok}, \texttt{gpt}, and \texttt{llama-4-mav}, which most consistently return scholars from the requested field. 
    For epoch-specific requests, the smallest \texttt{llama}~variant (\texttt{llama-3.3-8b}) yields the highest factuality. 
    For seniority-specific requests, the medium-sized \texttt{gpt}~model (\texttt{gpt-oss-20b}) performs best on average.
    }
    \label{app:fig:rq1:model:factuality}
\end{figure*}

\para{Factuality beyond author identity.}
Beyond verifying that recommended names correspond to real authors, we evaluate whether models satisfy the \texttt{criteria} specified in the prompt (\Cref{app:fig:prompt}). Author-level factuality can be assessed for all tasks, as each response necessarily contains recommended names. In contrast, the \texttt{field}, \texttt{epoch}, and \texttt{seniority}~tasks impose additional factual requirements. For the \texttt{field}~task, we verify whether the recommended author has published in the requested APS journal category (CMMP or PER). For the \texttt{epoch}~task, we verify whether the author’s publication years overlap with the requested epoch (1950-1960 or 2000-2010). For the \texttt{seniority}~task, we verify whether the author’s academic age, inferred from the span of publication years, satisfies the requested career stage ($\leq$10 years for early career, $\geq$20 years for senior).
\Cref{app:fig:rq1:model:factuality} reports average factuality scores per model. Author-level factuality is averaged across all tasks, including \texttt{top\_k}\ and \texttt{twins}, whereas field-, epoch-, and seniority-level factuality are computed only for their respective tasks. Across models, author factuality is consistently high, indicating that most systems reliably return real scholars. Factuality with respect to epoch is also relatively strong. %
In contrast, factuality for field and seniority is substantially lower and more variable.
This suggests that temporal constraints are easier to satisfy than topical or career-stage constraints.
While most models exhibit this behavior, a small number of exceptions emerge. For example, \texttt{gemma-3-12b}, \texttt{llama-4-scout}, \texttt{llama-3.3-8b}, and \texttt{qwen3}, \texttt{30b} and \texttt{235b}, achieve higher epoch factuality than author factuality, indicating that among the few factual authors temporal cues are likely satisfied. Overall, these results show that high author factuality does not guarantee correctness with respect to more specific scholarly attributes, exposing a clear trade-off between name validity and deeper factual grounding.

\begin{figure*}[t]
    \centering
    \begin{subfigure}{1.\textwidth}
        \includegraphics[width=\textwidth]{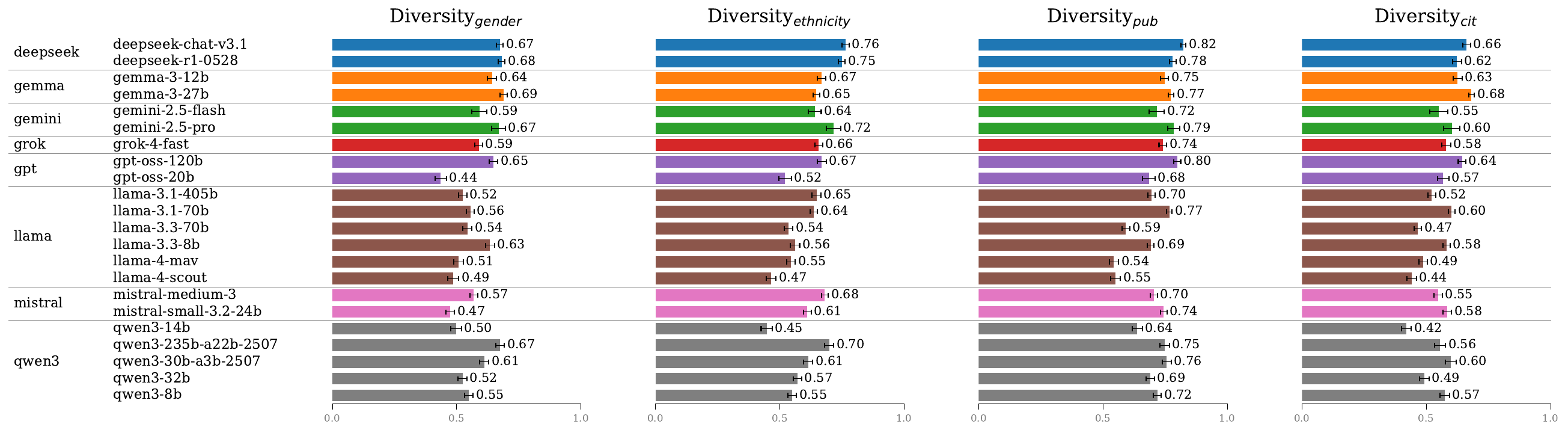}
        \caption{Diversity}
        \label{app:fig:rq1:model:social:diversity}
    \end{subfigure}
    \hfill
    \begin{subfigure}{1.\textwidth}
        \includegraphics[width=\textwidth]{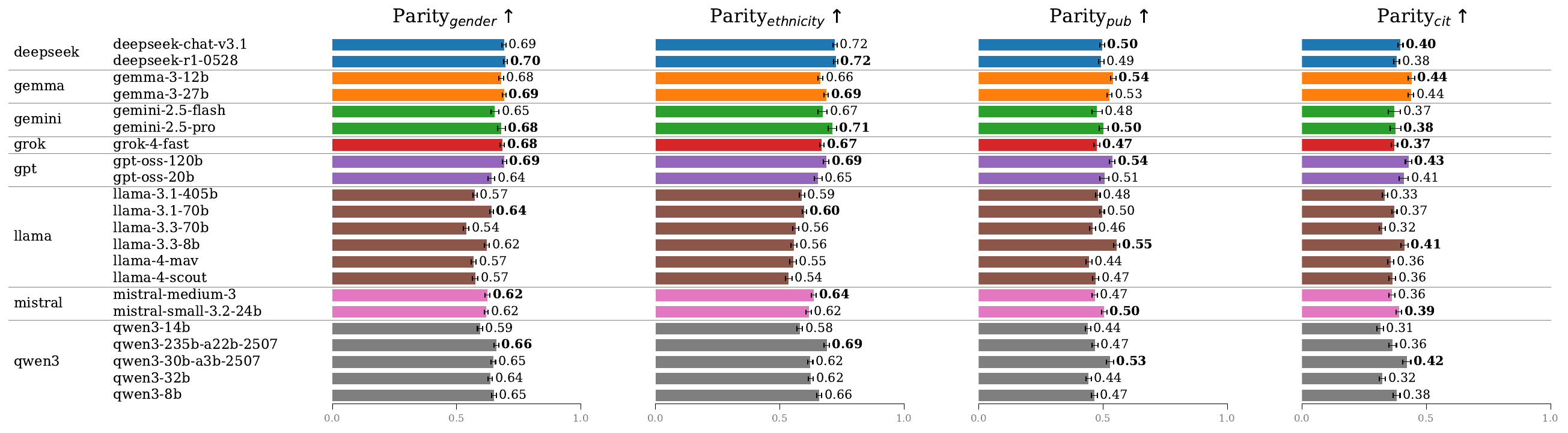}
        \caption{Parity}
        \label{app:fig:rq1:model:social:parity}
    \end{subfigure}
    \caption{Baseline social-benchmark performance by model.
    (a) Mean diversity of recommendations across all four attributes. \texttt{deepseek}~produces the most diverse recommendations on average, while \texttt{llama}~models exhibit the lowest diversity.
    (b) Mean parity of recommendations. \texttt{deepseek}~attains the highest parity for gender and ethnicity. 
    For parity with respect to scholarly prominence, measured by publication and citation strata, \texttt{llama-3.3-8b}~performs best, followed by \texttt{gpt-oss-120b}~and \texttt{gemma-3-12b}.
    }
    \label{app:fig:rq1:model:social}
\end{figure*}

\para{Diversity and parity beyond gender.}
While \Cref{app:fig:rq1:model} reports diversity and parity with respect to gender, \Cref{app:fig:rq1:model:social} extends this analysis to ethnicity and scholarly prominence, measured by publications and citations, across all models. 
Results for diversity (\Cref{app:fig:rq1:model:social:diversity}) and parity (\Cref{app:fig:rq1:model:social:parity}) show consistent patterns across dimensions.
\texttt{deepseek}~models achieve the highest diversity across all attributes considered, including gender, ethnicity, publications, and citations. 
This advantage partially extends to parity, where \texttt{deepseek}~leads for gender and ethnicity but not for publication- or citation-based parity. 
\texttt{gemma}, \texttt{gemini}, and \texttt{mistral}~models follow with also high performance, while \texttt{llama}~models are among the least diverse and consistently attain the lowest parity scores.
Across all attributes, variation in diversity and parity is driven primarily by model family rather than by differences between model versions within the same family. 
This indicates that social representativeness is a family-level property, in contrast to several technical quality metrics that vary more strongly across individual models.

\begin{figure*}[t]
    \centering
    \includegraphics[width=1.0\linewidth]{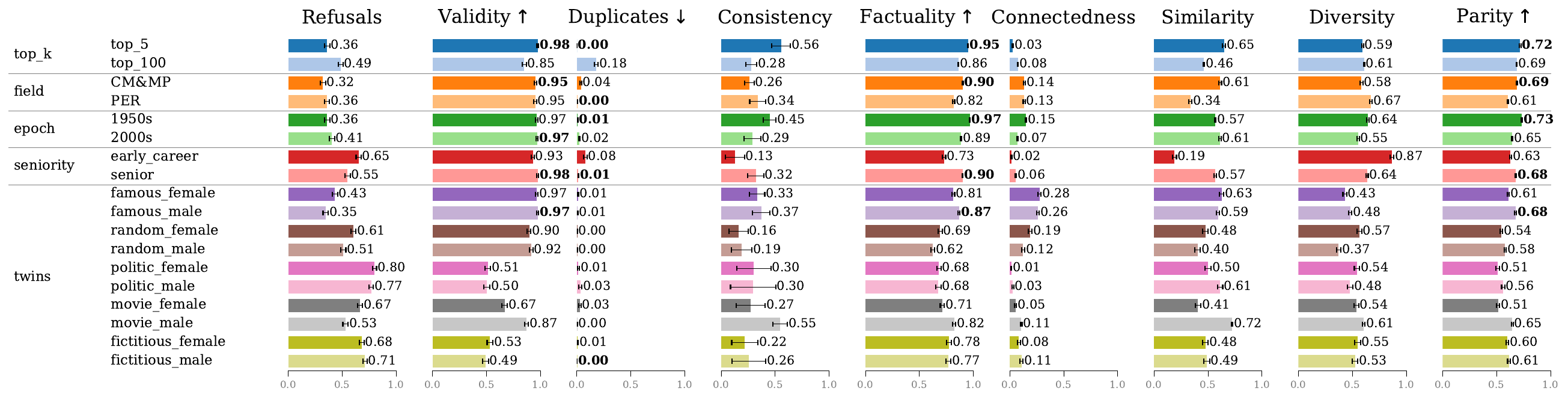}
    \caption{Baseline benchmark performance by task. 
    We report mean metric values ($\pm95\%$ CI) for each task. 
    Columns cover technical reliability metrics and social representativeness.
    Bars are color-coded by model task parameter.
    Results illustrate variation in difficulty across task parameters. 
    For example, increasing \texttt{top-k} reduces validity and increases duplicates, while twin-identification of politicians and fictitious names exhibit lower validity and higher refusals.}
    \label{app:fig:rq1:task}
\end{figure*}

\begin{figure*}[t]
    \centering
    \includegraphics[width=1.\linewidth]{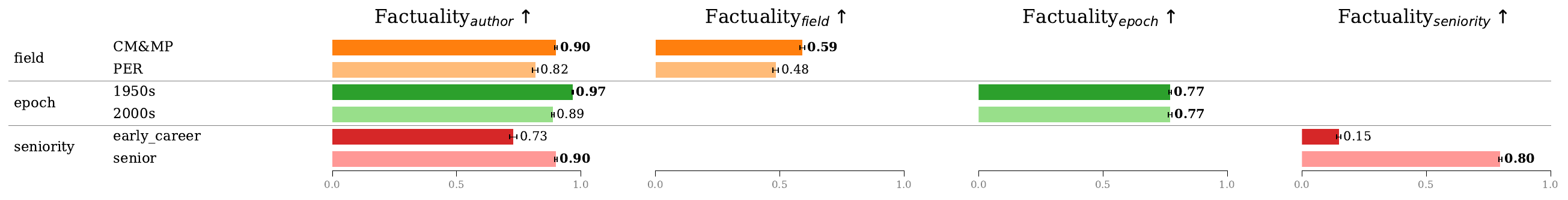}
    \caption{Task-level factuality. 
    LLMs are more likely to recommend real scientists than ensure correctness with respect to the requested field, epoch, or seniority, indicating increasing difficulty as constraints move from identity to attributes.
    }
    \label{app:fig:rq1:param:factuality}
\end{figure*}

\para{Task-level performance patterns.}
\Cref{app:fig:rq1:task} indicates that performance varies by task, with \texttt{twins}~showing the largest deviations across evaluation metrics.
Classical retrieval tasks, including \texttt{top\_k}, \texttt{field}-based queries, \texttt{epoch}, and \texttt{seniority}, consistently achieve higher validity, factuality, connectedness, diversity, and parity. In contrast, \texttt{twins} systematically lead to higher refusal rates and sharp drops across all quality and representation metrics, confirming that they are substantially more challenging and often ill-posed from the model perspective.
Within the set of classical retrieval tasks, performance is not uniform. \textit{Early-career} prompts yield the lowest factuality and similarity among retrieved authors, despite producing more diverse recommendations. This suggests that models struggle to anchor recommendations for less established scholars, even though they broaden the candidate set. By contrast, recommending experts from the \textit{1950s} and \textit{top-5} queries produce the most accurate outputs, with high factuality and strong parity, indicating that models perform best when targeting well-defined, historically established cohorts.
Prompting for \texttt{twins}~shows additional nuances. Twins of \textit{famous} individuals achieve substantially higher factuality than twins of \textit{random} individuals. They also lead to the highest connectedness, similarity, and parity among all twins variants. This aligns with prior findings that LLMs more reliably represent prominent scientists \cite{barolo2025whose, sandnes2024can, liu2025unequal}. These results indicate that models can distinguish between highly visible scientists and less prominent ones. When prompted with famous individuals, models tend to recommend authors who are not only similar in bibliometric terms but also closer in the coauthorship network, suggesting a stronger reliance on well-internalized scientific communities.
Overall, the task-level analysis shows that task formulation contributes to performance differences. Standard retrieval tasks (\texttt{top\_k}, \texttt{field}, \texttt{epoch}, \texttt{seniority}) yield accurate and balanced recommendations, whereas the \texttt{twins}~task exposes systematic limitations, with outcomes mediated by the prominence of the referenced individual.

\para{Task-level factuality beyond author identity.}
Despite author factuality being consistently high across tasks, accuracy with respect to field, epoch, and seniority is systematically lower. This indicates that identifying real scientists is substantially easier for LLMs than satisfying attribute-level constraints.
As shown in \Cref{app:fig:rq1:param:factuality}, field factuality is weaker than author factuality, with particularly low accuracy for \textit{PER}, the smallest APS subfield and the one with the highest proportion of women (32\% \cite{barolo2025whose}). 
Seniority exhibits the largest discrepancy: models are markedly more accurate at recommending \textit{senior} scholars than \textit{early-career} researchers, suggesting a strong bias toward established scientists.
For the \texttt{epoch}~task, epoch factuality is comparable between the \textit{1950s} and \textit{2000s} prompts. However, these prompts differ in author factuality, with recommendations for the \textit{1950s} more likely to correspond to real scientists. 
These patterns show multi-criteria factuality remains a challenge for LLMs.

\begin{figure*}[p]
    \centering
    \includegraphics[width=1.0\linewidth]{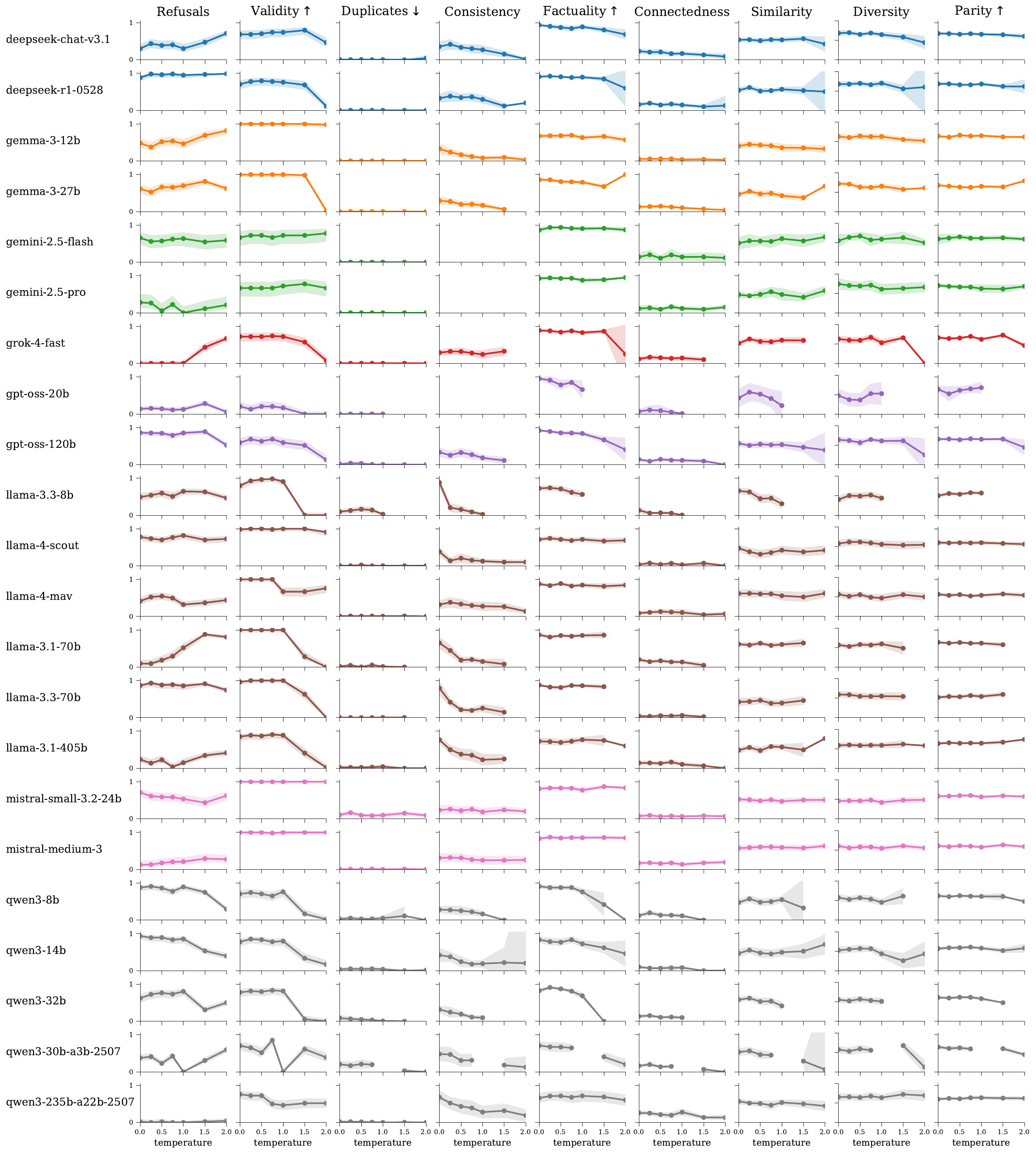}
    \caption{Effect of temperature on performance per model across all tasks.
    Mean metric values ($\pm$95\% CI) across temperature settings for each model, computed from the temporal analysis.
    Across models, higher temperatures systematically increase refusals with consistent declines in validity, consistency, factuality, and connectedness.
    In contrast, social representativeness metrics (similarity, diversity, and parity) remain largely insensitive to temperature.
    Duplicate rates are low for most models.
    }
    \label{app:fig:rq2:temperature}
\end{figure*}

\subsection{AQ2: End-user Interventions}
\label{app:sec:rq2}

\para{Varying temperature.}
The per-model analysis in~\Cref{app:fig:rq2:temperature} provides a finer-grained view of temperature as an intervention, complementing the infrastructure-level trends shown in~\Cref{fig:rq2:temperature}. While the aggregated results suggest smooth and largely monotonic effects of temperature, the disaggregated view shows that most individual models follow the same trends, with only a small number of exceptions. For the majority of models, increasing temperature consistently reduces validity, consistency, factuality, and connectedness, and increases refusals. Deviations from these patterns are limited to a few models that exhibit weaker sensitivity or delayed threshold effects.
For example, \texttt{llama-3} models maintain high validity up to a temperature of 1.0, beyond which validity declines sharply, whereas \texttt{gemma-3-12b}\ maintains high validity across the full temperature range.
Refusal rates increase for most models, though the onset and magnitude vary widely. 
In contrast, similarity, diversity, and parity remain largely stable across temperatures at the model level, confirming that the weak temperature sensitivity of social representativeness observed in the aggregated analysis is not an artifact of averaging. 
Overall, these results indicate that temperature control acts as a coarse intervention: its aggregate effects are predictable and largely consistent across models, but model-specific differences in sensitivity and threshold behavior limit its usefulness as a precise mechanism for steering performance.

\begin{figure*}[p]
    \centering
    \includegraphics[width=1\linewidth]{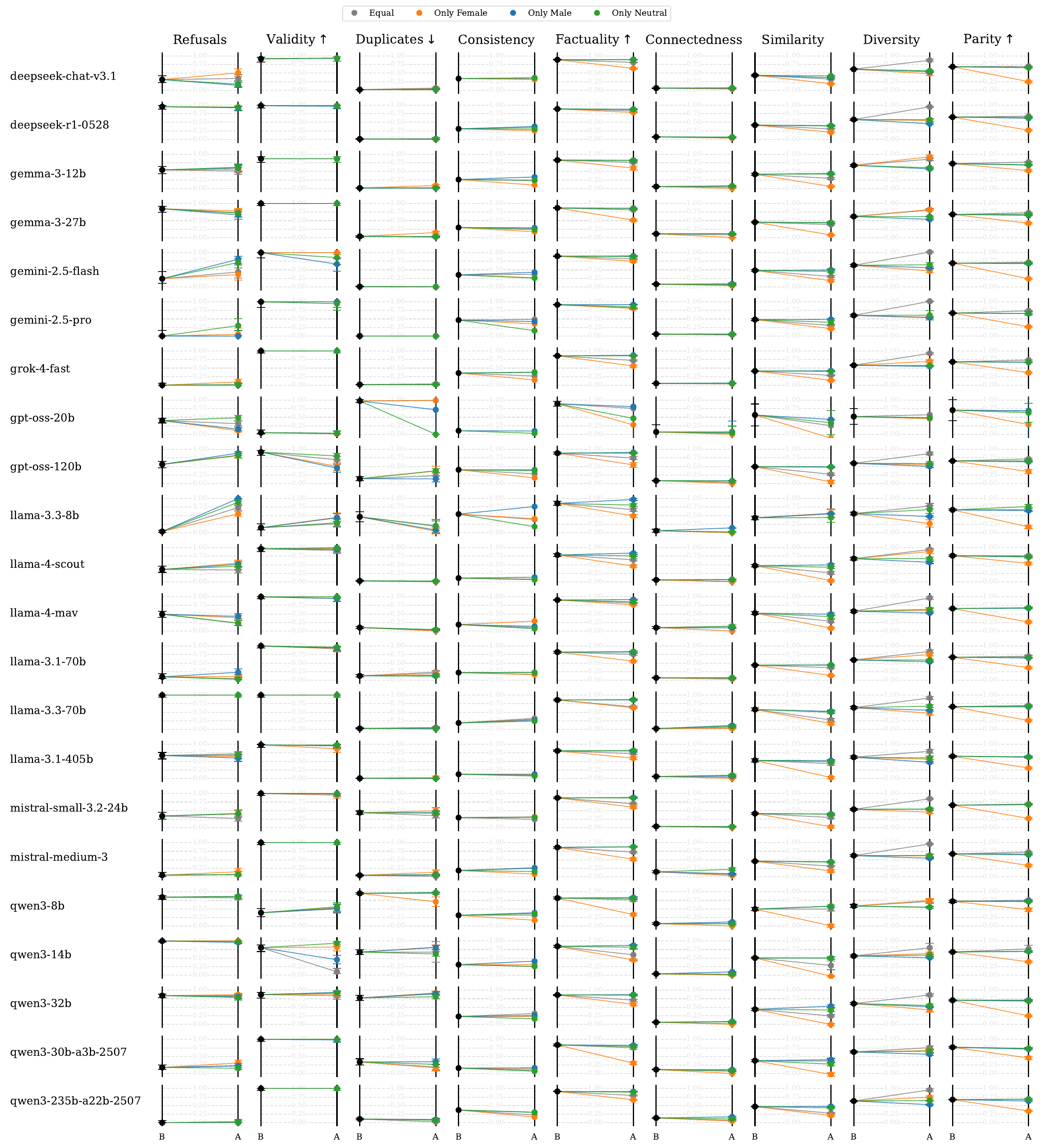}
    \caption{Model-level performance under gender-constrained prompting for top-100 expert recommendations.
    Mean values ($\pm 95\%$ CI) before (B) and after (A) gender-constrained prompting. 
    Colored points indicate mean performance under different prompting conditions (equal representation across all genders, female-only, male-only, neutral-names only), with lines showing changes relative to the (no intervention) baseline prompt.
    }
    \label{app:fig:rq2:biased_prompt:model:gender}
\end{figure*}

\begin{figure*}[p]
    \centering
    \includegraphics[width=1\linewidth]{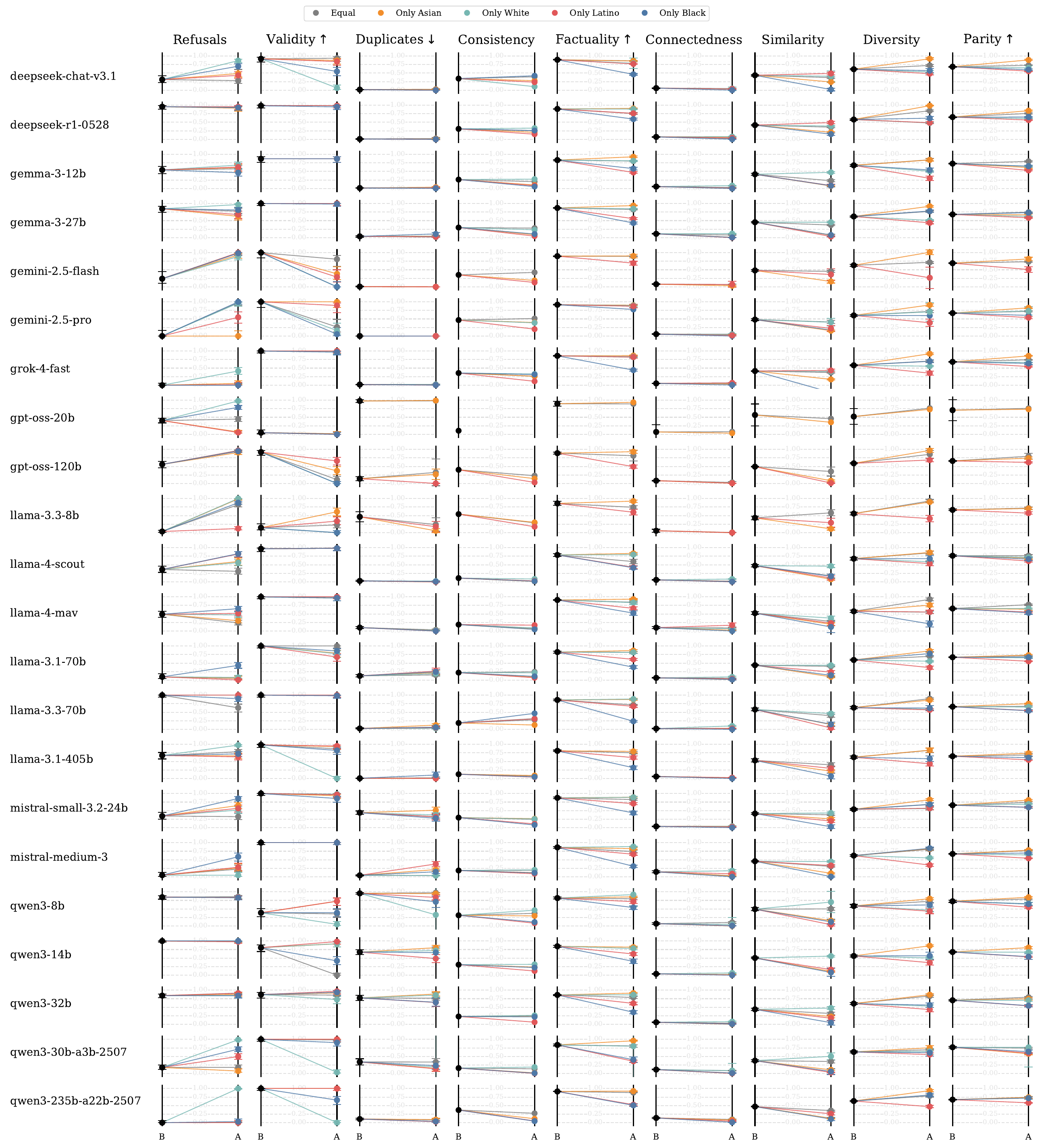}
    \caption{Model-level performance under ethnicity-constrained prompting for top-100 expert recommendations.
    Mean values ($\pm 95\%$ CI) before (B) and after (A) ethnicity-constrained prompting. 
    Colored points indicate mean performance under different prompting conditions (equal representation across all ethnicities, Asian-only, White-only, Latino-only, and Black only), with lines showing changes relative to the (no intervention) baseline prompt.
    }
    \label{app:fig:rq2:biased_prompt:model:ethnicity}
\end{figure*}

\begin{figure*}[p]
    \centering
    \includegraphics[width=1\linewidth]{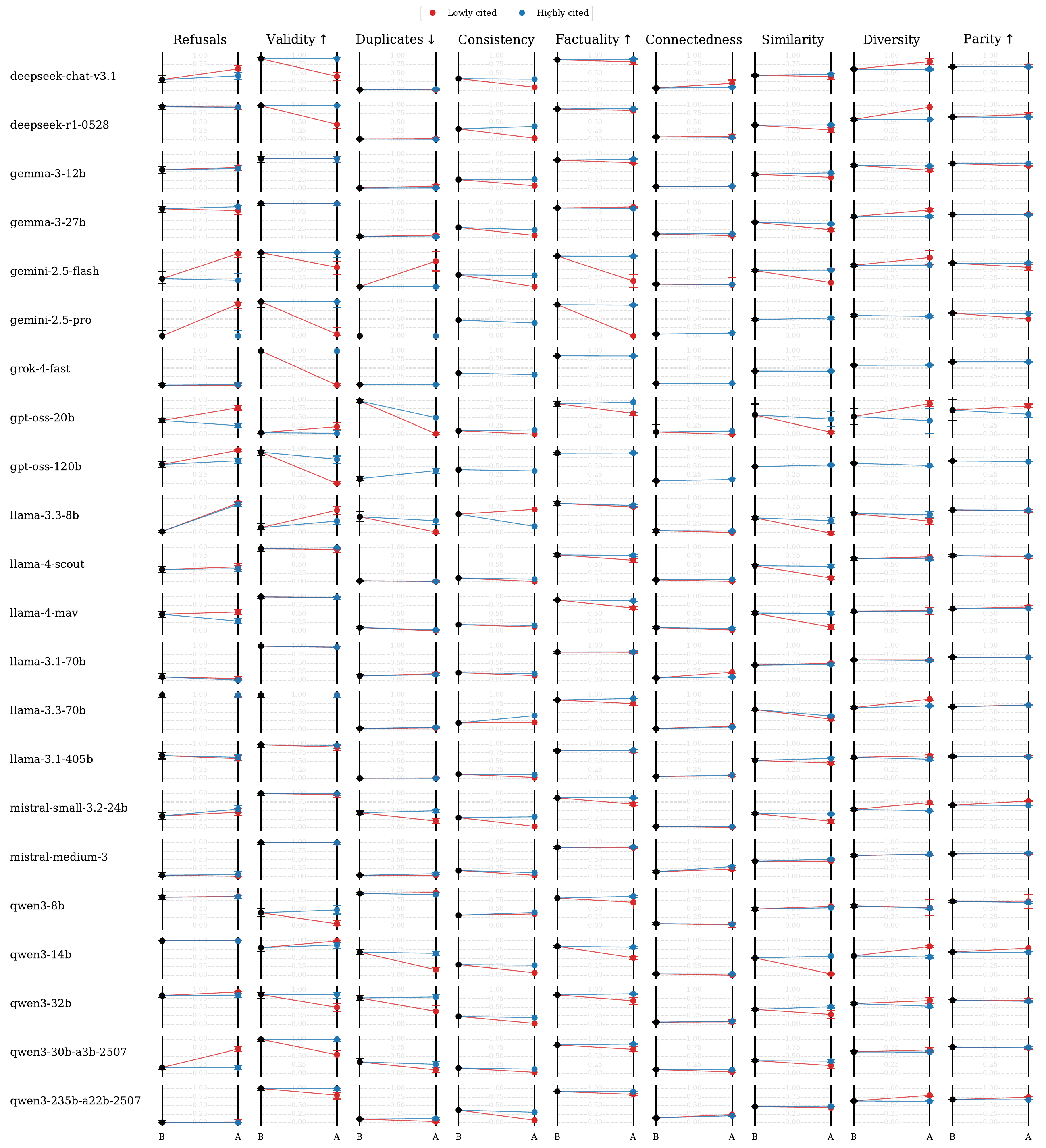}
    \caption{Model-level performance under citation-constrained prompting for top-100 expert recommendations.
    Mean values ($\pm 95\%$ CI) before (B) and after (A) citation-constrained prompting. 
    Colored points indicate mean performance under different prompting conditions (lowly cited-only, highly cited-only), with lines showing changes relative to the (no intervention) baseline prompt.
    }
    \label{app:fig:rq2:biased_prompt:model:citations}
\end{figure*}

\begin{figure*}[p]
    \centering
    \includegraphics[width=1\linewidth]{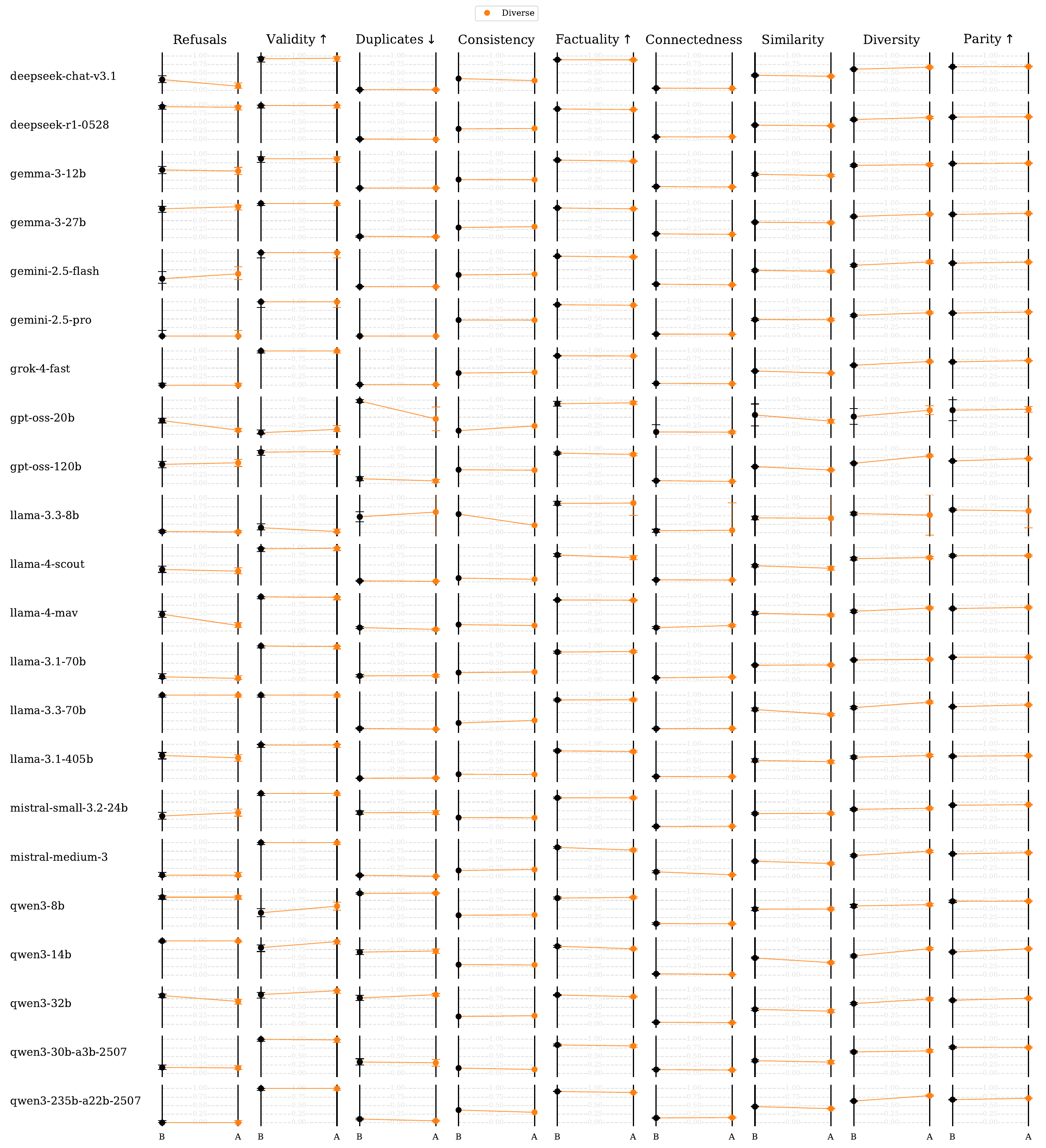}
    \caption{Model-level performance under general diversity-constrained prompting for top-100 expert recommendations.
    Mean values ($\pm 95\%$ CI) before (B) and after (A) general diversity-constrained prompting. 
    Lines show changes relative to the (no intervention) baseline prompt.
    }
    \label{app:fig:rq2:biased_prompt:model:general}
\end{figure*}

\begin{figure*}[t]
    \centering
    \begin{subfigure}[t]{1.\textwidth}
        \centering
        \includegraphics[width=1.\textwidth]{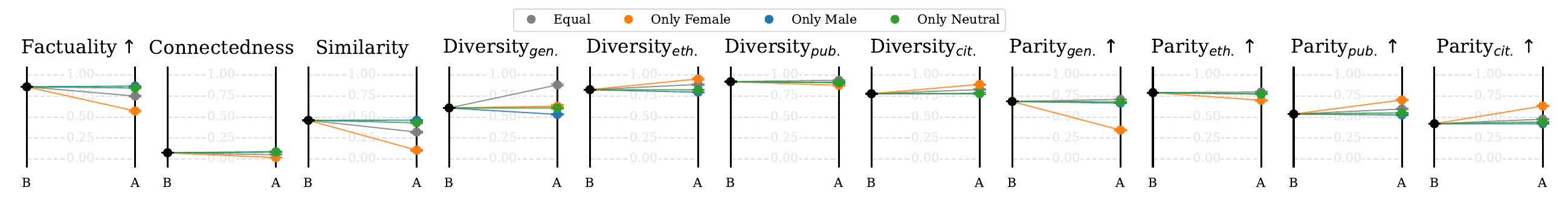}
        \caption{Gender-constrained}
        \label{app:fig:rq2:biased_prompt:social:gender}
    \end{subfigure}
    \hfill
    \begin{subfigure}[t]{1.\textwidth}
        \centering
        \includegraphics[width=1.\textwidth]{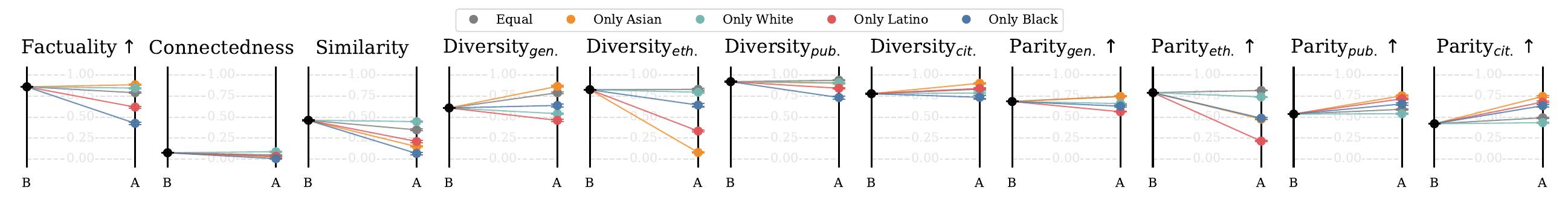}
        \caption{Ethnicity-constrained}
        \label{app:fig:rq2:biased_prompt:social:ethnicity}
    \end{subfigure}
    \hfill
    \begin{subfigure}[t]{1.\textwidth}
        \centering
        \includegraphics[width=1.\textwidth]{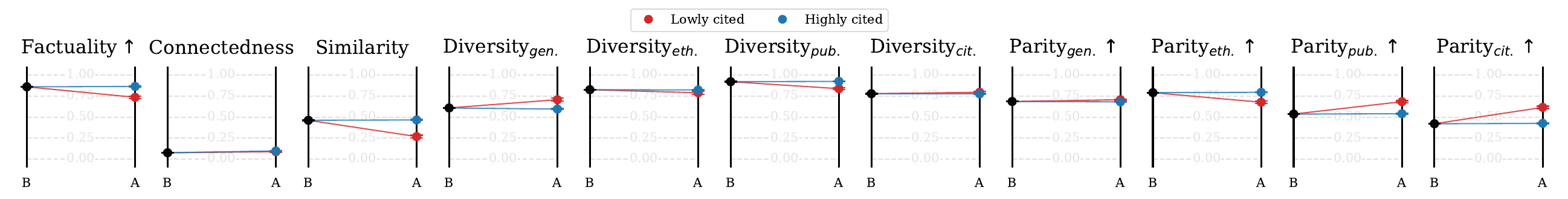}
        \caption{Citations-constrained}
        \label{app:fig:rq2:biased_prompt:social:citations}
    \end{subfigure}
    \hfill
    \begin{subfigure}[t]{1.\textwidth}
        \centering
        \includegraphics[width=1.\textwidth]{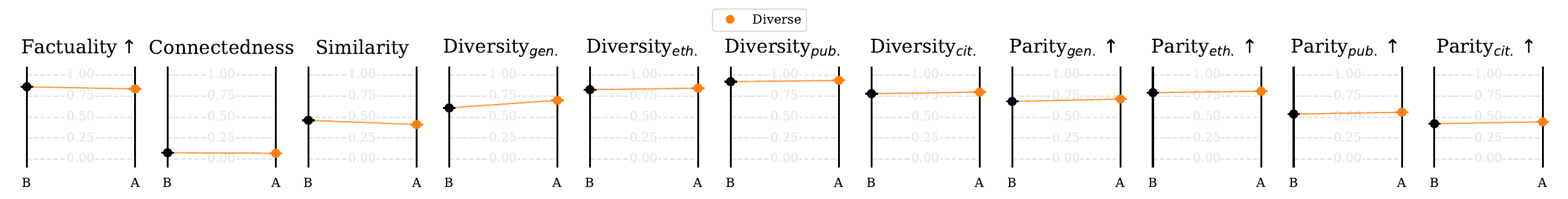}
        \caption{General diversity-constrained}
        \label{app:fig:rq2:biased_prompt:social:general}
    \end{subfigure}
    \caption{Trade-offs induced by constrained prompting in social representativeness for top-100 expert recommendations.
    Mean values ($\pm95\%$ CI) before (B) and after (A) constrained prompting.
    (a-c) The lowest factuality scores occur when requesting recommendations restricted to only female, only Black, or lowly cited scholars. 
    When asked for equal representation by gender or ethnicity, models do increase diversity for those attributes. 
    (d) In contrast, requesting a generally diverse list of scientists does not ensure diversity across ethnicity, publications, or citations. It only improves gender diversity at the cost of factuality.
    }
    \label{app:fig:rq2:biased_prompt:social}
\end{figure*}

\begin{figure*}[t]
    \centering
    \begin{subfigure}{1.\textwidth}
        \includegraphics[width=1.\textwidth]{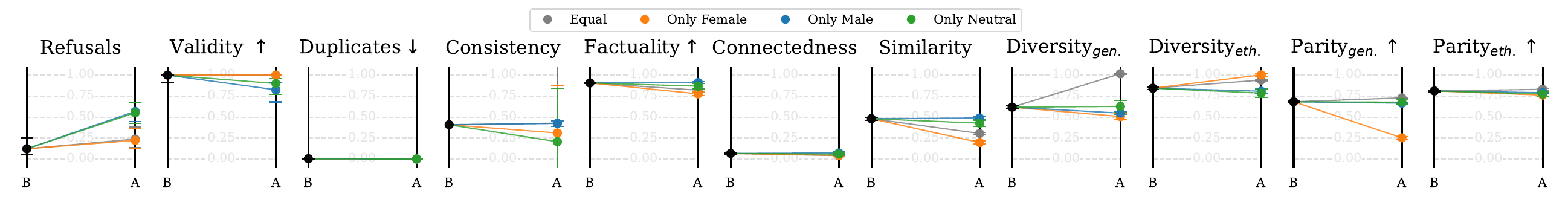}
        \caption{Constrained prompting only}
        \label{app:fig:constrained_rag:gender:cp}
    \end{subfigure}
    \hfill
    \begin{subfigure}{1.\textwidth}
        \includegraphics[width=1.\textwidth]{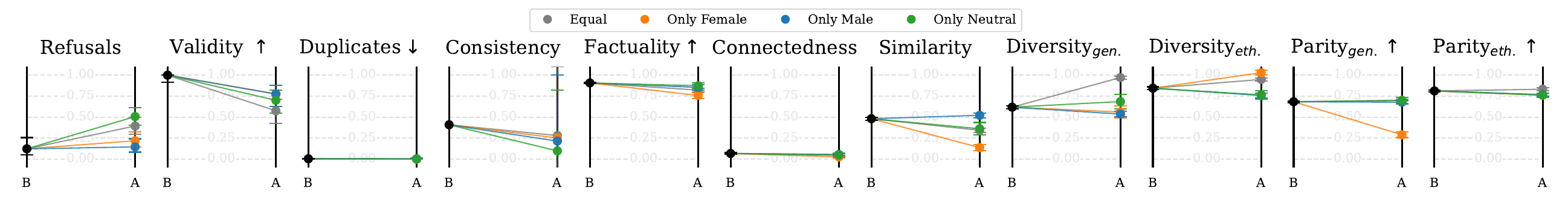}
        \caption{Constrained prompting with RAG}
        \label{app:fig:constrained_rag:gender:cpr}
    \end{subfigure}
    \caption{\textbf{Effects of gender-constrained prompting, with and without RAG (top-100 tasks, \texttt{gemini}).}
    Results are for top-100 expert recommendation lists generated with \texttt{gemini}~models.
    Each panel shows the mean metric value ($\pm$95\% CI) before (B) and after (A) applying a constraint that requests a single target gender distribution:
     \textbf{(a, top)} applies constrained prompting alone; \textbf{(b, bottom)} combines constrained prompting with RAG. 
     Adding RAG to constrained prompting reduces refusals but also lowers validity and consistency, while leaving duplicates, factuality, and the social metrics largely unchanged.
    }
    \label{app:fig:constrained_rag:gender}
\end{figure*}

\begin{figure*}[t]
    \centering
    \begin{subfigure}{1.\textwidth}
        \includegraphics[width=1.\textwidth]{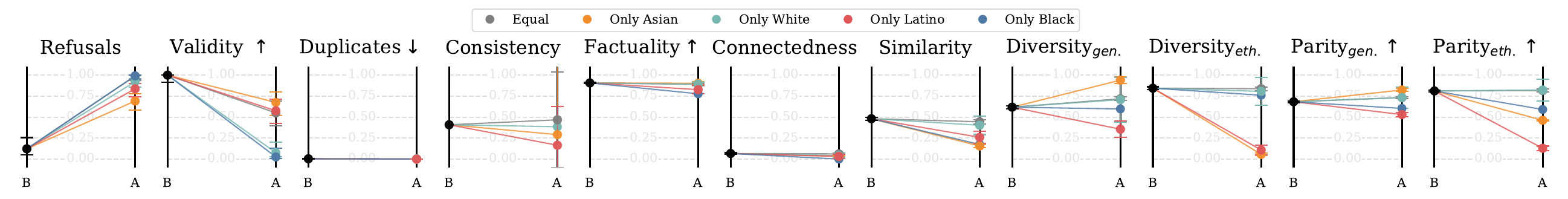}
        \caption{Constrained prompting only}
        \label{app:fig:constrained_rag:ethnicity:cp}
    \end{subfigure}
    \hfill
    \begin{subfigure}{1.\textwidth}
        \includegraphics[width=1.\textwidth]{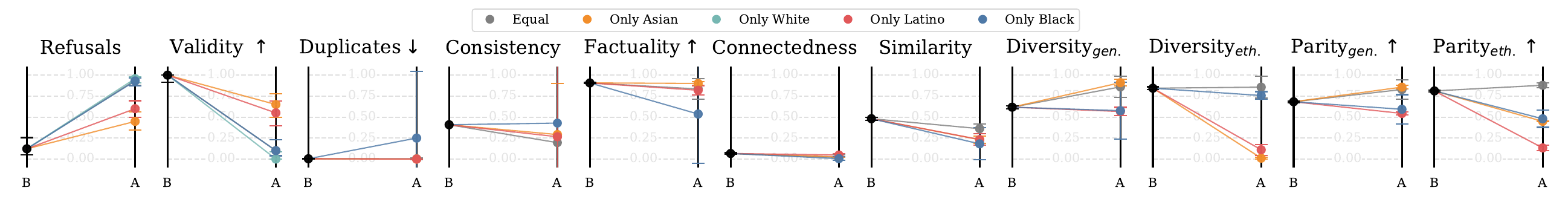}
        \caption{Constrained prompting with RAG}
        \label{app:fig:constrained_rag:ethnicity:cpr}
    \end{subfigure}
    \caption{\textbf{Effects of ethnicity-constrained prompting, with and without RAG (top-100 tasks, \texttt{gemini}).}
    Results are for top-100 expert recommendation lists generated with \texttt{gemini}~models.
    Each panel shows the mean metric value ($\pm$95\% CI) before (B) and after (A) applying a constraint that requests a single target ethnicity distribution:
     \textbf{(a, top)} applies constrained prompting alone; \textbf{(b, bottom)} combines constrained prompting with RAG. 
    Adding RAG to constrained prompting produces target-specific effects. It reduces refusals for Latino-only and Asian-only prompts but raises them for White-only, and increases duplicates for Black-only. Factuality drops for Black-only, while gender diversity improves for Equal and Latino-only targets.
    }
    
    \label{app:fig:constrained_rag:ethnicity}
\end{figure*}

\begin{figure*}[t]
    \centering
    \begin{subfigure}{1.\textwidth}
        \includegraphics[width=1.\textwidth]{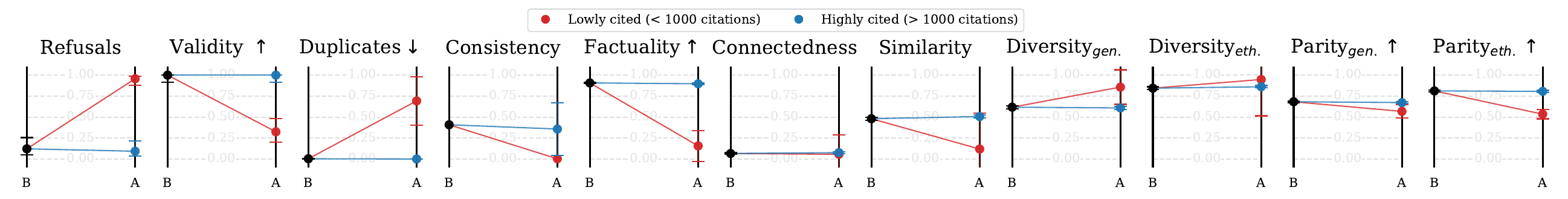}
        \caption{Constrained prompting only}
        \label{app:fig:constrained_rag:citations:cp}
    \end{subfigure}
    \hfill
    \begin{subfigure}{1.\textwidth}
        \includegraphics[width=1.\textwidth]{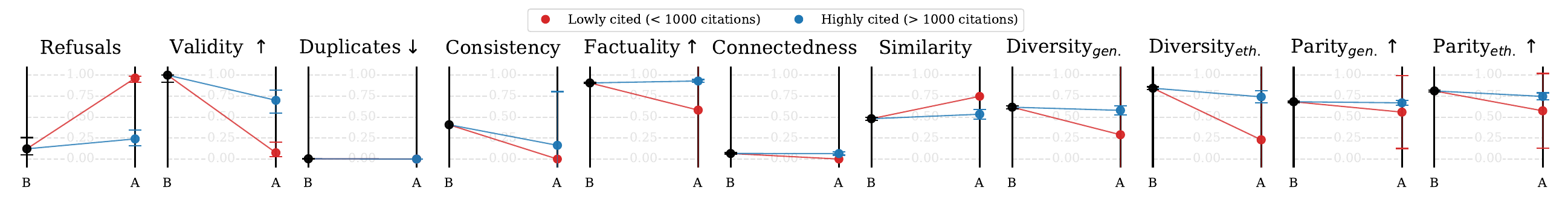}
        \caption{Constrained prompting with RAG}
        \label{app:fig:constrained_rag:citations:cpr}
    \end{subfigure}
    \caption{\textbf{Effects of citations-constrained prompting, with and without RAG (top-100 tasks, \texttt{gemini}).}
    Results are for top-100 expert recommendation lists generated with \texttt{gemini}~models.
    Each panel shows the mean metric value ($\pm$95\% CI) before (B) and after (A) applying a constraint requesting a single target citation-based distribution:
     \textbf{(a, top)} applies constrained prompting alone; \textbf{(b, bottom)} combines constrained prompting with RAG. 
    Adding RAG increases refusals for highly cited recommendations, reduces validity for both groups and consistency for highly cited recommendations. It also raises within-list similarity among lowly cited recommendations, at the cost of lower ethnicity and gender diversity within that group.
    }
    \label{app:fig:constrained_rag:citations}
\end{figure*}

\begin{figure*}[t]
    \centering
    \begin{subfigure}{1.\textwidth}
        \includegraphics[width=1.\textwidth]{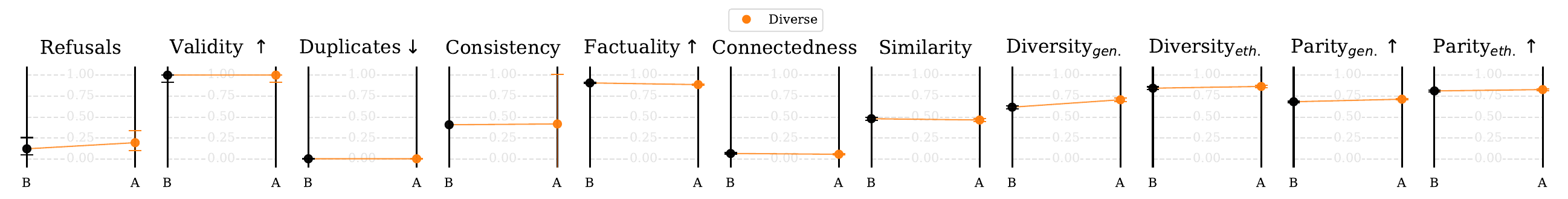}
        \caption{Constrained prompting only}
        \label{app:fig:constrained_rag:diverse:cp}
    \end{subfigure}
    \hfill
    \begin{subfigure}{1.\textwidth}
        \includegraphics[width=1.\textwidth]{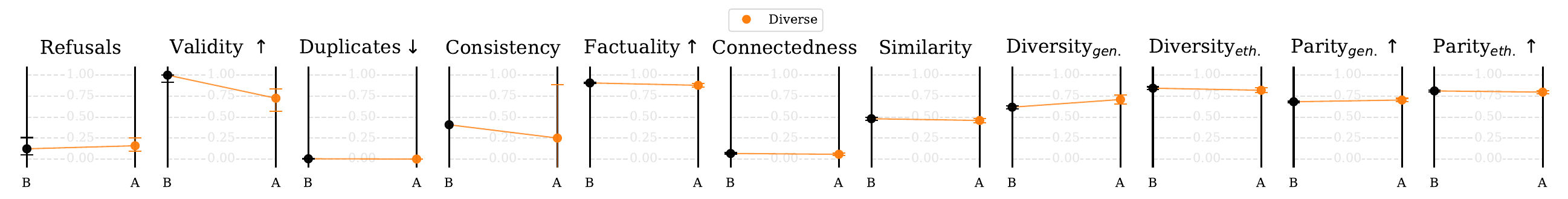}
        \caption{Constrained prompting with RAG}
        \label{app:fig:constrained_rag:diverse:cpr}
    \end{subfigure}
    \caption{\textbf{Effects of general diversity-constrained prompting, with and without RAG (top-100 tasks, \texttt{gemini}).}
    Results are for top-100 expert recommendation lists generated with \texttt{gemini}~models.
    Each panel shows the mean metric value ($\pm$95\% CI) before (B) and after (A) applying a constraint that requests a balanced gender distribution:
     \textbf{(a, top)} applies constrained prompting alone; \textbf{(b, bottom)} combines constrained prompting with RAG. 
     Adding RAG to constrained prompting reduces validity, while leaving most other metrics largely unchanged. Consistency also decreases on average, although the confidence interval is wide.
     }
    \label{app:fig:constrained_rag:diverse}
\end{figure*}

\para{Constrained prompting.}
\Cref{app:fig:rq2:biased_prompt:model:gender} reports model-level performance under gender-constrained prompting, complementing aggregated results (\Cref{seq:rq2}). 
Across models, technical quality metrics remain largely stable, with deviations limited to a small subset. 
In contrast, constrained prompting primarily reshapes social representativeness, with effects determined by the constraint direction.
Requests for \textit{female-only} recommendations consistently reduce factuality, similarity, and parity across all models. The parity decline is expected given the low base rate of women in APS data (\Cref{app:sec:gt:demo}); enforcing single-gender lists therefore violates statistical parity by construction rather than correcting it.
We observe similar trade-offs for ethnicity-, prominence-, and diversity-constrained prompts (\Cref{app:fig:rq2:biased_prompt:model:ethnicity,app:fig:rq2:biased_prompt:model:citations,app:fig:rq2:biased_prompt:model:general}). Ethnicity constraints increase refusal rates and reduce validity and factuality, with the largest factuality drop occurring for \textit{Black-only} prompts, indicating increased hallucination beyond APS coverage. 
Prominence constraints raise refusals and diversity when targeting \textit{lowly cited} scholars, at the cost of reduced validity, factuality, and similarity.
Aggregated results (\Cref{app:fig:rq2:biased_prompt:social:gender,app:fig:rq2:biased_prompt:social:ethnicity}) show that equal-representation constraints increase diversity only along the targeted dimension, with limited spillover to others. Citation-based constraints induce milder trade-offs, while generic diversity prompts fail to reliably improve diversity beyond gender and often reduce factuality.

Overall, constrained prompting does not uniformly improve social representativeness. 
Specific constraints enforce the requested composition but introduce predictable trade-offs, whereas broad diversity prompts lack generalizability across social dimensions.

\para{Constrained prompting and RAG.}
\Cref{app:fig:constrained_rag:diverse,app:fig:constrained_rag:gender,app:fig:constrained_rag:ethnicity,app:fig:constrained_rag:citations} show that combining RAG with constrained prompting rarely improves the representational outcomes achieved by the constraints themselves. Across constraint types, the dominant effect of RAG is a reduction in technical quality, particularly validity and occasionally consistency, while changes in factuality and social representation metrics are generally modest. Although some target-specific effects are observed, they are neither consistent across constraints nor large enough to offset the decline in technical performance. 
For general diversity constraints (\Cref{app:fig:constrained_rag:diverse}), RAG primarily lowers validity with limited effects on representation metrics. For gender constraints (\Cref{app:fig:constrained_rag:gender}), it similarly reduces validity and consistency while preserving most diversity and parity outcomes. For ethnicity constraints (\Cref{app:fig:constrained_rag:ethnicity}), RAG introduces more target-specific changes, including shifts in refusals, duplicates, and factuality for some groups, but without consistent representational gains. For citation-based constraints ((\Cref{app:fig:constrained_rag:citations}), RAG lowers validity and consistency and increases similarity among recommendations for low-citation targets.
These results reinforce the finding that RAG does not meaningfully strengthen constrained prompting and instead introduces additional quality trade-offs.

\begin{figure*}[t]
    \centering
    \begin{subfigure}{1.\textwidth}
        \includegraphics[width=1.\textwidth]{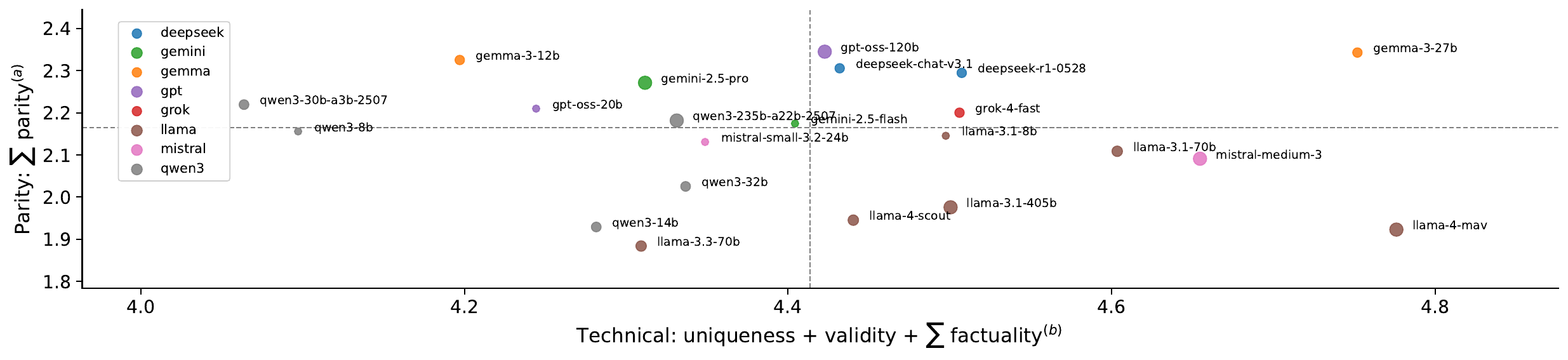}
        \caption{Baseline}
        \label{app:fig:quadrants:baseline}
    \end{subfigure}
    \hfill
    \begin{subfigure}{1.\textwidth}
        \includegraphics[width=1.\textwidth]{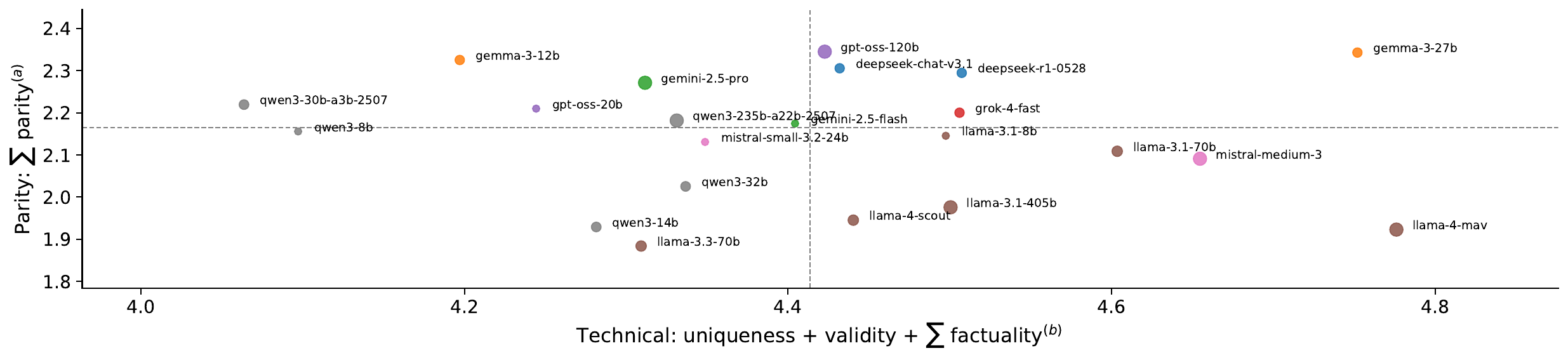}
        \caption{Temperature variation}
        \label{app:fig:quadrants:temperature}
    \end{subfigure}
    \hfill
    \begin{subfigure}{1.\textwidth}
        \includegraphics[width=1.\textwidth]{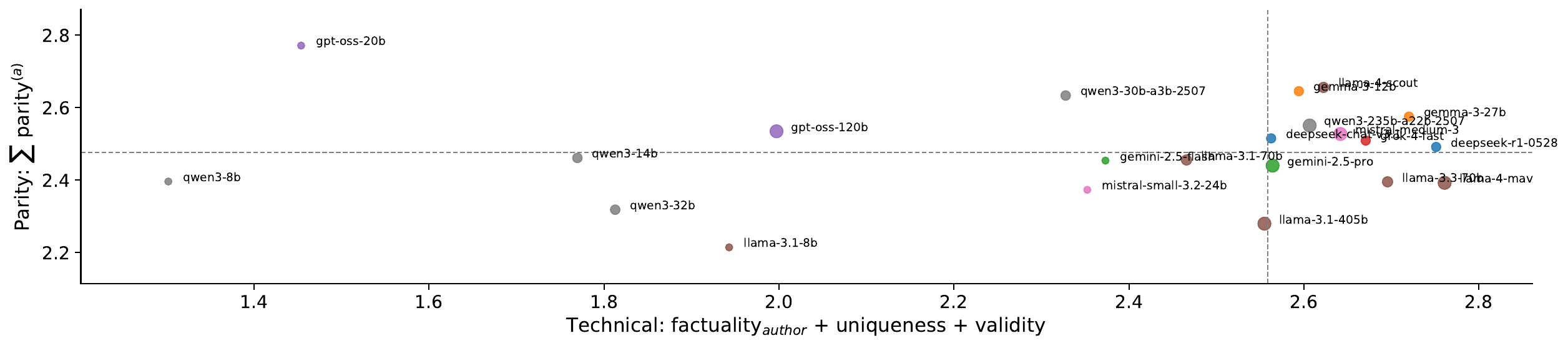}
        \caption{Constrained prompting}
        \label{app:fig:quadrants:constrained}
    \end{subfigure}
    \hfill
    \begin{subfigure}{1.\textwidth}
        \includegraphics[width=1.\textwidth]{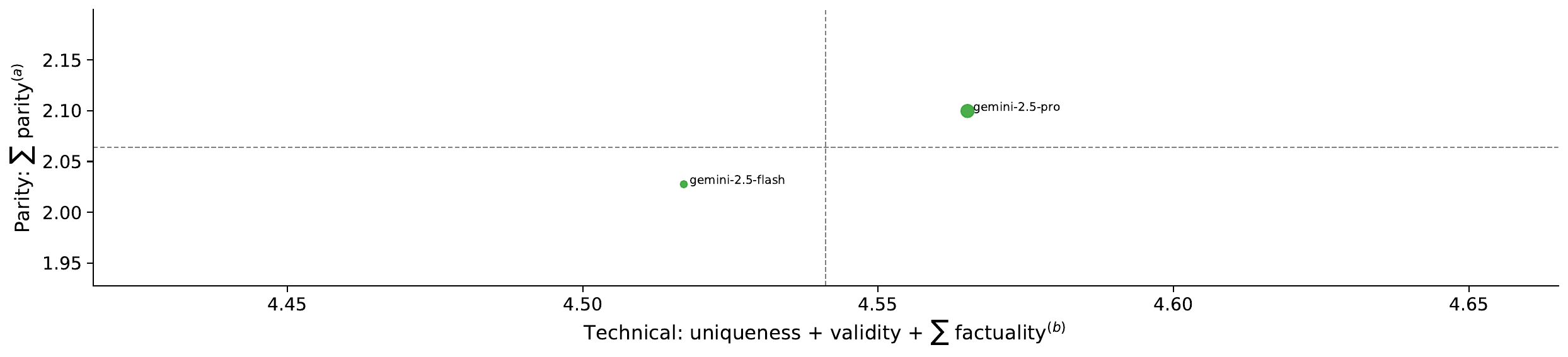}
        \caption{RAG}
        \label{app:fig:quadrants:rag}
    \end{subfigure}
    \hfill
    \caption{\textbf{Socio-technical trade-offs across inference-time configurations.}
    Each point represents a model. The x-axis measures technical quality and the y-axis parity as social representativeness; dashed lines indicate panel medians. Panels (a), (b), and (d) report averages across all tasks, whereas panel (c) reports results for the top-100 outputs under constrained prompting.
    In panels (a) and (b), \texttt{gemma-3-27b}~achieves strong performance on both dimensions, \texttt{llama-4-mav}~attains the highest technical quality, and \texttt{gpt-oss-120b}~the highest parity.
    In panel (c), no model simultaneously leads both dimensions: \texttt{llama-4-mav}~still achieves the highest technical quality, followed by \texttt{deepseek-r1-0528}, whereas \texttt{gpt-oss-20b}~attains the highest parity despite relatively low technical performance.
    In panel (d), \texttt{gemini-2.5-pro}{} outperforms \texttt{gemini-2.5-flash}{} in both dimensions under RAG, reversing their baseline ordering in technical quality observed in panel (a).
}
\Description{Four scatter plots showing model-level socio-technical trade-offs under baseline, temperature variation, constrained prompting, and RAG settings. The x-axis reports aggregated technical quality and the y-axis aggregated parity. Dashed lines denote panel medians. Model rankings and the balance between technical and social performance vary across inference-time configurations.}
    \label{app:fig:quadrants}
\end{figure*}

\subsection{Socio-Technical Trade-off}
\label{app:sec:tradeoffs}
\Cref{app:fig:quadrants} provides a joint view of technical and social performance that complements the metric-by-metric analyses presented throughout the paper. While individual metrics are necessary to diagnose specific failure modes, aggregated views are useful when a deployment scenario prioritizes overall performance rather than isolated dimensions. In this setting, higher validity, higher uniqueness (fewer duplicates), higher factuality, and higher parity are unambiguously desirable, whereas other considerations depend on context and application requirements. 
When technical and social objectives are jointly prioritized, the preferred model is the one occupying the upper-right region of the socio-technical plane, corresponding to high socio-technical performance.

Here, social performance denotes the sum of per-model mean parity scores across demographic attributes, including perceived gender, perceived ethnicity, publication prominence, and citation prominence. Technical performance denotes the sum of per-model mean scores for validity, uniqueness (i.e., one minus the duplicate rate), and factuality across criteria (author, field, epoch, and seniority). Importantly, although these aggregates are computed as per-model means, they are not directly comparable across inference-time configurations because the underlying sets of requests differ in size and composition due to differences in the evaluated tasks and models.
\Cref{app:fig:quadrants:baseline} reports all models on all tasks using the selected temperature for each model; 
\Cref{app:fig:quadrants:temperature}  reports all models on all tasks across multiple temperature values; 
\Cref{app:fig:quadrants:constrained} reports all models at their selected temperature on the top-100 tasks under constrained prompting; and 
\Cref{app:fig:quadrants:rag} reports \texttt{gemini}~models at their selected temperature on all tasks under RAG. Consequently, the aggregate scores are useful for characterizing socio-technical trade-offs within each inference-time configuration, but should not be interpreted as directly comparable across panels. This limitation is particularly relevant when contrasting constrained prompting and RAG, since panel (c) is restricted to the top-100 tasks whereas panel (d) includes all tasks. %

Model rankings along these axes depend on the inference-time configuration. Temperature variation largely preserves the baseline ordering, indicating that stochasticity alone does not materially alter the socio-technical frontier. Under these settings, \texttt{gemma-3-27b}{} emerges as the strongest joint performer, despite being a medium-sized, open-weight, non-reasoning model. In contrast, \texttt{llama}~models consistently rank lower on social performance, even though \texttt{llama-4-mav}~attains the highest technical scores, highlighting a pronounced trade-off. Within the \texttt{gemini}~family, \texttt{gemini-2.5-flash}{} outperforms \texttt{gemini-2.5-pro}~on technical metrics at baseline, while the reverse holds for social performance.

The socio-technical frontier changes under constrained prompting and retrieval-augmented generation. Constrained prompting, applied only on the top 100 tasks, systematically increases social performance but reduces technical quality across models, reflecting a redistribution rather than a uniform improvement. Within the \texttt{gemini}~family, both constrained prompting and RAG are associated with stronger relative technical performance of \texttt{gemini-2.5-pro}~compared to \texttt{gemini-2.5-flash}, illustrating how inference-time interventions can shift the socio-technical frontier and alter which models are preferred under joint optimization criteria.
A more appropriate comparison between constrained prompting, RAG, and their combination is provided in~\Cref{seq:rq2}.


\begin{thebibliography}{67}


\ifx \showCODEN    \undefined \def \showCODEN     #1{\unskip}     \fi
\ifx \showISBNx    \undefined \def \showISBNx     #1{\unskip}     \fi
\ifx \showISBNxiii \undefined \def \showISBNxiii  #1{\unskip}     \fi
\ifx \showISSN     \undefined \def \showISSN      #1{\unskip}     \fi
\ifx \showLCCN     \undefined \def \showLCCN      #1{\unskip}     \fi
\ifx \shownote     \undefined \def \shownote      #1{#1}          \fi
\ifx \showarticletitle \undefined \def \showarticletitle #1{#1}   \fi
\ifx \showURL      \undefined \def \showURL       {\relax}        \fi
\providecommand\bibfield[2]{#2}
\providecommand\bibinfo[2]{#2}
\providecommand\natexlab[1]{#1}
\providecommand\showeprint[2][]{arXiv:#2}

\bibitem[Ali et~al\mbox{.}(2024)]%
        {ali2024automated}
\bibfield{author}{\bibinfo{person}{Nurshat~Fateh Ali},
  \bibinfo{person}{Md~Mahdi Mohtasim}, \bibinfo{person}{Shakil Mosharrof},
  {and} \bibinfo{person}{T~Gopi Krishna}.} \bibinfo{year}{2024}\natexlab{}.
\newblock \showarticletitle{Automated literature review using nlp techniques
  and llm-based retrieval-augmented generation}. In
  \bibinfo{booktitle}{\emph{2024 International Conference on Innovations in
  Science, Engineering and Technology (ICISET)}}. IEEE, \bibinfo{pages}{1--6}.
\newblock


\bibitem[Altmäe et~al\mbox{.}(2023)]%
        {altmae2023artificial}
\bibfield{author}{\bibinfo{person}{Signe Altmäe}, \bibinfo{person}{Alberto
  Sola-Leyva}, {and} \bibinfo{person}{Andres Salumets}.}
  \bibinfo{year}{2023}\natexlab{}.
\newblock \showarticletitle{Artificial intelligence in scientific writing: a
  friend or a foe?}
\newblock \bibinfo{journal}{\emph{Reproductive BioMedicine Online}}
  \bibinfo{volume}{47}, \bibinfo{number}{1} (\bibinfo{year}{2023}),
  \bibinfo{pages}{3--9}.
\newblock
\showISSN{1472-6483}
\href{https://doi.org/10.1016/j.rbmo.2023.04.009}{doi:\nolinkurl{10.1016/j.rbmo.2023.04.009}}


\bibitem[{American Physical Society}(2024)]%
        {aps_datasets}
\bibfield{author}{\bibinfo{person}{{American Physical Society}}.}
  \bibinfo{year}{2024}\natexlab{}.
\newblock \bibinfo{title}{APS Data Sets for Research}.
\newblock \bibinfo{howpublished}{\url{https://journals.aps.org/datasets}}.
\newblock
\newblock
\shownote{Accessed: 2024-10-12}.


\bibitem[Ang and Yang(2025)]%
        {ang2025whosenamecomesup}
\bibfield{author}{\bibinfo{person}{Yi~Zhe Ang} {and}
  \bibinfo{person}{Liuhuaying Yang}.} \bibinfo{year}{2025}\natexlab{}.
\newblock \bibinfo{title}{Whose Name Comes Up? An Interactive Visualization for
  Scholar Recommendation}.
\newblock \bibinfo{howpublished}{\url{https://vis.csh.ac.at/whosenamecomesup}}.
\newblock
\newblock
\shownote{Accessed: 2026-05-31}.


\bibitem[Balog et~al\mbox{.}(2009)]%
        {balog2009language}
\bibfield{author}{\bibinfo{person}{Krisztian Balog}, \bibinfo{person}{Leif
  Azzopardi}, {and} \bibinfo{person}{Maarten {de Rijke}}.}
  \bibinfo{year}{2009}\natexlab{}.
\newblock \showarticletitle{A language modeling framework for expert finding}.
\newblock \bibinfo{journal}{\emph{Information Processing \& Management}}
  \bibinfo{volume}{45}, \bibinfo{number}{1} (\bibinfo{year}{2009}),
  \bibinfo{pages}{1--19}.
\newblock
\showISSN{0306-4573}
\href{https://doi.org/10.1016/j.ipm.2008.06.003}{doi:\nolinkurl{10.1016/j.ipm.2008.06.003}}


\bibitem[Barolo et~al\mbox{.}(2025)]%
        {barolo2025whose}
\bibfield{author}{\bibinfo{person}{Daniele Barolo}, \bibinfo{person}{Chiara
  Valentin}, \bibinfo{person}{Fariba Karimi}, \bibinfo{person}{Luis
  Galárraga}, \bibinfo{person}{Gonzalo~G. Méndez}, {and}
  \bibinfo{person}{Lisette Espín-Noboa}.} \bibinfo{year}{2025}\natexlab{}.
\newblock \showarticletitle{Whose Name Comes Up? I: Auditing LLM-Based Scholar
  Recommendations}.
\newblock  (\bibinfo{year}{2025}).
\newblock
\href{https://doi.org/10.48550/arXiv.2506.00074}{doi:\nolinkurl{10.48550/arXiv.2506.00074}}


\bibitem[Bolukbasi et~al\mbox{.}(2016)]%
        {bolukbasi2016man}
\bibfield{author}{\bibinfo{person}{Tolga Bolukbasi}, \bibinfo{person}{Kai-Wei
  Chang}, \bibinfo{person}{James Zou}, \bibinfo{person}{Venkatesh Saligrama},
  {and} \bibinfo{person}{Adam Kalai}.} \bibinfo{year}{2016}\natexlab{}.
\newblock \showarticletitle{Man is to computer programmer as woman is to
  homemaker? debiasing word embeddings}. In
  \bibinfo{booktitle}{\emph{Proceedings of the 30th International Conference on
  Neural Information Processing Systems}} (Barcelona, Spain)
  \emph{(\bibinfo{series}{NIPS'16})}. \bibinfo{publisher}{Curran Associates
  Inc.}, \bibinfo{address}{Red Hook, NY, USA}, \bibinfo{pages}{4356–4364}.
\newblock
\showISBNx{9781510838819}


\bibitem[Chen et~al\mbox{.}(2025)]%
        {chen2025chineseecomqa}
\bibfield{author}{\bibinfo{person}{Haibin Chen}, \bibinfo{person}{Kangtao Lv},
  \bibinfo{person}{Chengwei Hu}, \bibinfo{person}{Yanshi Li},
  \bibinfo{person}{Yujin Yuan}, \bibinfo{person}{Yancheng He},
  \bibinfo{person}{Xingyao Zhang}, \bibinfo{person}{Langming Liu},
  \bibinfo{person}{Shilei Liu}, \bibinfo{person}{Wenbo Su}, {and}
  \bibinfo{person}{Bo Zheng}.} \bibinfo{year}{2025}\natexlab{}.
\newblock \showarticletitle{ChineseEcomQA: A Scalable E-commerce Concept
  Evaluation Benchmark for Large Language Models}. In
  \bibinfo{booktitle}{\emph{Proceedings of the 31st ACM SIGKDD Conference on
  Knowledge Discovery and Data Mining V.2 (KDD '25)}} (Toronto, ON, Canada).
  \bibinfo{publisher}{Association for Computing Machinery},
  \bibinfo{address}{New York, NY, USA}, \bibinfo{numpages}{11}~pages.
\newblock
\href{https://doi.org/10.1145/3711896.3737374}{doi:\nolinkurl{10.1145/3711896.3737374}}


\bibitem[Christen(2012)]%
        {christen2012data}
\bibfield{author}{\bibinfo{person}{Peter Christen}.}
  \bibinfo{year}{2012}\natexlab{}.
\newblock \showarticletitle{The Data Matching Process}.
\newblock In \bibinfo{booktitle}{\emph{Data Matching: Concepts and Techniques
  for Record Linkage, Entity Resolution, and Duplicate Detection}}.
  \bibinfo{publisher}{Springer Berlin Heidelberg}, \bibinfo{address}{Berlin,
  Heidelberg}, \bibinfo{pages}{23--35}.
\newblock
\showISBNx{978-3-642-31164-2}
\href{https://doi.org/10.1007/978-3-642-31164-2_2}{doi:\nolinkurl{10.1007/978-3-642-31164-2_2}}


\bibitem[Chugunova et~al\mbox{.}(2026)]%
        {CHUGUNOVA2026105381}
\bibfield{author}{\bibinfo{person}{Marina Chugunova}, \bibinfo{person}{Dietmar
  Harhoff}, \bibinfo{person}{Katharina Hölzle}, \bibinfo{person}{Verena
  Kaschub}, \bibinfo{person}{Sonal Malagimani}, \bibinfo{person}{Ulrike
  Morgalla}, {and} \bibinfo{person}{Robert Rose}.}
  \bibinfo{year}{2026}\natexlab{}.
\newblock \showarticletitle{Who uses AI in research, and for what? Large-scale
  survey evidence from Germany}.
\newblock \bibinfo{journal}{\emph{Research Policy}} \bibinfo{volume}{55},
  \bibinfo{number}{2} (\bibinfo{year}{2026}), \bibinfo{pages}{105381}.
\newblock
\showISSN{0048-7333}
\href{https://doi.org/10.1016/j.respol.2025.105381}{doi:\nolinkurl{10.1016/j.respol.2025.105381}}


\bibitem[Chung and Park(2012)]%
        {chung2012web}
\bibfield{author}{\bibinfo{person}{Chung~Joo Chung} {and}
  \bibinfo{person}{Han~Woo Park}.} \bibinfo{year}{2012}\natexlab{}.
\newblock \showarticletitle{Web visibility of scholars in media and
  communication journals}.
\newblock \bibinfo{journal}{\emph{Scientometrics}} \bibinfo{volume}{93},
  \bibinfo{number}{1} (\bibinfo{year}{2012}), \bibinfo{pages}{207--215}.
\newblock


\bibitem[Dai et~al\mbox{.}(2023)]%
        {dai2023uncovering}
\bibfield{author}{\bibinfo{person}{Sunhao Dai}, \bibinfo{person}{Ninglu Shao},
  \bibinfo{person}{Haiyuan Zhao}, \bibinfo{person}{Weijie Yu},
  \bibinfo{person}{Zihua Si}, \bibinfo{person}{Chen Xu},
  \bibinfo{person}{Zhongxiang Sun}, \bibinfo{person}{Xiao Zhang}, {and}
  \bibinfo{person}{Jun Xu}.} \bibinfo{year}{2023}\natexlab{}.
\newblock \showarticletitle{Uncovering ChatGPT’s Capabilities in Recommender
  Systems}. In \bibinfo{booktitle}{\emph{Proceedings of the 17th ACM Conference
  on Recommender Systems}} (Singapore, Singapore)
  \emph{(\bibinfo{series}{RecSys '23})}. \bibinfo{publisher}{Association for
  Computing Machinery}, \bibinfo{address}{New York, NY, USA},
  \bibinfo{pages}{1126–1132}.
\newblock
\showISBNx{9798400702419}
\href{https://doi.org/10.1145/3604915.3610646}{doi:\nolinkurl{10.1145/3604915.3610646}}


\bibitem[Di~Palma(2023)]%
        {diPalma2023retrieval}
\bibfield{author}{\bibinfo{person}{Dario Di~Palma}.}
  \bibinfo{year}{2023}\natexlab{}.
\newblock \showarticletitle{Retrieval-augmented Recommender System: Enhancing
  Recommender Systems with Large Language Models}. In
  \bibinfo{booktitle}{\emph{Proceedings of the 17th ACM Conference on
  Recommender Systems}} (Singapore, Singapore) \emph{(\bibinfo{series}{RecSys
  '23})}. \bibinfo{publisher}{Association for Computing Machinery},
  \bibinfo{address}{New York, NY, USA}, \bibinfo{pages}{1369–1373}.
\newblock
\showISBNx{9798400702419}
\href{https://doi.org/10.1145/3604915.3608889}{doi:\nolinkurl{10.1145/3604915.3608889}}


\bibitem[Espín-Noboa and Barolo(2025)]%
        {llmscholarbench2025data}
\bibfield{author}{\bibinfo{person}{Lisette Espín-Noboa} {and}
  \bibinfo{person}{Daniele Barolo}.} \bibinfo{year}{2025}\natexlab{}.
\newblock \bibinfo{booktitle}{\emph{{LLMScholarBench -- Benchmark \&
  Intervention Audits (datasets)}}}.
\newblock
\href{https://doi.org/10.5281/zenodo.20417106}{doi:\nolinkurl{10.5281/zenodo.20417106}}


\bibitem[Espín-Noboa and Barolo(2026)]%
        {llmscholarbench2025}
\bibfield{author}{\bibinfo{person}{Lisette Espín-Noboa} {and}
  \bibinfo{person}{Daniele Barolo}.} \bibinfo{year}{2026}\natexlab{}.
\newblock \bibinfo{booktitle}{\emph{{LLMScholarBench}: A Benchmark for Auditing
  LLM-Based Scholar Recommendation}}.
\newblock
\href{https://doi.org/10.5281/zenodo.20415692}{doi:\nolinkurl{10.5281/zenodo.20415692}}


\bibitem[Fan(2026)]%
        {fan2026comparison}
\bibfield{author}{\bibinfo{person}{Zhuhao Fan}.}
  \bibinfo{year}{2026}\natexlab{}.
\newblock \showarticletitle{A Comparison of Open-Source and Proprietary Large
  Language Models}.
\newblock \bibinfo{journal}{\emph{Internet Economics XIX}}
  (\bibinfo{year}{2026}), \bibinfo{pages}{18}.
\newblock


\bibitem[Feng et~al\mbox{.}(2025)]%
        {feng2025citybench}
\bibfield{author}{\bibinfo{person}{Jie Feng}, \bibinfo{person}{Jun Zhang},
  \bibinfo{person}{Tianhui Liu}, \bibinfo{person}{Xin Zhang},
  \bibinfo{person}{Tianjian Ouyang}, \bibinfo{person}{Junbo Yan},
  \bibinfo{person}{Yuwei Du}, \bibinfo{person}{Siqi Guo}, {and}
  \bibinfo{person}{Yong Li}.} \bibinfo{year}{2025}\natexlab{}.
\newblock \showarticletitle{CityBench: Evaluating the Capabilities of Large
  Language Models for Urban Tasks}. In \bibinfo{booktitle}{\emph{Proceedings of
  the 31st ACM SIGKDD Conference on Knowledge Discovery and Data Mining V.2
  (KDD '25)}} (Toronto, ON, Canada). \bibinfo{publisher}{Association for
  Computing Machinery}, \bibinfo{address}{New York, NY, USA},
  \bibinfo{numpages}{12}~pages.
\newblock
\href{https://doi.org/10.1145/3711896.3737375}{doi:\nolinkurl{10.1145/3711896.3737375}}


\bibitem[Guo et~al\mbox{.}(2025)]%
        {guo2025large}
\bibfield{author}{\bibinfo{person}{Yanzhu Guo}, \bibinfo{person}{Simone Conia},
  \bibinfo{person}{Zelin Zhou}, \bibinfo{person}{Min Li},
  \bibinfo{person}{Saloni Potdar}, {and} \bibinfo{person}{Henry Xiao}.}
  \bibinfo{year}{2025}\natexlab{}.
\newblock \showarticletitle{Do large language models have an English accent?
  evaluating and improving the naturalness of multilingual LLMs}. In
  \bibinfo{booktitle}{\emph{Proceedings of the 63rd Annual Meeting of the
  Association for Computational Linguistics (Volume 1: Long Papers)}}.
  \bibinfo{publisher}{Association for Computational Linguistics},
  \bibinfo{pages}{3823--3838}.
\newblock
\urldef\tempurl%
\url{https://aclanthology.org/2025.acl-long.193.pdf}
\showURL{%
\tempurl}


\bibitem[Jiang et~al\mbox{.}(2025a)]%
        {jiang2025beyond}
\bibfield{author}{\bibinfo{person}{Chumeng Jiang}, \bibinfo{person}{Jiayin
  Wang}, \bibinfo{person}{Weizhi Ma}, \bibinfo{person}{Charles L.~A. Clarke},
  \bibinfo{person}{Shuai Wang}, \bibinfo{person}{Chuhan Wu}, {and}
  \bibinfo{person}{Min Zhang}.} \bibinfo{year}{2025}\natexlab{a}.
\newblock \showarticletitle{Beyond Utility: Evaluating LLM as Recommender}. In
  \bibinfo{booktitle}{\emph{Proceedings of the ACM on Web Conference 2025}}
  (Sydney NSW, Australia) \emph{(\bibinfo{series}{WWW '25})}.
  \bibinfo{publisher}{Association for Computing Machinery},
  \bibinfo{address}{New York, NY, USA}, \bibinfo{pages}{3850–3862}.
\newblock
\showISBNx{9798400712746}
\href{https://doi.org/10.1145/3696410.3714759}{doi:\nolinkurl{10.1145/3696410.3714759}}


\bibitem[Jiang et~al\mbox{.}(2025b)]%
        {jiang2025hibench}
\bibfield{author}{\bibinfo{person}{Zhuohang Jiang}, \bibinfo{person}{Pangjing
  Wu}, \bibinfo{person}{Ziran Liang}, \bibinfo{person}{Peter~Q. Chen},
  \bibinfo{person}{Xu Yuan}, \bibinfo{person}{Ye Jia},
  \bibinfo{person}{Jiancheng Tu}, \bibinfo{person}{Chen Li},
  \bibinfo{person}{Peter H.~F. Ng}, {and} \bibinfo{person}{Qing Li}.}
  \bibinfo{year}{2025}\natexlab{b}.
\newblock \showarticletitle{HiBench: Benchmarking LLMs Capability on
  Hierarchical Structure Reasoning}. In \bibinfo{booktitle}{\emph{Proceedings
  of the 31st ACM SIGKDD Conference on Knowledge Discovery and Data Mining V.2
  (KDD '25)}} (Toronto, ON, Canada). \bibinfo{publisher}{Association for
  Computing Machinery}, \bibinfo{address}{New York, NY, USA},
  \bibinfo{numpages}{11}~pages.
\newblock
\href{https://doi.org/10.1145/3711896.3737378}{doi:\nolinkurl{10.1145/3711896.3737378}}


\bibitem[Johns and Dye(2019)]%
        {johns2019gender}
\bibfield{author}{\bibinfo{person}{Brendan~T Johns} {and}
  \bibinfo{person}{Melody Dye}.} \bibinfo{year}{2019}\natexlab{}.
\newblock \showarticletitle{Gender bias at scale: Evidence from the usage of
  personal names}.
\newblock \bibinfo{journal}{\emph{Behavior research methods}}
  \bibinfo{volume}{51}, \bibinfo{number}{4} (\bibinfo{year}{2019}),
  \bibinfo{pages}{1601--1618}.
\newblock


\bibitem[Kaplan et~al\mbox{.}(2020)]%
        {kaplan2020scaling}
\bibfield{author}{\bibinfo{person}{Jared Kaplan}, \bibinfo{person}{Sam
  McCandlish}, \bibinfo{person}{Tom Henighan}, \bibinfo{person}{Tom~B. Brown},
  \bibinfo{person}{Benjamin Chess}, \bibinfo{person}{Rewon Child},
  \bibinfo{person}{Scott Gray}, \bibinfo{person}{Alec Radford},
  \bibinfo{person}{Jeffrey Wu}, {and} \bibinfo{person}{Dario Amodei}.}
  \bibinfo{year}{2020}\natexlab{}.
\newblock \showarticletitle{Scaling Laws for Neural Language Models}.
\newblock  (\bibinfo{year}{2020}).
\newblock
\href{https://doi.org/10.48550/arXiv.2001.08361}{doi:\nolinkurl{10.48550/arXiv.2001.08361}}


\bibitem[Kojima et~al\mbox{.}(2022)]%
        {kojima2022large}
\bibfield{author}{\bibinfo{person}{Takeshi Kojima},
  \bibinfo{person}{Shixiang~Shane Gu}, \bibinfo{person}{Machel Reid},
  \bibinfo{person}{Yutaka Matsuo}, {and} \bibinfo{person}{Yusuke Iwasawa}.}
  \bibinfo{year}{2022}\natexlab{}.
\newblock \showarticletitle{Large language models are zero-shot reasoners}. In
  \bibinfo{booktitle}{\emph{Proceedings of the 36th International Conference on
  Neural Information Processing Systems}} (New Orleans, LA, USA)
  \emph{(\bibinfo{series}{NIPS '22})}. \bibinfo{publisher}{Curran Associates
  Inc.}, \bibinfo{address}{Red Hook, NY, USA}, Article
  \bibinfo{articleno}{1613}, \bibinfo{numpages}{15}~pages.
\newblock
\showISBNx{9781713871088}
\urldef\tempurl%
\url{https://dl.acm.org/doi/10.5555/3600270.3601883}
\showURL{%
\tempurl}


\bibitem[Kong et~al\mbox{.}(2022)]%
        {kong2022influence}
\bibfield{author}{\bibinfo{person}{Hyunsik Kong}, \bibinfo{person}{Samuel
  Martin-Gutierrez}, {and} \bibinfo{person}{Fariba Karimi}.}
  \bibinfo{year}{2022}\natexlab{}.
\newblock \showarticletitle{Influence of the first-mover advantage on the
  gender disparities in physics citations}.
\newblock \bibinfo{journal}{\emph{Communications Physics}} \bibinfo{volume}{5},
  \bibinfo{number}{1} (\bibinfo{date}{Oct.} \bibinfo{year}{2022}),
  \bibinfo{pages}{243}.
\newblock
\showISSN{2399-3650}
\href{https://doi.org/10.1038/s42005-022-00997-x}{doi:\nolinkurl{10.1038/s42005-022-00997-x}}


\bibitem[Lahoti et~al\mbox{.}(2023)]%
        {lahoti2023improving}
\bibfield{author}{\bibinfo{person}{Preethi Lahoti}, \bibinfo{person}{Nicholas
  Blumm}, \bibinfo{person}{Xiao Ma}, \bibinfo{person}{Raghavendra
  Kotikalapudi}, \bibinfo{person}{Sahitya Potluri}, \bibinfo{person}{Qijun
  Tan}, \bibinfo{person}{Hansa Srinivasan}, \bibinfo{person}{Ben Packer},
  \bibinfo{person}{Ahmad Beirami}, \bibinfo{person}{Alex Beutel}, {and}
  \bibinfo{person}{Jilin Chen}.} \bibinfo{year}{2023}\natexlab{}.
\newblock \showarticletitle{Improving Diversity of Demographic Representation
  in Large Language Models via Collective-Critiques and Self-Voting}. In
  \bibinfo{booktitle}{\emph{Proceedings of the 2023 Conference on Empirical
  Methods in Natural Language Processing}}. \bibinfo{publisher}{Association for
  Computational Linguistics}, \bibinfo{address}{Singapore},
  \bibinfo{pages}{10383--10405}.
\newblock
\href{https://doi.org/10.18653/v1/2023.emnlp-main.643}{doi:\nolinkurl{10.18653/v1/2023.emnlp-main.643}}


\bibitem[Laohaprapanon et~al\mbox{.}(2022)]%
        {ethnicolr}
\bibfield{author}{\bibinfo{person}{Suriyan Laohaprapanon},
  \bibinfo{person}{Gaurav Sood}, {and} \bibinfo{person}{Bashar Naji}.}
  \bibinfo{year}{2022}\natexlab{}.
\newblock \bibinfo{title}{{ethnicolr: Predict Race and Ethnicity From Name}}.
\newblock
\urldef\tempurl%
\url{https://github.com/appeler/ethnicolr}
\showURL{%
\tempurl}


\bibitem[Lerman et~al\mbox{.}(2022)]%
        {lerman2022gendered}
\bibfield{author}{\bibinfo{person}{Kristina Lerman}, \bibinfo{person}{Yulin
  Yu}, \bibinfo{person}{Fred Morstatter}, {and} \bibinfo{person}{Jay Pujara}.}
  \bibinfo{year}{2022}\natexlab{}.
\newblock \showarticletitle{Gendered citation patterns among the scientific
  elite}.
\newblock \bibinfo{journal}{\emph{Proceedings of the National Academy of
  Sciences}} \bibinfo{volume}{119}, \bibinfo{number}{40}
  (\bibinfo{year}{2022}), \bibinfo{pages}{e2206070119}.
\newblock
\href{https://doi.org/10.1073/pnas.2206070119}{doi:\nolinkurl{10.1073/pnas.2206070119}}


\bibitem[Lewis et~al\mbox{.}(2020)]%
        {lewis2020rag}
\bibfield{author}{\bibinfo{person}{Patrick Lewis}, \bibinfo{person}{Ethan
  Perez}, \bibinfo{person}{Aleksandra Piktus}, \bibinfo{person}{Fabio Petroni},
  \bibinfo{person}{Vladimir Karpukhin}, \bibinfo{person}{Naman Goyal},
  \bibinfo{person}{Heinrich K\"{u}ttler}, \bibinfo{person}{Mike Lewis},
  \bibinfo{person}{Wen-tau Yih}, \bibinfo{person}{Tim Rockt\"{a}schel},
  \bibinfo{person}{Sebastian Riedel}, {and} \bibinfo{person}{Douwe Kiela}.}
  \bibinfo{year}{2020}\natexlab{}.
\newblock \showarticletitle{Retrieval-augmented generation for
  knowledge-intensive NLP tasks}. In \bibinfo{booktitle}{\emph{Proceedings of
  the 34th International Conference on Neural Information Processing Systems}}
  (Vancouver, BC, Canada) \emph{(\bibinfo{series}{NIPS '20})}.
  \bibinfo{publisher}{Curran Associates Inc.}, \bibinfo{address}{Red Hook, NY,
  USA}, Article \bibinfo{articleno}{793}, \bibinfo{numpages}{16}~pages.
\newblock
\showISBNx{9781713829546}


\bibitem[Li et~al\mbox{.}(2025a)]%
        {li2024banishing}
\bibfield{author}{\bibinfo{person}{Johnny Li}, \bibinfo{person}{Saksham
  Consul}, \bibinfo{person}{Eda Zhou}, \bibinfo{person}{James Wong},
  \bibinfo{person}{Naila Farooqui}, \bibinfo{person}{Yuxin Ye},
  \bibinfo{person}{Nithyashree Manohar}, \bibinfo{person}{Zhuxiaona Wei},
  \bibinfo{person}{Tian Wu}, \bibinfo{person}{Ben Echols},
  \bibinfo{person}{Sharon Zhou}, {and} \bibinfo{person}{Gregory Diamos}.}
  \bibinfo{year}{2025}\natexlab{a}.
\newblock \showarticletitle{Banishing LLM Hallucinations Requires Rethinking
  Generalization}.
\newblock  (\bibinfo{year}{2025}).
\newblock
\href{https://doi.org/10.48550/arXiv.2406.17642}{doi:\nolinkurl{10.48550/arXiv.2406.17642}}


\bibitem[Li et~al\mbox{.}(2025b)]%
        {li2025exploring}
\bibfield{author}{\bibinfo{person}{Lujun Li}, \bibinfo{person}{Lama Sleem},
  \bibinfo{person}{Niccolo’ Gentile}, \bibinfo{person}{Geoffrey Nichil},
  {and} \bibinfo{person}{Radu State}.} \bibinfo{year}{2025}\natexlab{b}.
\newblock \showarticletitle{Exploring the Impact of Temperature on Large
  Language Models: Hot or Cold?}
\newblock \bibinfo{journal}{\emph{Procedia Computer Science}}
  \bibinfo{volume}{264} (\bibinfo{year}{2025}), \bibinfo{pages}{242--251}.
\newblock
\showISSN{1877-0509}
\href{https://doi.org/10.1016/j.procs.2025.07.135}{doi:\nolinkurl{10.1016/j.procs.2025.07.135}}
\newblock
\shownote{International Neural Network Society Workshop on Deep Learning
  Innovations and Applications 2025}.


\bibitem[Liang and Acuna(2021)]%
        {liang2021demographicx}
\bibfield{author}{\bibinfo{person}{L. Liang} {and} \bibinfo{person}{D.E.
  Acuna}.} \bibinfo{year}{2021}\natexlab{}.
\newblock \bibinfo{title}{demographicx: A Python package for estimating gender
  and ethnicity using deep learning transformers}.
\newblock
  \bibinfo{howpublished}{\url{https://github.com/sciosci/demographicx}}.
\newblock


\bibitem[Liao et~al\mbox{.}(2024)]%
        {liao2024llms}
\bibfield{author}{\bibinfo{person}{Zhehui Liao}, \bibinfo{person}{Maria
  Antoniak}, \bibinfo{person}{Inyoung Cheong}, \bibinfo{person}{Evie Yu-Yen
  Cheng}, \bibinfo{person}{Ai-Heng Lee}, \bibinfo{person}{Kyle Lo},
  \bibinfo{person}{Joseph~Chee Chang}, {and} \bibinfo{person}{Amy~X. Zhang}.}
  \bibinfo{year}{2024}\natexlab{}.
\newblock \showarticletitle{LLMs as Research Tools: A Large Scale Survey of
  Researchers' Usage and Perceptions}.
\newblock  (\bibinfo{year}{2024}).
\newblock
\href{https://doi.org/10.48550/arXiv.2411.05025}{doi:\nolinkurl{10.48550/arXiv.2411.05025}}


\bibitem[Liu et~al\mbox{.}(2023)]%
        {liu2023pretrainpromptpredict}
\bibfield{author}{\bibinfo{person}{Pengfei Liu}, \bibinfo{person}{Weizhe Yuan},
  \bibinfo{person}{Jinlan Fu}, \bibinfo{person}{Zhengbao Jiang},
  \bibinfo{person}{Hiroaki Hayashi}, {and} \bibinfo{person}{Graham Neubig}.}
  \bibinfo{year}{2023}\natexlab{}.
\newblock \showarticletitle{Pre-train, Prompt, and Predict: A Systematic Survey
  of Prompting Methods in Natural Language Processing}.
\newblock \bibinfo{journal}{\emph{ACM Comput. Surv.}} \bibinfo{volume}{55},
  \bibinfo{number}{9}, Article \bibinfo{articleno}{195} (\bibinfo{date}{Jan.}
  \bibinfo{year}{2023}), \bibinfo{numpages}{35}~pages.
\newblock
\showISSN{0360-0300}
\href{https://doi.org/10.1145/3560815}{doi:\nolinkurl{10.1145/3560815}}


\bibitem[Liu et~al\mbox{.}(2025)]%
        {liu2025unequal}
\bibfield{author}{\bibinfo{person}{Yixuan Liu}, \bibinfo{person}{Abel Elekes},
  \bibinfo{person}{Jianglin Lu}, \bibinfo{person}{Rodrigo Dorantes-Gilardi},
  {and} \bibinfo{person}{Albert-Laszlo Barabasi}.}
  \bibinfo{year}{2025}\natexlab{}.
\newblock \showarticletitle{Unequal Scientific Recognition in the Age of
  {LLM}s}. In \bibinfo{booktitle}{\emph{Findings of the Association for
  Computational Linguistics: EMNLP 2025}}. \bibinfo{publisher}{Association for
  Computational Linguistics}, \bibinfo{address}{Suzhou, China},
  \bibinfo{pages}{23558--23568}.
\newblock
\showISBNx{979-8-89176-335-7}
\href{https://doi.org/10.18653/v1/2025.findings-emnlp.1279}{doi:\nolinkurl{10.18653/v1/2025.findings-emnlp.1279}}


\bibitem[MacNell et~al\mbox{.}(2015)]%
        {macnell2015s}
\bibfield{author}{\bibinfo{person}{Lillian MacNell}, \bibinfo{person}{Adam
  Driscoll}, {and} \bibinfo{person}{Andrea~N Hunt}.}
  \bibinfo{year}{2015}\natexlab{}.
\newblock \showarticletitle{What’s in a name: Exposing gender bias in student
  ratings of teaching}.
\newblock \bibinfo{journal}{\emph{Innovative Higher Education}}
  \bibinfo{volume}{40}, \bibinfo{number}{4} (\bibinfo{year}{2015}),
  \bibinfo{pages}{291--303}.
\newblock


\bibitem[Maries et~al\mbox{.}(2024)]%
        {Maries_2025}
\bibfield{author}{\bibinfo{person}{Alexandru Maries},
  \bibinfo{person}{Yangquiting Li}, {and} \bibinfo{person}{Chandralekha
  Singh}.} \bibinfo{year}{2024}\natexlab{}.
\newblock \showarticletitle{Challenges faced by women and persons excluded
  because of their ethnicity and race in physics learning environments: review
  of the literature and recommendations for departments and instructors}.
\newblock \bibinfo{journal}{\emph{Reports on Progress in Physics}}
  \bibinfo{volume}{88}, \bibinfo{number}{1} (\bibinfo{date}{dec}
  \bibinfo{year}{2024}), \bibinfo{pages}{015901}.
\newblock
\href{https://doi.org/10.1088/1361-6633/ad91c4}{doi:\nolinkurl{10.1088/1361-6633/ad91c4}}


\bibitem[Marvin et~al\mbox{.}(2024)]%
        {Ggaliwango2024}
\bibfield{author}{\bibinfo{person}{Ggaliwango Marvin},
  \bibinfo{person}{Nakayiza Hellen}, \bibinfo{person}{Daudi Jjingo}, {and}
  \bibinfo{person}{Joyce Nakatumba-Nabende}.} \bibinfo{year}{2024}\natexlab{}.
\newblock \showarticletitle{Prompt Engineering in Large Language Models}. In
  \bibinfo{booktitle}{\emph{Data Intelligence and Cognitive Informatics}}.
  \bibinfo{publisher}{Springer Nature Singapore}, \bibinfo{address}{Singapore},
  \bibinfo{pages}{387--402}.
\newblock
\showISBNx{978-981-99-7962-2}


\bibitem[Merton(1968)]%
        {merton1968matthew}
\bibfield{author}{\bibinfo{person}{Robert~K. Merton}.}
  \bibinfo{year}{1968}\natexlab{}.
\newblock \showarticletitle{The Matthew Effect in Science}.
\newblock \bibinfo{journal}{\emph{Science}} \bibinfo{volume}{159},
  \bibinfo{number}{3810} (\bibinfo{year}{1968}), \bibinfo{pages}{56--63}.
\newblock
\href{https://doi.org/10.1126/science.159.3810.56}{doi:\nolinkurl{10.1126/science.159.3810.56}}


\bibitem[Meyer et~al\mbox{.}(2023)]%
        {meyer2023chatgpt}
\bibfield{author}{\bibinfo{person}{Jesse~G. Meyer}, \bibinfo{person}{Ryan~J.
  Urbanowicz}, \bibinfo{person}{Patrick C.~N. Martin}, \bibinfo{person}{Karen
  O{'}Connor}, \bibinfo{person}{Ruowang Li}, \bibinfo{person}{Pei-Chen Peng},
  \bibinfo{person}{Tiffani~J. Bright}, \bibinfo{person}{Nicholas Tatonetti},
  \bibinfo{person}{Kyoung~Jae Won}, \bibinfo{person}{Graciela
  Gonzalez-Hernandez}, {and} \bibinfo{person}{Jason~H. Moore}.}
  \bibinfo{year}{2023}\natexlab{}.
\newblock \showarticletitle{ChatGPT and large language models in academia:
  opportunities and challenges}.
\newblock \bibinfo{journal}{\emph{BioData Mining}} \bibinfo{volume}{16},
  \bibinfo{number}{1} (\bibinfo{date}{July} \bibinfo{year}{2023}),
  \bibinfo{pages}{20}.
\newblock
\showISSN{1756-0381}
\href{https://doi.org/10.1186/s13040-023-00339-9}{doi:\nolinkurl{10.1186/s13040-023-00339-9}}


\bibitem[Naddaf(2026)]%
        {naddafmore}
\bibfield{author}{\bibinfo{person}{Miryam Naddaf}.}
  \bibinfo{year}{2026}\natexlab{}.
\newblock \showarticletitle{More than half of researchers now use AI for peer
  review—often against guidance}.
\newblock \bibinfo{journal}{\emph{Nature}} \bibinfo{volume}{649},
  \bibinfo{number}{8096} (\bibinfo{year}{2026}), \bibinfo{pages}{273--274}.
\newblock
\href{https://doi.org/10.1038/d41586-025-04066-5}{doi:\nolinkurl{10.1038/d41586-025-04066-5}}


\bibitem[Nakano et~al\mbox{.}(2022)]%
        {nakano2021webgpt}
\bibfield{author}{\bibinfo{person}{Reiichiro Nakano}, \bibinfo{person}{Jacob
  Hilton}, \bibinfo{person}{Suchir Balaji}, \bibinfo{person}{Jeff Wu},
  \bibinfo{person}{Long Ouyang}, \bibinfo{person}{Christina Kim},
  \bibinfo{person}{Christopher Hesse}, \bibinfo{person}{Shantanu Jain},
  \bibinfo{person}{Vineet Kosaraju}, \bibinfo{person}{William Saunders},
  \bibinfo{person}{Xu Jiang}, \bibinfo{person}{Karl Cobbe},
  \bibinfo{person}{Tyna Eloundou}, \bibinfo{person}{Gretchen Krueger},
  \bibinfo{person}{Kevin Button}, \bibinfo{person}{Matthew Knight},
  \bibinfo{person}{Benjamin Chess}, {and} \bibinfo{person}{John Schulman}.}
  \bibinfo{year}{2022}\natexlab{}.
\newblock \showarticletitle{WebGPT: Browser-assisted question-answering with
  human feedback}.
\newblock  (\bibinfo{year}{2022}).
\newblock
\href{https://doi.org/10.48550/arXiv.2112.09332}{doi:\nolinkurl{10.48550/arXiv.2112.09332}}


\bibitem[Paruschke and Philipps(2025)]%
        {paruschke2025hidden}
\bibfield{author}{\bibinfo{person}{Laura Paruschke} {and} \bibinfo{person}{Axel
  Philipps}.} \bibinfo{year}{2025}\natexlab{}.
\newblock \showarticletitle{Hidden in the light: Scientists’ online presence
  on institutional websites and professional networking sites}.
\newblock \bibinfo{journal}{\emph{Journal of Information Science}}
  \bibinfo{volume}{51}, \bibinfo{number}{2} (\bibinfo{year}{2025}),
  \bibinfo{pages}{324--337}.
\newblock


\bibitem[Pierson et~al\mbox{.}(2025)]%
        {pierson2023use}
\bibfield{author}{\bibinfo{person}{Emma Pierson}, \bibinfo{person}{Divya
  Shanmugam}, \bibinfo{person}{Rajiv Movva}, \bibinfo{person}{Jon Kleinberg},
  \bibinfo{person}{Monica Agrawal}, \bibinfo{person}{Mark Dredze},
  \bibinfo{person}{Kadija Ferryman}, \bibinfo{person}{Judy~Wawira Gichoya},
  \bibinfo{person}{Dan Jurafsky}, \bibinfo{person}{Pang~Wei Koh},
  {et~al\mbox{.}}} \bibinfo{year}{2025}\natexlab{}.
\newblock \showarticletitle{Using large language models to promote health
  equity}.
\newblock \bibinfo{journal}{\emph{NEJM AI}} \bibinfo{volume}{2},
  \bibinfo{number}{2} (\bibinfo{year}{2025}), \bibinfo{pages}{8}.
\newblock
\href{https://doi.org/10.1056/AIp2400889}{doi:\nolinkurl{10.1056/AIp2400889}}


\bibitem[Pop et~al\mbox{.}(2024)]%
        {pop2024rethinking}
\bibfield{author}{\bibinfo{person}{Florin Pop}, \bibinfo{person}{Judd
  Rosenblatt}, \bibinfo{person}{Diogo~Schwerz de Lucena}, {and}
  \bibinfo{person}{Michael Vaiana}.} \bibinfo{year}{2024}\natexlab{}.
\newblock \showarticletitle{Rethinking harmless refusals when fine-tuning
  foundation models}.
\newblock  (\bibinfo{year}{2024}).
\newblock
\href{https://doi.org/10.48550/arXiv.2406.19552}{doi:\nolinkurl{10.48550/arXiv.2406.19552}}


\bibitem[Priem et~al\mbox{.}(2022)]%
        {priem2022openalex}
\bibfield{author}{\bibinfo{person}{Jason Priem}, \bibinfo{person}{Heather
  Piwowar}, {and} \bibinfo{person}{Richard Orr}.}
  \bibinfo{year}{2022}\natexlab{}.
\newblock \showarticletitle{OpenAlex: A fully-open index of scholarly works,
  authors, venues, institutions, and concepts}.
\newblock  (\bibinfo{year}{2022}).
\newblock
\href{https://doi.org/10.48550/arXiv.2205.01833}{doi:\nolinkurl{10.48550/arXiv.2205.01833}}


\bibitem[Raj et~al\mbox{.}(2025)]%
        {raj2024breaking}
\bibfield{author}{\bibinfo{person}{Chahat Raj}, \bibinfo{person}{Anjishnu
  Mukherjee}, \bibinfo{person}{Aylin Caliskan}, \bibinfo{person}{Antonios
  Anastasopoulos}, {and} \bibinfo{person}{Ziwei Zhu}.}
  \bibinfo{year}{2025}\natexlab{}.
\newblock \showarticletitle{Breaking Bias, Building Bridges: Evaluation and
  Mitigation of Social Biases in LLMs via Contact Hypothesis}. In
  \bibinfo{booktitle}{\emph{Proceedings of the 2024 AAAI/ACM Conference on AI,
  Ethics, and Society}} (San Jose, California, USA)
  \emph{(\bibinfo{series}{AIES '24})}. \bibinfo{publisher}{AAAI Press},
  \bibinfo{pages}{1180–1189}.
\newblock


\bibitem[Reimers and Gurevych(2019)]%
        {reimers-2019-sentence-bert}
\bibfield{author}{\bibinfo{person}{Nils Reimers} {and} \bibinfo{person}{Iryna
  Gurevych}.} \bibinfo{year}{2019}\natexlab{}.
\newblock \showarticletitle{{Sentence-BERT}: {Sentence embeddings using siamese
  bert-networks}}. In \bibinfo{booktitle}{\emph{Proceedings of the 2019
  conference on empirical methods in natural language processing and the 9th
  international joint conference on natural language processing
  (EMNLP-IJCNLP)}}. \bibinfo{publisher}{Association for Computational
  Linguistics}, \bibinfo{pages}{3982--3992}.
\newblock


\bibitem[Rosa and Mensah(2016)]%
        {rosa2016educational}
\bibfield{author}{\bibinfo{person}{Katemari Rosa} {and}
  \bibinfo{person}{Felicia~Moore Mensah}.} \bibinfo{year}{2016}\natexlab{}.
\newblock \showarticletitle{Educational pathways of Black women physicists:
  Stories of experiencing and overcoming obstacles in life}.
\newblock \bibinfo{journal}{\emph{Phys. Rev. Phys. Educ. Res.}}
  \bibinfo{volume}{12} (\bibinfo{date}{Aug} \bibinfo{year}{2016}),
  \bibinfo{pages}{020113}.
\newblock
Issue 2.
\href{https://doi.org/10.1103/PhysRevPhysEducRes.12.020113}{doi:\nolinkurl{10.1103/PhysRevPhysEducRes.12.020113}}


\bibitem[Samoilenko and Yasseri(2014)]%
        {samoilenko2014distorted}
\bibfield{author}{\bibinfo{person}{Anna Samoilenko} {and} \bibinfo{person}{Taha
  Yasseri}.} \bibinfo{year}{2014}\natexlab{}.
\newblock \showarticletitle{The distorted mirror of Wikipedia: a quantitative
  analysis of Wikipedia coverage of academics}.
\newblock \bibinfo{journal}{\emph{EPJ data science}} \bibinfo{volume}{3},
  \bibinfo{number}{1} (\bibinfo{year}{2014}), \bibinfo{pages}{1}.
\newblock


\bibitem[Sancheti et~al\mbox{.}(2024)]%
        {sancheti2024llm}
\bibfield{author}{\bibinfo{person}{Prateek Sancheti},
  \bibinfo{person}{Kamalakar Karlapalem}, {and} \bibinfo{person}{Kavita
  Vemuri}.} \bibinfo{year}{2024}\natexlab{}.
\newblock \showarticletitle{LLM Driven Web Profile Extraction for Identical
  Names}. In \bibinfo{booktitle}{\emph{Companion Proceedings of the ACM Web
  Conference 2024}} (Singapore, Singapore) \emph{(\bibinfo{series}{WWW '24})}.
  \bibinfo{publisher}{Association for Computing Machinery},
  \bibinfo{address}{New York, NY, USA}, \bibinfo{pages}{1616–1625}.
\newblock
\showISBNx{9798400701726}
\href{https://doi.org/10.1145/3589335.3651946}{doi:\nolinkurl{10.1145/3589335.3651946}}


\bibitem[Sandnes(2024)]%
        {sandnes2024can}
\bibfield{author}{\bibinfo{person}{Frode~Eika Sandnes}.}
  \bibinfo{year}{2024}\natexlab{}.
\newblock \showarticletitle{Can we identify prominent scholars using ChatGPT?}
\newblock \bibinfo{journal}{\emph{Scientometrics}} \bibinfo{volume}{129},
  \bibinfo{number}{1} (\bibinfo{date}{Jan.} \bibinfo{year}{2024}),
  \bibinfo{pages}{713--718}.
\newblock
\showISSN{1588-2861}
\href{https://doi.org/10.1007/s11192-023-04882-4}{doi:\nolinkurl{10.1007/s11192-023-04882-4}}


\bibitem[Sax et~al\mbox{.}(2016)]%
        {sax2016women}
\bibfield{author}{\bibinfo{person}{Linda~J. Sax}, \bibinfo{person}{Kathleen~J.
  Lehman}, \bibinfo{person}{Ram\'on~S. Barthelemy}, {and}
  \bibinfo{person}{Gloria Lim}.} \bibinfo{year}{2016}\natexlab{}.
\newblock \showarticletitle{Women in physics: A comparison to science,
  technology, engineering, and math education over four decades}.
\newblock \bibinfo{journal}{\emph{Phys. Rev. Phys. Educ. Res.}}
  \bibinfo{volume}{12} (\bibinfo{date}{Aug} \bibinfo{year}{2016}),
  \bibinfo{pages}{020108}.
\newblock
Issue 2.
\href{https://doi.org/10.1103/PhysRevPhysEducRes.12.020108}{doi:\nolinkurl{10.1103/PhysRevPhysEducRes.12.020108}}


\bibitem[Schick et~al\mbox{.}(2023)]%
        {schick2023toolformer}
\bibfield{author}{\bibinfo{person}{Timo Schick}, \bibinfo{person}{Jane
  Dwivedi{-}Yu}, \bibinfo{person}{Roberto Dess{\`i}}, \bibinfo{person}{Roberta
  Raileanu}, \bibinfo{person}{Maria Lomeli}, \bibinfo{person}{Luke
  Zettlemoyer}, \bibinfo{person}{Nicola Cancedda}, {and}
  \bibinfo{person}{Thomas Scialom}.} \bibinfo{year}{2023}\natexlab{}.
\newblock \showarticletitle{Toolformer: Language Models Can Teach Themselves to
  Use Tools}. In \bibinfo{booktitle}{\emph{Advances in Neural Information
  Processing Systems (NeurIPS)}}.
\newblock


\bibitem[Shah(2025)]%
        {shah2024prompt}
\bibfield{author}{\bibinfo{person}{Chirag Shah}.}
  \bibinfo{year}{2025}\natexlab{}.
\newblock \showarticletitle{From Prompt Engineering to Prompt Science with
  Humans in the Loop}.
\newblock \bibinfo{journal}{\emph{Commun. ACM}} \bibinfo{volume}{68},
  \bibinfo{number}{6} (\bibinfo{date}{June} \bibinfo{year}{2025}),
  \bibinfo{pages}{54–61}.
\newblock
\showISSN{0001-0782}
\href{https://doi.org/10.1145/3709599}{doi:\nolinkurl{10.1145/3709599}}


\bibitem[Troshin et~al\mbox{.}(2025)]%
        {troshin2025control}
\bibfield{author}{\bibinfo{person}{Sergey Troshin}, \bibinfo{person}{Wafaa
  Mohammed}, \bibinfo{person}{Yan Meng}, \bibinfo{person}{Christof Monz},
  \bibinfo{person}{Antske Fokkens}, {and} \bibinfo{person}{Vlad Niculae}.}
  \bibinfo{year}{2025}\natexlab{}.
\newblock \showarticletitle{Control the Temperature: Selective Sampling for
  Diverse and High-Quality LLM Outputs}.
\newblock  (\bibinfo{year}{2025}).
\newblock
\href{https://doi.org/10.48550/arXiv.2510.01218}{doi:\nolinkurl{10.48550/arXiv.2510.01218}}


\bibitem[Vargas-Parada(2025)]%
        {vargaslarge}
\bibfield{author}{\bibinfo{person}{Laura Vargas-Parada}.}
  \bibinfo{year}{2025}\natexlab{}.
\newblock \showarticletitle{Large language models are biased-local initiatives
  are fighting for change}.
\newblock \bibinfo{journal}{\emph{Nature}} (\bibinfo{year}{2025}).
\newblock
\href{https://doi.org/10.1038/d41586-025-03891-y}{doi:\nolinkurl{10.1038/d41586-025-03891-y}}


\bibitem[V{\'a}s{\'a}rhelyi and Horv{\'a}t(2023)]%
        {vasarhelyi2023benefits}
\bibfield{author}{\bibinfo{person}{Orsolya V{\'a}s{\'a}rhelyi} {and}
  \bibinfo{person}{Em{\H{o}}ke-{\'A}gnes Horv{\'a}t}.}
  \bibinfo{year}{2023}\natexlab{}.
\newblock \showarticletitle{Who Benefits from Altmetrics? The Effect of Team
  Gender Composition on the Link Between Online Visibility and Citation
  Impact}. In \bibinfo{booktitle}{\emph{Proceedings of the 19th International
  Conference of the International Society for Scientometrics and Informetrics
  (ISSI 2023)}}. \bibinfo{address}{Bloomington, Indiana, USA}.
\newblock


\bibitem[Vlasceanu and Amodio(2022)]%
        {vlasceanu2022propagation}
\bibfield{author}{\bibinfo{person}{Mihaela Vlasceanu} {and}
  \bibinfo{person}{David~M. Amodio}.} \bibinfo{year}{2022}\natexlab{}.
\newblock \showarticletitle{Propagation of societal gender inequality by
  internet search algorithms}.
\newblock \bibinfo{journal}{\emph{Proceedings of the National Academy of
  Sciences}} \bibinfo{volume}{119}, \bibinfo{number}{29}
  (\bibinfo{year}{2022}), \bibinfo{pages}{e2204529119}.
\newblock


\bibitem[von Hippel and Buck(2023)]%
        {vonHippel2023improve}
\bibfield{author}{\bibinfo{person}{Paul~T. von Hippel} {and}
  \bibinfo{person}{Stephanie Buck}.} \bibinfo{year}{2023}\natexlab{}.
\newblock \showarticletitle{Improve academic search engines to reduce scholars'
  biases}.
\newblock \bibinfo{journal}{\emph{Nature Human Behaviour}} \bibinfo{volume}{7},
  \bibinfo{number}{2} (\bibinfo{year}{2023}), \bibinfo{pages}{157--158}.
\newblock


\bibitem[Waltman and van Eck(2012)]%
        {waltman2012inconsistency}
\bibfield{author}{\bibinfo{person}{Ludo Waltman} {and}
  \bibinfo{person}{Nees~Jan van Eck}.} \bibinfo{year}{2012}\natexlab{}.
\newblock \showarticletitle{The inconsistency of the h-index}.
\newblock \bibinfo{journal}{\emph{Journal of the American Society for
  Information Science and Technology}} \bibinfo{volume}{63},
  \bibinfo{number}{2} (\bibinfo{year}{2012}), \bibinfo{pages}{406--415}.
\newblock
\href{https://doi.org/10.1002/asi.21678}{doi:\nolinkurl{10.1002/asi.21678}}


\bibitem[Wang et~al\mbox{.}(2024)]%
        {wang2024human}
\bibfield{author}{\bibinfo{person}{Xinru Wang}, \bibinfo{person}{Hannah Kim},
  \bibinfo{person}{Sajjadur Rahman}, \bibinfo{person}{Kushan Mitra}, {and}
  \bibinfo{person}{Zhengjie Miao}.} \bibinfo{year}{2024}\natexlab{}.
\newblock \showarticletitle{Human-llm collaborative annotation through
  effective verification of llm labels}. In
  \bibinfo{booktitle}{\emph{Proceedings of the 2024 CHI Conference on Human
  Factors in Computing Systems}}. \bibinfo{pages}{1--21}.
\newblock


\bibitem[Wang et~al\mbox{.}(2017)]%
        {wang2017efficient}
\bibfield{author}{\bibinfo{person}{Yaoshu Wang}, \bibinfo{person}{Jianbin Qin},
  {and} \bibinfo{person}{Wei Wang}.} \bibinfo{year}{2017}\natexlab{}.
\newblock \showarticletitle{Efficient Approximate Entity Matching Using
  Jaro-Winkler Distance}. In \bibinfo{booktitle}{\emph{Web Information Systems
  Engineering -- WISE 2017}}. \bibinfo{publisher}{Springer International
  Publishing}, \bibinfo{address}{Cham}, \bibinfo{pages}{231--239}.
\newblock
\showISBNx{978-3-319-68783-4}


\bibitem[Wold et~al\mbox{.}(1987)]%
        {wold1987principal}
\bibfield{author}{\bibinfo{person}{Svante Wold}, \bibinfo{person}{Kim
  Esbensen}, {and} \bibinfo{person}{Paul Geladi}.}
  \bibinfo{year}{1987}\natexlab{}.
\newblock \showarticletitle{Principal component analysis}.
\newblock \bibinfo{journal}{\emph{Chemometrics and Intelligent Laboratory
  Systems}} \bibinfo{volume}{2}, \bibinfo{number}{1} (\bibinfo{year}{1987}),
  \bibinfo{pages}{37--52}.
\newblock
\showISSN{0169-7439}
\href{https://doi.org/10.1016/0169-7439(87)80084-9}{doi:\nolinkurl{10.1016/0169-7439(87)80084-9}}
\newblock
\shownote{Proceedings of the Multivariate Statistical Workshop for Geologists
  and Geochemists}.


\bibitem[Yao et~al\mbox{.}(2023)]%
        {yao2023react}
\bibfield{author}{\bibinfo{person}{Shunyu Yao}, \bibinfo{person}{Jeffrey Zhao},
  \bibinfo{person}{Dian Yu}, \bibinfo{person}{Nan Du}, \bibinfo{person}{Izhak
  Shafran}, \bibinfo{person}{Karthik Narasimhan}, {and} \bibinfo{person}{Yuan
  Cao}.} \bibinfo{year}{2023}\natexlab{}.
\newblock \showarticletitle{ReAct: Synergizing Reasoning and Acting in Language
  Models}.
\newblock  (\bibinfo{year}{2023}).
\newblock
\href{https://doi.org/10.48550/arXiv.2210.03629}{doi:\nolinkurl{10.48550/arXiv.2210.03629}}


\bibitem[Zhang et~al\mbox{.}(2024)]%
        {bao2023large}
\bibfield{author}{\bibinfo{person}{Jizhi Zhang}, \bibinfo{person}{Keqin Bao},
  \bibinfo{person}{Yang Zhang}, \bibinfo{person}{Wenjie Wang},
  \bibinfo{person}{Fuli Feng}, {and} \bibinfo{person}{Xiangnan He}.}
  \bibinfo{year}{2024}\natexlab{}.
\newblock \showarticletitle{Large Language Models for Recommendation:
  Progresses and Future Directions}. In \bibinfo{booktitle}{\emph{Companion
  Proceedings of the ACM Web Conference 2024}} (Singapore, Singapore)
  \emph{(\bibinfo{series}{WWW '24})}. \bibinfo{publisher}{Association for
  Computing Machinery}, \bibinfo{address}{New York, NY, USA},
  \bibinfo{pages}{1268–1271}.
\newblock
\showISBNx{9798400701726}
\href{https://doi.org/10.1145/3589335.3641247}{doi:\nolinkurl{10.1145/3589335.3641247}}


\bibitem[Zhou et~al\mbox{.}(2022)]%
        {zhou2022prompt}
\bibfield{author}{\bibinfo{person}{Chunting Zhou}, \bibinfo{person}{Junxian
  He}, \bibinfo{person}{Xuezhe Ma}, \bibinfo{person}{Taylor Berg-Kirkpatrick},
  {and} \bibinfo{person}{Graham Neubig}.} \bibinfo{year}{2022}\natexlab{}.
\newblock \showarticletitle{Prompt Consistency for Zero-Shot Task
  Generalization}.
\newblock  (\bibinfo{date}{Dec.} \bibinfo{year}{2022}),
  \bibinfo{pages}{2613--2626}.
\newblock
\href{https://doi.org/10.18653/v1/2022.findings-emnlp.192}{doi:\nolinkurl{10.18653/v1/2022.findings-emnlp.192}}


\bibitem[Zhu et~al\mbox{.}(2023)]%
        {zhu2024hot}
\bibfield{author}{\bibinfo{person}{Yuqi Zhu}, \bibinfo{person}{Jia Li},
  \bibinfo{person}{Ge Li}, \bibinfo{person}{YunFei Zhao}, \bibinfo{person}{Jia
  Li}, \bibinfo{person}{Zhi Jin}, {and} \bibinfo{person}{Hong Mei}.}
  \bibinfo{year}{2023}\natexlab{}.
\newblock \showarticletitle{Hot or Cold? Adaptive Temperature Sampling for Code
  Generation with Large Language Models}.
\newblock
\href{https://doi.org/10.48550/arXiv.2309.02772}{doi:\nolinkurl{10.48550/arXiv.2309.02772}}


\end{thebibliography}
\end{document}